\documentclass[useAMS,usenatbib,aas_macros]{mn2e} 
\usepackage{subfigure}
\usepackage{graphicx}
\usepackage{epsfig,times,aas_macros}
\usepackage{amsbsy}
\usepackage{amsmath}
\usepackage{amssymb}
\usepackage[T1]{fontenc}
\usepackage{aecompl}

\newcommand{\equ}[1]{eq.~(\ref{eq:#1})}
\newcommand{\equs}[1]{eqs.~(\ref{eq:#1})}

\newcommand{\equnp}[1]{eq.~\ref{eq:#1}}
\newcommand{\se}[1]{\S\ref{sec:#1}}
\newcommand{\fig}[1]{Fig.~\ref{fig:#1}}

\newcommand{\Fig}[1]{Figure~\ref{fig:#1}}

\newcommand{\tab}[1]{Table~\ref{tab:#1}}
\newcommand{\be}{\begin{equation}}
\newcommand{\ee}{\end{equation}}
\newcommand{\bea}{\begin{eqnarray}}
\newcommand{\eea}{\end{eqnarray}}

\newcommand{\msun}{{\rm M}_\odot}
\newcommand{\Msun}{M_\odot}

\newcommand{\ifm}[1]{\relax\ifmmode#1\else$\mathsurround=0pt #1$\fi}
\newcommand{\kms}{\ifmmode\,{\rm km}\,{\rm s}^{-1}\else km$\,$s$^{-1}$\fi}

\newcommand{\kpc}{\,{\rm kpc}}
\newcommand{\pc}{\,{\rm pc}}
\newcommand{\Gyr}{\,{\rm Gyr}}

\newcommand{\Myr}{\,{\rm Myr}}

\newcommand{\ltsima}{$\; \buildrel < \over \sim \;$}
\newcommand{\lsim}{\lower.5ex\hbox{\ltsima}}
\newcommand{\gtsima}{$\; \buildrel > \over \sim \;$}
\newcommand{\gsim}{\lower.5ex\hbox{\gtsima}}
\newcommand{\prop}{\propto}

\def\la{\langle}
\def\ra{\rangle}

\def\omm{\Omega_{\rm m}}
\def\oml{\Omega_{\Lambda}}
\def\omb{\Omega_{\rm b}}

\def\cmc{\,{\rm cm}^{-3}}
\def\cms{\,{\rm cm}^{-2}}

\def\Mv{M_{\rm v}}
\def\Rv{R_{\rm v}}
\def\Vv{V_{\rm v}}
\def\Tv{T_{\rm v}}
\def\Ts{T_{\rm s}}
\def\Mg{M_{\rm g}}

\def\Rd{R_{\rm d}}

\def\Ms{M_*}

\def\tv{t_{\rm v}}
\def\tc{t_{\rm torq}}

\def\rd0{r_{{\rm d}0}}
\def\Vdi0{V_{{\rm d}0}}

\def\Jd{J_{\rm d}}
\def\Jneg{J_{-}}
\def\torb{t_{\rm orb}}

\usepackage{color}

\begin{document}

\setlength{\voffset}{-2.2cm}

\large
%Large

\pagerange{\pageref{firstpage}--\pageref{lastpage}} \pubyear{2002}

\title[Angular-momentum buildup in galaxies]
{ 
Four phases of angular-momentum buildup in high-z galaxies:\\ 
from cosmic-web streams through an extended ring to disc and bulge
}

\author[Danovich et al.]
{\parbox[t]{\textwidth}
{
Mark Danovich$^1$ \thanks{E-mail: mark.danovich@gmail.com},
Avishai Dekel$^1$\thanks{E-mail: dekel@huji.ac.il},
Oliver Hahn$^2$,
Daniel Ceverino$^3$,
Joel Primack$^4$
}
\\ \\ 
$^1$Center for Astrophysics and Planetary Science, Racah Institute of Physics,
The Hebrew University, Jerusalem 91904 Israel;\\
$^2$Institute for Astronomy, Department of Physics, ETH Zürich, CH-8093 Zurich,
Switzerland\\
$^3$Departamento de Fisica Teorica, Universidad Autonoma de Madrid, 
Madrid E-28049, Spain\\
$^4$Department of Physics, University of California, Santa Cruz, CA, 95064, USA
}
\date{}

\pagerange{\pageref{firstpage}--\pageref{lastpage}} \pubyear{0000}

\maketitle

\label{firstpage}

\begin{abstract}
We study the angular-momentum (AM) buildup in high-$z$ massive galaxies using high-resolution cosmological simulations.  The AM originates in 
co-planar streams of cold gas and merging galaxies tracing cosmic-web
filaments, and it undergoes four phases of evolution.  (I) Outside the halo 
virial radius ($\Rv\!\sim\!100\kpc$), the elongated streams gain AM by tidal 
torques with a specific AM (sAM) $\sim\!1.7$ times the dark-matter (DM) spin
due to the gas' higher quadrupole moment.  This AM is expressed as stream 
impact parameters, from $\sim\!0.3\Rv$ to counter rotation. (II) In
the outer halo, while the incoming DM mixes with the existing halo of lower sAM
to a spin $\lambda_{\rm dm}\!\sim\!0.04$, the cold streams transport the AM to
the inner halo such that their spin in the halo is $\sim\!3\lambda_{\rm dm}$.
(III) Near pericenter, the streams dissipate into an irregular 
rotating ring extending to $\sim\!0.3\Rv$ and tilted relative to the inner 
disc.  Torques exerted partly by the disc make the ring gas lose AM, spiral in,
and settle into the disc within one orbit.  The ring is observable with 30\% 
probability as a damped Lyman-$\alpha$ absorber.  (IV) Within the disc,
$<\!0.1\Rv$, torques associated with violent disc instability drive AM out and 
baryons into a central bulge, while outflows remove low-spin gas,
introducing certain sensitivity to feedback strength. 
Despite the different AM histories of gas and DM, the disc spin is comparable 
to the DM-halo spin. 
Counter rotation can strongly affect disc evolution.
\end{abstract}


%%%%%%%%%%%%%%%%%%%%%%%%%%%%% 1
\section{Introduction}
\label{sec:intro}

% 1
\smallskip
The acquisition of angular momentum (AM)
by galaxies is of fundamental importance for
the understanding of galaxy properties and evolution. 
Being supported by rotation, the size, structure and dynamics of galactic 
discs, as well as the star formation in them,
are predominantly determined by their AM and its distribution.
In turn, the formation of spheroids is driven by AM losses.

% TTT 2
\smallskip
The basic understanding of the origin of AM in galaxies 
is that, as the proto-galactic density perturbations grow, they
acquire AM through the gravitational tidal torques exerted by neighboring 
perturbations \citep{hoyle51,peebles69,doroshkevich70}.
This has been clarified and put together by \citet{white84} 
as the standard linear tidal-torque theory (TTT).
According to TTT, the AM of the proto-haloes, dominated by dark matter, 
grow linearly with time until they reach maximum expansion followed by 
collapse to virial equilibrium
\citep[quantified by simulations, e.g.,][]{porciani02a,porciani02b}.
The AM is predicted from the initial conditions by the asymmetric product 
of the tidal tensor $T_{ij}$, 
determined by the neighboring fluctuation field, and the inertia tensor 
$I_{ij}$, characterizing the Lagrangian spatial distribution of the 
mass in the proto-halo. 

% lambda 3
\smallskip
The total AM of the halo is characterized by the dimensionless spin parameter
\citep[as defined by][]{bullock01_j}
\be
\lambda = \frac{j}{\sqrt{2}\Rv\Vv} \, ,
\label{eq:lambda}
\ee
where $j\!=\!J/M$ is the magnitude of the specific angular momentum (sAM), 
namely the AM per unit mass within the halo, 
and $\Rv$ and $\Vv$ are the virial radius and velocity of 
the halo, respectively.\footnote{In the Einstein-deSitter (EdS) phase, 
the virial radius and velocity are related to the virial mass $\Mv$ as 
${\Mv}_{12}\!\simeq\!(1+z)_3^{-3/2} 
{\Vv}_{200}^3\!\simeq\! (1+z)_3^3 {\Rv}_{100}^3$,
where ${\Mv}_{12}\!=\!\Mv/10^{12}\msun$, ${\Vv}_{200}\!=\!\Vv/200\kms$,
${\Rv}_{100}\!=\!\Rv/100\kpc$, and $(1+z)_3\!=\!(1+z)/3$.}
To within the factor of $\sqrt{2}$,\footnote{The factor
$\sqrt{2}$ makes this spin parameter coincide with the earlier definition
by \citet{peebles69}, $\lambda\!=\!J |E|^{1/2} G^{-1} M^{-5/2}$ 
(where $E$ is the
energy), in the case of a singular isothermal sphere truncated at $\Rv$.}  
$\lambda$ measures the AM with respect to the maximum possible AM
had all the mass been in a coherent circular orbit with a velocity 
$\Vv$ at $\Rv$.
TTT predicts that the typical value of $\lambda$ for dark-matter haloes
is universal, independent of halo mass and redshift (see below). 
Cosmological N-body simulations \citep{bullock01_j} confirm this universality,
and reveal that the distribution of $\lambda$ among haloes is well fitted by 
a log-normal distribution, with a mean $\la \lambda \ra \!=\! 0.035 \pm 0.005$ 
and a standard deviation $\sigma \!=\! 0.50 \pm 0.03$ in natural log.
A complementary way to view the AM history of DM haloes is
through their buildup by mergers \citep[e.g.,][]{maller_somer02}.

% baryons 4
\smallskip
In the standard picture, the baryons were expected to initially follow the
dark-matter mass distribution and share the same sAM distribution.
This seemed natural in the case of a joint semi-spherical infall from outside 
the halo, and where the gas within the halo goes through a phase where it is
pressure-supported after being heated to the virial temperature by a virial 
shock \citep{ro77,silk77,binney77}.
The hot gas is then assumed to cool radiatively, contract to the bottom of the
spherical (or at least cylindrically symmetric) potential well while 
conserving AM, and form
a central rotation-supported disc, whose density profile is dictated by the
original distribution of AM within the halo 
\citep{fe80,fall83,mmw98,bullock01_j}.  % wr78?
This has been the basis for most analytic and semi-analytic modeling of galaxy
formation.
The predictions based on the simplified models are in the ball park of observed
AM in low-redshift galaxies \citep[e.g.,][]{romanowsky12,fall13}.

% new paradigm:  cosmic web, cold streams  5
\smallskip
The developing new picture takes into account the fact that massive galaxies, 
in their most active phase of assembly and star formation at redshifts 
$z\!=\!1\!-\!4$, 
form at the nodes of the cosmic web, where filaments intersect.
This is a generic feature for massive galaxies at high redshift, as they
emerge from high-sigma peaks of the density field.
The evolution of the baryons in this framework has been tracked by 
high-resolution cosmological hydro-gravitational simulations, 
and understood by simple physics
\citep{bd03,keres05,db06,cattaneo06,keres09,ocvirk08,dekel09}.
The baryons flow into the galaxies as streams that initially follow the 
central cords of the dark-matter filaments. 
The streams of cold gas, at a few times $10^4$K, which at high redshift are
narrow and dense compared to the halo, penetrate cold through 
the virial radius \citep{dekel13}, even if the halo is massive enough 
to support a shock in the rest of the virial shell
\citep[$\Mv \gtrsim 10^{12}\msun$,][]{db06}.
In massive galaxies at high redshift, the streams also bring in  
the gas-rich merging galaxies, as a population of ex-situ 
clumps with a broad range of masses \citep[e.g.,][]{dekel09}.  

% messy region 6
\smallskip
In the inner halo, typically inside $\sim\!0.3\Rv$, the simulations reveal that
the streams enter a ``messy" interface region where they show a complex 
structure and kinematics \citep[][Fig.~3]{cdb10}
before they eventually join the central disc, 
typically inside a radius smaller than $0.1\Rv$.
The nature and origin of this messy region is not yet understood,
and it may involve break up and heating \citep{nelson13},
potentially by a combination of hydrodynamical, thermal and gravitational
instabilities, shocks within the supersonic streams, and interactions of 
streams with other streams, with substructures, with the inner disc, and with
outflows.
The way the coherently inflowing supersonic streams eventually join the 
rotating disc is an open question of great interest, which we attempt to 
explore in this paper.

% VDI, spheroids 7
\smallskip
Being continuously fed by cold streams, the gas-rich disc tends to develop 
a violent
disc instability (VDI), where the disc is turbulent and highly perturbed,
with large transient features and in-situ giant clumps 
%\citep{dsc09,agertz09,cdb10}. 
\citep{noguchi99,immeli04a, % immeli04b,
genzel06,bournaud07c,genzel08,
dsc09,agertz09,cdb10,ceverino12,mandelker14}.
Torques within the disc drive gas inflow and inward clump migration,
which compactify the disc to drive central star formation in the form of a 
``blue nugget" \citep{barro13,barro14a,db14,zolotov15},
grow a massive rotating stellar bulge 
\citep{genzel06,genzel08,bournaud11,cdb10,ceverino14_e},
and feed the central black hole \citep{bournaud11}. 

% AM by streams 8
\smallskip
The acquisition of AM by galaxies within this developing framework for galaxy
formation is the challenge that we attempt to address here,
following preliminary attempts \citep{pichon11,danovich12,codis12,stewart13}.
The details of how galaxies are fed by elongated streams 
that originate in the cosmic web provide an explicit setting where
TTT can be realized, possibly in a modified form. 
Then, the very non-spherical inflow through streams, their counter-rotating
component, the interaction in the 
messy region at the greater disc environment, and the VDI processes in the 
disc, should all play major roles in the chain of events that lead to
the AM buildup of galaxies, and they should be thoroughly investigated. 
In particular, one wonders to what extent the final AM in the disc is expected
to differ from that implied by the simplified model based on spherical gas
contraction conserving AM.

% Danovich 12
\smallskip
\citet{danovich12} used 1kpc-resolution cosmological hydro simulations
with hundreds of massive galaxies to study the way they are fed
at high-$z$ by streams, focusing on the halo scales. 
The simulations reveal that the streams tend to be confined to a plane, the
``stream plane", which extends from inside the halo to a few virial radii.
The stream plane is weakly (anti-) correlated with the disc plane, consistent
with TTT (see below). 
On average, 70\% of the influx through the virial radius is found to be
in narrow streams,
67\% of it in one dominant stream, and 95\% in 3 major streams.
It turns out that typically 87\% of the AM is carried by one dominant stream,
the one with the highest product of mass inflow rate and impact parameter.
This explains why the plane defined by the AM at $\Rv$ 
%(the ``virial AM plane")
is only weakly correlated with the stream plane that involves the other 
streams.
The virial AM plane itself is also found to be only weakly correlated with 
the inner
disc plane, indicating a significant exchange of AM in the inner halo.
This happens within the greater environment of the disc, the ``AM sphere" 
of radius $\sim\!0.3\Rv$, which roughly coincides with the messy region 
seen in the simulations \citep[][Fig.~3]{cdb10}.

% AM in 4 steps
\smallskip
We propose here that the AM buildup in high-redshift galaxies
consists of four phases 
that represent both stages of a time sequence and different 
spatial zones.\footnote{We note that while the phases and the zones are
correlated, there is not always a one-to-one correspondence between them, 
and the boundaries between zones and between phases are not always sharp.}
Phase I is the linear TTT phase, occurring in the cosmic web outside the halo,
where the thin, elongated cold streams acquire more AM than the dark matter, 
expressed as impact parameters. 
Phase II is the transport phase, where the streams flow with a rather constant
velocity and mass-inflow rate and roughly along straight
lines down to $\sim\!0.3\Rv$.
Phase III is the strong-torque phase, in a tilted, inflowing, outer ``ring" at 
$(0.1\!-\!0.3)\Rv$.
Finally, 
phase IV involves the inner disc and the growing bulge, where the final 
rearrangement of AM occurs due to VDI torques and feedback-driven outflows.

% simulations
\smallskip
Here, we address this picture of AM buildup using a large suite of 
high-resolution zoom-in AMR 
hydro-cosmological simulations of massive galaxies at high redshift. 
The maximum resolution for the gas is $\sim\!50\pc$.
We focus here on the average behavior in the studied halo mass range 
($10^{11.5}\!-\!10^{12.5}$ at $z\!=\!2$) and redshift range ($z\!=\!4\!-\!1.5$).
We do not attempt to address the possible mass dependence and do not 
highlight the redshift dependence within these ranges -- those are deferred to
future work. Here we are after the generic behavior common to massive galaxies 
at high redshift, being fed by streams from the cosmic web.

% outline
\smallskip
The paper is organized as follows. 
In \se{method} we describe the cosmological simulations and the analysis of AM
in them. 
In \se{I} to \se{IV} we address each of the four phases,
following the different components of baryons and dark matter with a special
focus on the cold gas that dominates the supply to the growing disc.
In \se{feedback} we provide a preliminary discussion of the robustness of 
our proposed scenario of AM buildup to variations in the assumed feedback 
based on simulations with stronger feedback.
In \se{conc} we summarize and discuss the proposed picture, its theoretical
implications and its observational predictions.

%%%%%%%%%%%%%%%%%%%%%%%%%%%%%  2
\section{Method}
\label{sec:method}

\subsection{The cosmological simulations}  
\label{sec:sim}  % 2.1

% 1
Our cosmological simulations utilize the hydro-gravitational code ART
\citep{krav97,krav03} which uses adaptive mesh refinement (AMR)
to follow the Eulerian gas dynamics. The code implements sub-grid models of
the key physical processes relevant for galaxy formation. 
These include gas cooling by atomic hydrogen and helium, molecular hydrogen 
and metals, and  
photo-ionization heating by a UV background with partial self shielding.
Star formation is stochastic in cells with gas temperature $T\leq 10^4$K and
densities $n_H\geq 1 \cmc$, at a rate consistent with the Kennicutt-Schmidt
law \citep{kennicutt98}.
Stellar mass loss and metal enrichment of the ISM are included.
Feedback from stellar winds and supernovae is implemented by 
injecting thermal energy to the neighboring gas at a constant rate. 

% 2
\smallskip
The cosmological model adopted in the simulation is the standard 
$\Lambda$CDM model with the WMAP5 cosmological parameters 
($\omm\!=\!0.27$, $\oml\!=\!0.73$, $\omb\!=\!0.045$, $h\!=\!0.7$,
$\sigma_8\!=\!0.82$) 
\citep{komatsu08_wmap5}.
Individual haloes were selected at $z\!=\!1$ from an $N$-body dark-matter-only 
simulation of a large cosmological box. 
%$20$ and $40h^{-1}{\rm Mpc}$, and which 
Each halo and its environment were re-simulated at higher resolution with 
gas and the associated baryonic processes.
The dark matter in each halo, out to a few virial radii,
is typically represented by  
$\sim\! 10^7$ particles of mass $6.6\times 10^5 \Msun$ each. 
The particles representing stars have a minimum mass of $10^4 \Msun$.
The AMR cells in the dense regions are refined to a minimum size in the range 
$35-70 {\rm pc}$ at all times. 
The adaptive refinement algorithm is such that
a cell is divided to 8 cells once it contains a mass in stars and dark-matter 
more than $2\times 10^6 \Msun$, or a gas mass larger than 
$1.5\times 10^6 \Msun$.
The force resolution is $1\!-\!2$ grid cells of the maximum resolution.
Artificial fragmentation on the cell size is prevented by introducing
a pressure floor, which ensures that the Jeans scale is resolved by at least
7 cells \citep[see][]{cdb10}.
More details concerning the simulations are provided in 
\citet{ceverino+k09}, \citet{cdb10}, \citet{ceverino12},
\citet{dekel13} and \citet{mandelker14}.

\smallskip
Our sample consists of 29 galaxies in the redshift range $z\!=\!4\!-\!1.5$, 
listed in \tab{gals}. 
The virial masses at $z\!=\!2$ are in the range 
$10^{11.5}\!-\!10^{12.5}\msun$. 
The haloes were selected not to undergo a major major at or near $z\!\sim\!1$.
This turned out to eliminate about 10\% of the haloes in the selected mass 
range.
More details concerning the sample are listed in Table 1 of \citet{dekel13}
and discussed there.

\smallskip
\begin{table}
   \vspace{0.5 cm}
   \centering
   \begin{tabular}{c | c | c | c | c}
     \hline
     Galaxy&$\Mv$&$\Rv$&$\Rd$& $a_{\rm final}$ \\
           &$[10^{12}\msun]$&$[{\rm kpc}]$&$[{\rm kpc}]$& \\
     \hline
     MW1 &1.32&142&8.56&0.40 \\
     MW2 &0.95&109&4.46&0.34 \\
     MW3 &1.28&140&6.57&0.40 \\
     MW4 &1.99&164&5.73&0.40 \\
     MW5 &3.51&155&8.37&0.31 \\
     MW6 &0.78&119&6.50&0.40 \\
     MW7 &0.49&102&6.63&0.40 \\
     MW8 &0.29& 85&8.23&0.40 \\
     MW9 &0.34& 90&3.97&0.40 \\
     MW10&1.41&145&4.99&0.40 \\
     MW11&0.61&110&4.50&0.40 \\
     MW12&1.79&191&13.47&0.39 \\
     \hline
     SFG1&1.98&154&10.73&0.38 \\
     SFG4&1.31&191&10.16&0.38 \\
     SFG5&1.74&155&6.33&0.40 \\
     SFG8&1.42&128&7.64&0.35 \\
     SFG9&3.54&198&9.39&0.40 \\
     \hline
     VL01&1.89&149&3.53&0.37 \\
     VL02&0.93&126&9.99&0.40 \\
     VL03&1.16&114&6.71&0.32 \\
     VL04&1.20&137&4.23&0.40 \\
     VL05&2.08&165&12.23&0.40 \\
     VL06&0.82&121&6.46&0.40 \\
     VL07&1.63&130&6.00&0.34 \\
     VL08&1.13&135&5.91&0.40 \\
     VL09&0.54& 91&3.21&0.34 \\
     VL10&1.09&133&3.34&0.40 \\
     VL11&1.98&163&3.29&0.40 \\
     VL12&1.04&131&5.60&0.40 \\
         \hline
   \end{tabular}
\caption{
The sample of simulated galaxies used in the main analysis of the current
paper.
Specified are the galaxy name, the halo virial mass, virial radius and disc
radius at $a\!=\!0.34$ ($z\!=\!1$), 
and the expansion factor at the last snapshot 
used in the analysis, where $a\!=\!(1+z)^{-1}$.
More details are provided in \citet[][Table 1]{dekel13}.
}
\label{tab:gals}
\end{table}

\smallskip
 \begin{table}
   \vspace{0.5 cm}
   \centering
   \begin{tabular}{ccccccc}
     \hline
    Galaxy&$\Mv$&$\Ms$&$\Mg$&$\Rv$&$\Rd$&$z$ \\
        &$\!\![10^{12}\msun]\!\!$&$\!\!\![10^{11}\msun]\!\!\!$&
                            $\!\![10^{11}\msun]\!\!\!$&
            $\![{\rm kpc}]\!$&$\![{\rm kpc}]\!$&$ $ \\
     \hline
     MW1 &1.16&1.14&0.65&129& 4.5&1.63 \\
     MW3 &0.36&0.26&0.31& 70& 5.3&2.33 \\
     MW8 &0.23&0.22&0.12& 61& 4.3&2.33 \\
     SFG1&1.27&1.41&0.60& 86&10.0&3.17 \\
     SFG8&1.06&1.23&0.50& 86& 3.6&2.92 \\
     SFG8&1.22&1.38&0.60& 98& 5.9&2.57 \\
     SFG8&1.35&1.56&0.77&114& 6.8&2.17 \\
     VL01&0.97&1.10&0.49& 91& 5.1&2.57 \\
     VL01&1.01&1.16&0.52& 95& 6.5&2.45 \\
     VL02&1.08&1.27&0.59&141& 9.7&1.33 \\
     VL11&1.98&2.43&1.17&163& 8.8&1.50 \\
     \hline  
     VEL3&0.19&0.07&0.13& 74& 9.4&1.50 \\  
     VEL7&1.49&1.23&0.77&169&18.1&1.17 \\                  
	 \hline
   \end{tabular}
\caption{
Snapshots of galaxies used in the analysis in specific sections and figures. 
Each snapshot is identified by the galaxy name and the redshift,
and the quantities quoted are the virial mass $\Mv$ and radius $\Rv$,
as well as the stellar and gas mass within the galaxy, $\Ms$ and $\Mg$, 
and the disk radius $\Rd$.
The two snapshots identified as VEL are from a new suite of simulations
with higher resolution, lower SFR efficiency
and stronger feedback, that are used only 
in one illustrative picture and not in the quantitative analysis.
}
\label{tab:snaps}
 \end{table}

\smallskip % feedback      
Several limitations of the simulations used here, mostly in terms of how star
formation and feedback are implemented, are described in \se{feedback}. 
In order to evaluate the possible sensitivity of our results to feedback
strength, we briefly refer there to preliminary results from an improved suite
of simulations, with higher resolution, lower SFR efficiency, and
stronger feedback. These simulations fit observations better,
and will be analyzed in detail in a future paper. 
We find that the results reported here are qualitatively robust, with
non-negligible effects limited to the AM distribution within the inner disc.
In the visual study of gas in the inner halo, \se{III},
we make a very limited use of two snapshots from the new simulations,
which are listed at the bottom of \tab{snaps}.

%---------------------- 2.2
\subsection{Angular momentum of different components}

% Angular momentum 
The instantaneous position of the origin of the frame of reference
and the velocity rest-frame for each galaxy  
were determined by the center of mass of the inner stellar particles. 
The relevant inner sphere was chosen to contain at least 20 stellar 
particles within a radius of $130\pc$ or larger, and it has been positioned 
through an iterative procedure that maximizes the stellar density within it.
The positions and velocities of the gas, stars and
dark-matter particles were all corrected to this origin and rest frame,
and the AM is always measured in this frame with respect to the galaxy center. 
In order to test robustness,
the analysis has been repeated with alternative definitions of rest-frame,
such as the center-of-mass velocity of the gas within $\Rv$, 
yielding differences of order 10\% in spin parameter.
When referring to AM changes due to torques, 
we ignore the fact that the sequence of instantaneous
rest-frames do not define a proper inertial frame.

% components
\smallskip
Throughout the paper we consider the AM of different mass components,
distinguishing dark matter from baryons, and considering cold gas, 
hot gas, and stars.
The gas is defined as ``hot" if its temperature is higher than $10^5$K and
higher than the temperature of the shock-heated gas, $\Ts\!=\! 3\Tv/8$,
where $\Tv$ is the virial temperature,
$\Tv \!=\! 1.44\times10^5{\rm K}\,{\Vv}_{200}^2$
\citep{db06}. 
For the masses and redshifts analyzed,
this threshold temperature is typically a few times $10^5$K. 
The gas is defined as ``cold" once the gas temperature is below $10^5$K,
and the gas density is above a number density threshold of
$100\,\bar n_H$, where 
$\bar n_H \!\simeq\! 2\times 10^{-7} {\rm H}\cmc (1+z)^3$ % actually used 1.9
is the mean cosmological number density of hydrogen.
This density threshold was chosen such that the cold streams are clearly 
revealed at large distances by visual inspection 
\citep[also adopted by][]{kimm11}.
This threshold eliminates only $4\%$ of the cold gas within the virial radius,
and $44\%$ outside it, so it is effective mostly outside the virial radius.
The gas fraction with a temperature in the range between $10^5$K and $\Ts$
(when $\Ts\!>\!10^5$K), namely the gas that falls between the cold and hot 
components as defined here, is on average $9\%$ within $(0.1\!-\!1)\,\Rv$ 
and $22\%$ at $(1\!-\!2)\,\Rv$.
We find that the results not sensitive to moderate variations in the above 
temperature and density criteria.
In limited cases specified below, dense cold gas in clumps is eliminated above 
$10^3\bar n_H$ in order to reduce wild fluctuations.
There is a practical upper limit for the gas density above $\sim\! 1 \cmc$,
the built-in density threshold for star formation in the simulations,
though this is not a strict limit because the star formation is stochastic.

%%%%%%%%%%%%%%%%%%%%%%%%%%%%% 3
\section{Phase I: tidal torques in the cosmic web}
\label{sec:I}

%-------------------- 3.1
\subsection{Summary of Tidal Torques Theory}

The first stage in the origin of the AM in galaxies takes place when the
streams are part of the cosmic web before they enter the halo, typically 
at a distance of two to a few virial radii from the halo center.
The tidal-torque theory (TTT) for the origin of halo AM
has been developed by \citet{white84} based on the pioneering work of
\citet{peebles69} using the Zel'dovich approximation in the quasi-linear
regime, and has been tested using cosmological N-body simulations
\citep{porciani02a,porciani02b}.
The basic ideas of TTT are robust and could be still valid, 
possibly with some modifications,
so it is worth summarizing here before we proceed to the analysis of the
simulations.

% TTT   T*I
\smallskip
According to TTT, 
the final AM of the halo is determined in the proto-halo stage in the 
linear regime, by the gravitational coupling of the tidal field generated by
neighboring fluctuations and the quadrupole moment of the Lagrangian volume 
of the proto-halo, $\Gamma$.  
The spatial dependence of the components of the AM vector $\vec J$
are given by the antisymmetric tensor product
\be
J_i\propto\epsilon_{ijk}T_{jl}I_{lk} \, .
\label{eq:L_space}
\ee
The tensor $T_{jl}$ is the tidal tensor,\footnote{The tensor as expressed 
here is commonly termed the ``deformation" tensor, and its traceless part is 
referred to as the ``tidal", or ``shear" tensor.}
\be
T_{jl}=-\frac{\partial^2\phi}{\partial q_j \partial q_l} \, ,
\label{eq:T}
\ee
where $\phi$ is the gravitational potential and $\vec q$ refers to the 
Lagrangian, comoving coordinates. 
The tidal tensor is evaluated at the center-of-mass of $\Gamma$,
using a second-order Taylor expansion of the potential.
The tensor $I_{lk}$ is the inertia tensor of $\Gamma$,
\be
I_{lk}=\bar\rho\, a \int_\Gamma q_l q_k d^3q \, ,
\label{eq:I}
\ee
where $\bar\rho$ is the mean density at the time when the universal
expansion factor is $a$.
The traceless part of $I_{lk}$ is the quadrupolar inertia tensor.
It follows from \equ{L_space} that the traceless parts of these tensors
can be used without changing the resultant AM.
We learn from \equ{L_space} that
a net AM is generated due to the deviation of the proto-halo from spherical
symmetry and the misalignment of the eigenvectors of these two tensors, which
is found in simulations to be non-zero but rather small.
This is why the halo is far from being supported by rotation,
with a spin parameter of only a few percent.

\smallskip
In the principle coordinates of the inertia tensor, the AM 
components are given by
\be
J_1\! \propto\! T_{23}(I_3\!-\!I_2), \, 
J_2\! \propto\! T_{13}(I_1\!-\!I_3), \, 
J_3\! \propto\! T_{12}(I_2\!-\!I_1),  
\label{eq:quadrupole}
\ee
where $I_i$ are the corresponding eigenvalues ordered from large to small.
This implies that $J_2$, the AM component along the intermediate axis of the 
inertia tensor of $\Gamma$, for which $|I_1-I_3|$ refers to the difference 
between the major and minor eigenvalues, has the highest 
probability to be the largest AM component
\citep[see applications to disc alignment with large-scale structure 
in][]{navarro04,hahn10}.

% time growth
\smallskip
In the linear regime, as the perturbation associated with the proto-halo grows,
TTT predicts that the AM of the proto-halo grows in proportion to $a^2 D(a)$, 
where $D(a)$ is the growing mode of linear perturbations. 
In the Einstein-deSitter (EdS) regime, this growth is linear with time,
\be
J_i \propto a^2 D(a) \propto t \, . 
\label{eq:L_time}
\ee
It has been confirmed in $N$-body simulations \citep{porciani02a,porciani02b} 
that the acquisition of AM significantly slows down once the proto-halo 
reaches maximum expansion, at twice its current virial radius 
according to the spherical collapse model, after which it 
turns around and collapses to the virialized halo within the virial radius.
After maximum expansion, the tidal torques become inefficient due to
the declining moments of inertia, so the AM remains roughly fixed 
during the collapse and virialization. 

% universality of lambda
\smallskip
From the above basics of TTT one can derive the universality of
the halo spin parameter in terms of halo mass $\Mv$ and expansion
factor at virialization $a_{\rm v}$, as follows.
The tidal tensor in \equ{L_space} scales as $\nabla^2\phi$ in Lagrangian
space. 
The time of maximum expansion, corresponding to an expansion factor 
$a_{\rm max}$,
can be identified as the time when the linearly growing density fluctuation  
of the proto-halo, $\delta$, reaches a value of order unity.
Based on linear theory of fluctuation growth,
one can write $\delta \prop D(a) \nabla^2\phi$. 
Therefore, when $\delta\! \sim\! 1$, one has 
$\nabla^2\phi \prop D(a_{\rm max})^{-1}$. 
In the EdS regime this is $D^{-1} \prop a_{\rm max}^{-1} \prop a_{\rm v}^{-1}$.
The inertia tensor in \equ{L_space}, per unit mass, scales like $R^2$, 
where $R$ is the comoving size of $\Gamma$, namely $R^2 \propto \Mv^{2/3}$. 
Put together using \equs{L_space} and (\ref{eq:L_time}), one obtains for the
sAM in the halo $j \propto a_{\rm v}^{1/2} \Mv^{2/3}$.
Indeed, one can deduce from the standard virial relations that the same
scaling applies for the combination $\Rv\Vv$ by which $j$ should be divided
in order to evaluate $\lambda$ in \equ{lambda}.
The dark-matter halo spin $\lambda$ is therefore predicted to be independent of 
both $\Mv$ and cosmological time.
This has been confirmed using cosmological $N$-body simulations
\citep{bullock01_j,bett07}.

\begin{figure*} %1
\centering
\includegraphics[width=\linewidth,height=0.35\textheight]
{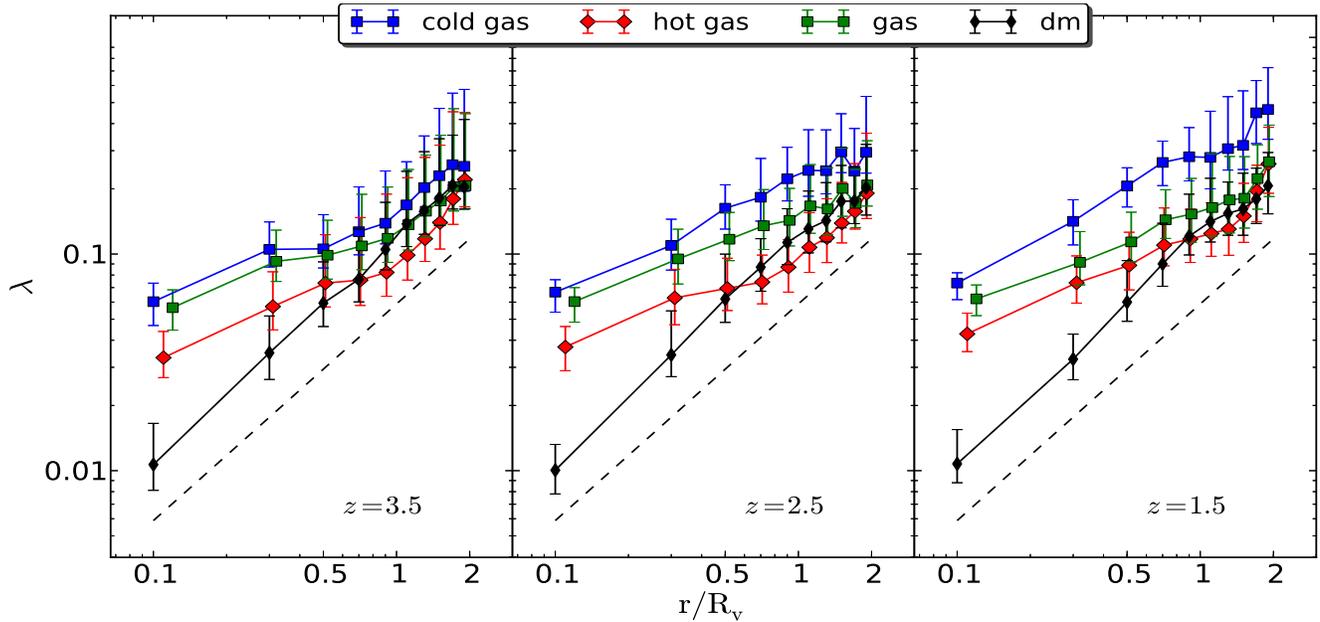}
\caption{
Radial profiles of the spin parameter,
$\lambda(r/\Rv)\!=\!j(r/\Rv)/(\sqrt{2}\Rv\Vv)$,
where $j$ is the specific AM in a shell of radius $r/\Rv$,
averaged over all haloes from \tab{gals} at three different redshifts.
The curves refer to the different mass components
of cold dense gas, hot gas, all the gas, and dark matter.
The sAM is computed in shells of thickness $0.1\Rv$.
Shown are the mean and standard deviation.
The dashed line are of slope unity.
In zone I and II,
the dark-matter spin profile is roughly linear with radius and self-similar in
time. The spin of cold gas is higher, by a factor of $1.3-2$ outside the halo
and by a factor of a few inside the halo.
}
\label{fig:spin_prof}
\end{figure*}

\begin{figure*} %2
\centering
\includegraphics[width=\linewidth]{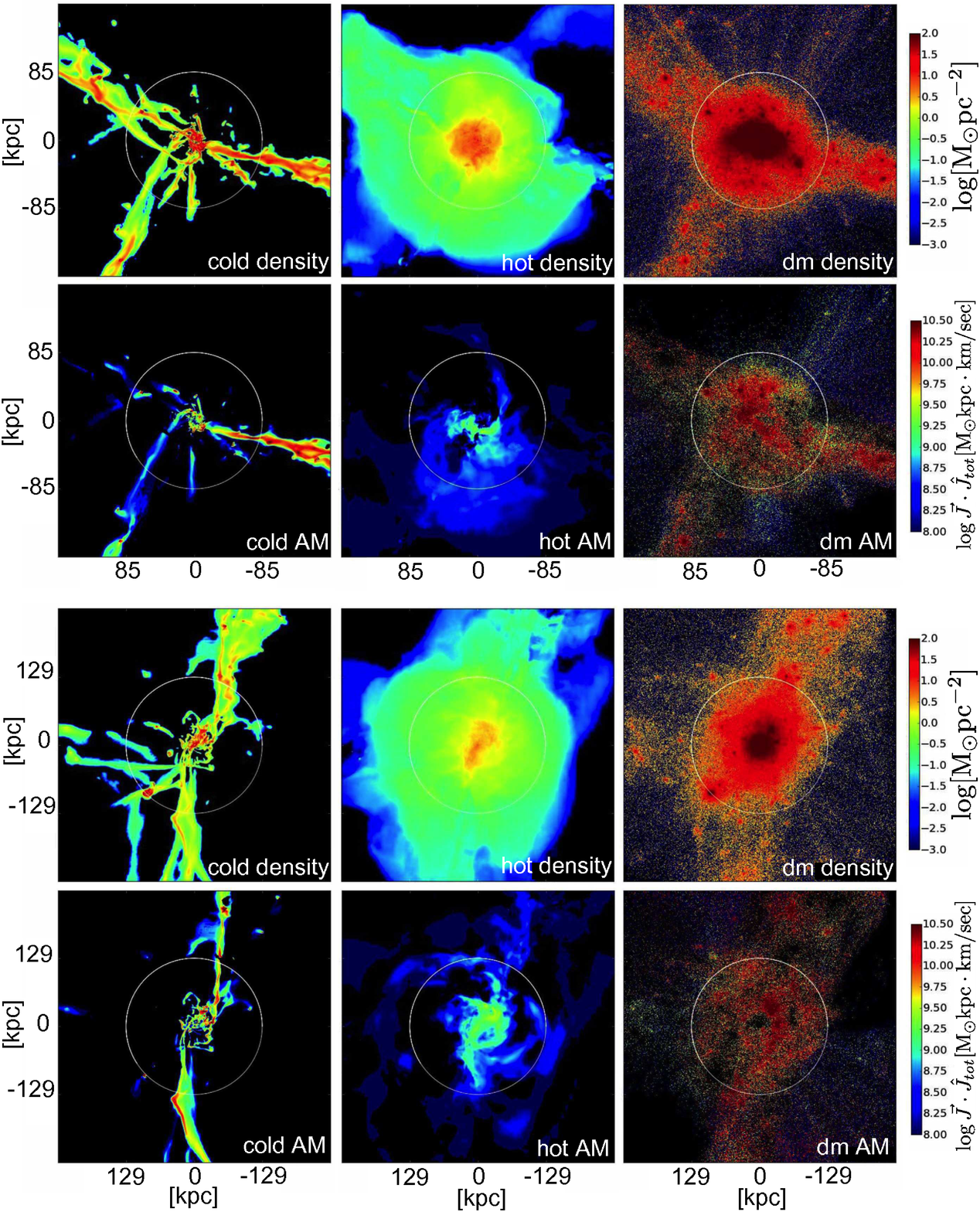}
\caption{
Surface density (top) and positive AM (bottom)
for the different mass components in two galaxies from \tab{snaps}
(top two rows SFG1 $z\!=\!3.17$, bottom two rows MW1 $z\!=\!1.63$). 
The components are cold gas (left), hot gas (middle) and dark matter (right).
The view is face-on in the disc plane.  
The cubic box is of side $2\Rv$, with $\Rv$ marked by the circle.
The positive AM component shown is in the direction of the total AM vector
of the given mass component within the halo. 
In zone I (and II)
the cold streams are much thiner and more coherent than the dark-matter
inflows.
}
\label{fig:proj_sfg1}
\end{figure*}

%----------- 3.2
\subsection{Dark-matter spin profile}

% profiles
\Fig{spin_prof} shows the spin-parameter radial profile (namely the sAM profile
$j(r)$ normalized by $\sqrt{2}\Rv\Vv$, \equnp{lambda}) 
averaged over all the simulated haloes. 
The sAM is computed in shells of thickness $0.1\Rv$ at
radii in the range $r\!=\!(0.1\!-\!2)\Rv$.
The profiles are shown for the different mass components at three different 
redshifts.

% DM profiles inside Rv
\smallskip
The dark-matter spin profiles show a universal linear shape at all times, 
roughly 
\be
j(r) \propto r \, , 
\ee
rising from $\lambda(r)\! \sim\! 0.011$ at $0.1\Rv$ to $0.22$ at $2\Rv$.
The self-similarity in time of the dark-matter spin distribution follows from 
TTT, as argued above.
The linear shape of the profile for the dark matter is consistent with 
earlier findings within virialized haloes using $N$-body simulations 
\citep{bullock01_j,bett10},
and with analytic predictions derived using a simple toy model that 
incorporates the combined effects of tidal stripping and dynamical friction 
on the merging sub-haloes that build the host halo as they spiral in, 
assuming a halo with a density profile
close to that of an isothermal sphere \citep{bullock01_j}. 
 
\smallskip
An interesting feature in \fig{spin_prof}
is that the same profile seems to extend from
inside the halo out to $2\Rv$ and possibly beyond. 
This may partly reflect the fact that the region where the dark matter has been
mixed actually extends beyond the formal virial radius defined by a given 
overdensity of $\sim\!200$ 
\citep{prada06}, so the tidal stripping and dynamical friction are active
already at larger radii, thus explaining the extension of the 
$j \propto r$ profile to beyond $\Rv$.
 
% Outside Rv TTT top hat
\smallskip
It is worth noting that a monotonically increasing $j(r)$ profile is expected 
already in the inflowing region outside $\Rv$ based on TTT.
This is because a shell that is now at an earlier stage of its infall
toward $\Rv$ started from a larger Lagrangian radius and thus had a larger 
quadrupole moment, and because it reached maximum expansion at a later time.
According to the spherical-collapse model, with the virial radius at half the
maximum expansion radius of the shell now crossing $\Rv$, 
a simple calculation shows that $j(2\Rv)/j(\Rv)\!\simeq\! 1.4$,
corresponding to an effective power-law fit of roughly $j(r)\!\propto\!r^{0.5}$ 
in the range $(1\!-\!2)\Rv$. 
This is an increase with $r$, but slightly shallower than what is seen in 
\fig{spin_prof}. 
However, there is indeed a hint for a shallowing of the profile outside 
$\Rv$ at the later redshifts. 
The deviation may hint for a partial failure of the spherical model to
account for the strong deviations from spherical symmetry associated with 
the inflowing dark-matter streams.
Alternatively, as suggested above,
it may reflect the effects of tidal stripping and dynamical
friction already outside the formal virial radius.  

\smallskip
We should point out that the rising $\lambda(r)$ outside the halo is not in 
conflict with the expected universal value for $\lambda$ within dark-matter 
haloes. 
This is because when a shell that is now at $r\!>\!\Rv$ will enter the virial 
radius with its larger $j$, the normalization factor $\Rv\Vv$ will also be 
larger due to the halo growth during that period, such that the obtained 
$\lambda$ within the halo will be the same as it is now.

\smallskip
We see in \fig{spin_prof} that
the spin parameter of the dark matter shell upon entry of the virial radius
is $\lambda\!\simeq\!0.13$ at all times. We will see that the average over the
whole halo is $\lambda\!\simeq\!0.04$, similar to the findings in 
previous studies \citep{bullock01_j,bett07}.

%------------------ 3.3
\subsection{Cold gas outside the halo}

% cold lambda higher than DM
The most interesting finding from 
\fig{spin_prof} is that outside the virial radius the spin parameter 
of the cold gas is higher than that of the dark matter, by a factor of $1.5-2$,
a factor that grows with cosmic time.
A similar phenomenon has been detected in other simulations
\citep{pichon11,kimm11,powell11,tillson12,stewart13}.
The spin parameter of the cold gas upon entry through the virial radius
varies from $\lambda\!\simeq\!0.2$ at $z\!=\!3.5$ to $\lambda\!\simeq\!0.3$ at
$z\!=\!1.5$, compared to the $0.13$ of the dark matter. 
This trend continues beyond $2\Rv$ (beyond the radii shown in the figure).
The origin of this, to be addressed next,
could in principle be due to gravitational torques, 
involving different quadrupole moments for the two components while the tidal
tensor acting on both is similar and so is its misalignment with the inertia 
tensors of the two components. Alternatively, the difference between the
spins of cold gas and dark matter could possibly be due to 
pressure torques that act on the gas only (see \se{pressure} below).

\begin{figure} %3
\centering
\includegraphics[width=1\linewidth]{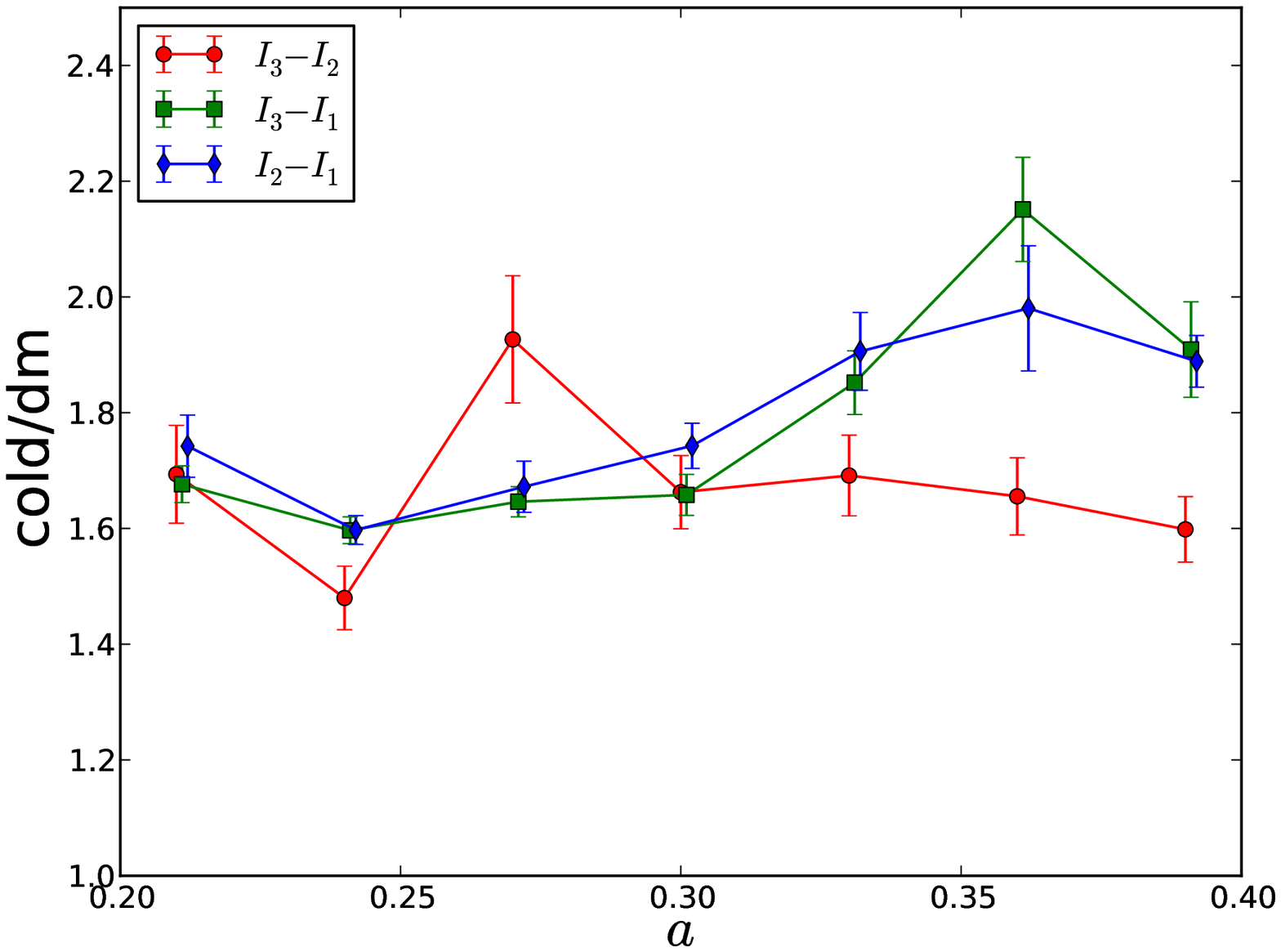}
\caption{
The ratio of quadrupole moments for cold gas and dark matter
in the shell $(1\!-\!2)\Rv$ averaged over all snapshots.
Shown as a function of expansion factor $a$ are the medians of the
ratios of the quantities $I_j-I_k$, the differences between the
eigenvalues of the inertia tensors that appear in \equ{quadrupole}.
The error bars refer to the errors of the mean.
The quadrupole moments of the cold gas in zone I (and II)
are larger than those of
the dark matter by a factor 1.5-2, reflecting the thin nature of the
cold streams compared to the thick filaments and the wide-angle distribution
of inflowing dark matter.
}
\label{fig:quad_ratio_evol}
\end{figure}

% cold thin, DM thick filaments
\smallskip
\Fig{proj_sfg1} shows maps of projected gas density and the corresponding
z-component sAM in the face-on disc frame for 
the different mass components in snapshots of two simulated galaxies.
One can clearly see that
the mass and AM are carried from well outside the halo into its central region
by elongated, rather straight streams. 
While the cold-gas streams are rather narrow, the 
dark-matter streams are much thicker, with a non-negligible component of
inflow from a wide angle between the large streams. 
% new
The gas has dissipatively contracted into the cords of the dark-matter
filaments at an earlier stage, but outside the halo the gas distribution is 
still elongated and possibly still in its expansion phase along the 
filament in the radial direction away from the galaxy.
While in the spherical collapse model there is a single maximum-expansion
event for each shell, here the relevant maximum expansion is along the 
filament, when the torques are most effective. 
Therefore, it is the quasi-linear Eulerian inertia tensor near this moment
that is relevant. \Fig{proj_sfg1} indicates that the
quadrupole moment of the cold gas outside the halo is significantly
larger than that of the dark matter.

% quadrupole of cold vs DM
\smallskip
\Fig{quad_ratio_evol} shows a crude estimate of the ratio of the quadrupole
moments for cold gas and dark matter outside the halo.
The instantaneous 
inertia tensor of each mass component is computed and diagonalized within 
the shell $(1\!-\!2)\Rv$ in each galaxy.
The eigenvectors of the two components are found to be well aligned in most
cases. 
Under the assumption that the tidal tensor that exerts torques on the two
components is practically the same, we conclude that the non-vanishing mutual 
misalignment of the tidal and inertia tensors 
\citep[studied for the dark matter in][]{porciani02a,porciani02b}
is similar for the two components, 
so the difference in AM should arise from the magnitude of the eigenvalues.
We use the eigenvalue differences $I_j-I_k$ as proxies for the quadrupole
moments. 
Shown in the figure are the average ratios of these differences 
for the cold gas and the dark matter. We learn that the quadrupole moment
of the cold gas is consistently higher than that of the dark matter
by a factor of $1.5-2$, which is increasing with cosmic time.
This implies that the sAM of the cold gas outside the halo 
should be higher by a similar factor, as indeed seen in \fig{spin_prof}.
% new
This indicates that TTT is crudely valid once applied to the large-scale mass
distribution in the elongated thin streams outside the halo where they are
near their maximum expansion. 
The validity of TTT is allowed in this quasi-linear regime by the fact 
that the Zel'dovich approximation used in the TTT analysis 
is a good approximation in this regime near pancake and filament 
configurations.
We can conclude that the higher spin parameter of the cold gas at the virial
radius predominantly originates from its larger quadrupole moment in the 
TTT phase outside the halo.
This confirms and explains the conjecture made by \citet{stewart13}.

% impact parameters
\smallskip
As a result of the tidal torques acting prior to maximum expansion, 
each of the cold streams, which originally tend to be pointing to the halo 
center, obtains a non-zero impact parameter. This can be interpreted as
generated by asymmetric flows from voids and sheets into the streams 
from the transverse directions to the streams \citep{pichon11,codis12}. 
The largest impact parameter is typically $b\!\sim\!0.3\Rv$.
Since the stream velocity is typically $\sim\!1.5\Vv$, this corresponds to 
$\lambda\!\sim\!0.3$ at pericenter for the streams that carries the largest AM.
The other streams acquire smaller impact
parameters, and sometime contribute AM in the opposite direction, as can be
seen in the two galaxies shown in \fig{proj_sfg1} for the streams that come 
from the left.
The mixing of AM from streams with large and small (or even opposite-sign)
impact parameters tends to lower the net spin brought into the inner halo
by the cold gas to $\lambda\!\sim\!0.1$.

\smallskip
One can see in \fig{proj_sfg1} that
beyond the virial radius the spin parameter of the hot gas is closer to the 
spin of the dark matter than that of the cold gas. This is consistent with the
wide-angle distribution of the hot gas that is closer to that of 
the dark-matter and thus has a similar quadrupole moment.

\smallskip
% new:
Our analysis indicates that TTT, once modified to account for the 
pre-collapse of
the gas to the cords of the filaments, provides a sensible qualitative
description of the initial acquisition of AM by the cold gas outside the halo
from the cosmological environment. For a more quantitative analysis one may
suspect that some of the additional assumptions made in standard TTT, such as
the second-order expansion of the potential about the halo centre,
may not be accurate enough given the deviations from spherical symmetry
associated with elongated streams that stretch to a few virial radii,
and should be modified.

%--------------- 3.4
\subsection{Gravitational versus pressure torques}
\label{sec:pressure}

The AM per unit volume of a fluid in a given position $\vec{r}$,
where the density is $\rho$ and the velocity is $\vec{v}$, is
\be
\vec l = \rho\, \vec{r} \times \vec{v} \, .
\ee
Its Eulerian rate of change, $\partial \vec l/\partial t$,
can be expressed in terms of the gravitational force and the pressure 
gradient (plus other
terms) using the continuity equation and Euler's equation of motion.
After some algebra, and subtracting off the advection term, one obtains
the Lagrangian torque, $d \vec l/d t$, as a sum of three terms:
\be
\vec\tau=\vec\tau_{\rm \phi}+\vec\tau_{\rm p}+\vec\tau_{s}
= 
  -\rho\, \vec{r}\!\times\!\vec\nabla \phi 
  -\vec{r}\!\times\!\vec\nabla P 
  -\vec{l}\ \vec\nabla \!\cdot\! \vec{v} \, ,
\label{eq:tau}
\ee
where $\phi$ and $P$ are the gravitational potential and the 
pressure at $\vec{r}$, respectively.

\smallskip
The first term $\vec\tau_\phi$ is the torque due to the gravitational 
force acting on all the mass components.
Since the torque is measured with respect to the disc center, a non-zero torque
requires a non-radial force component. Possible sources for such a torque
on large scales are the non-uniform mass distribution in the cosmic web
including neighboring haloes and subhaloes. Within the inner halo,
gravitational torques could result for example from dynamical friction acting
on the clumpy components by the smooth dark-matter distribution, 
or from the gravitational field generated by the disc.

\smallskip
The second term $\vec\tau_{\rm p}$ is the torque resulting from a pressure 
gradient, acting only on the gas.
The pressure torque may be important in the interface between hot and cold
gas where a pressure gradient may be present. 
We note that it may depend on the resolution,
but one may expect a smaller contribution of the pressure torques once
the boundary layer between cold and hot gas is better resolved, where
this layer may become unstable.

\smallskip  % the different terms
The third term is the strain 
associated with the viscosity that couples the
off-diagonal elements in the stress tensor to the hydro equations.
Being a vector proportional to the AM vector itself, the strain can only
change the magnitude of the AM.
Since the viscosity in the simulations is only a small numerical viscosity,
this term turns out to be negligible compared to the other two.

\begin{figure} %4
\centering
\includegraphics[width=1\linewidth]{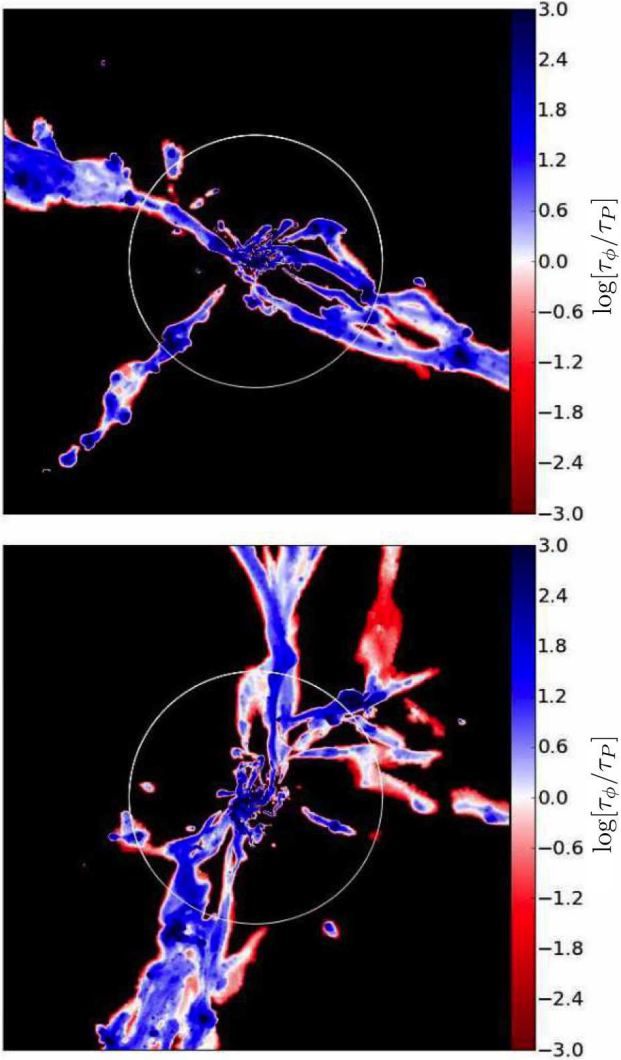}
\caption{
The ratio of gravitational to pressure torques acting on the cold gas,
$\log |\tau_{\phi}|/|\tau_{\rm p}|$.
Shown are projections in the direction of the stream plane as defined at $\Rv$
for SFG1 $z\!=\!3.17$ (top) and MW1 $z\!=\!1.63$ (bottom).
The circle marks the virial radius.
The gravitational torques dominate almost everywhere.
}
\label{fig:torque_ratio}
\end{figure}

\begin{figure} %5
\centering
\includegraphics[width=1\linewidth]{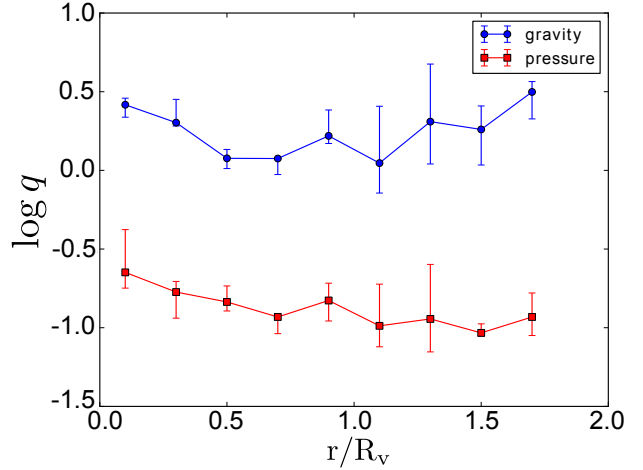}
\caption{
Profiles of gravitational and pressure torques,
$q_\phi$ and $q_{\rm p}$ (\equnp{q}),
acting on the cold gas, with dense clumps eliminated.
The vectors of torque $\tau$ and AM $l$ are integrated in shells of
thickness $0.1\Rv$, and $\tc\!=\!0.5\tv$.
These profiles are generated by stacking 6 snapshots from \tab{snaps}
(MW1 $z\!=\!1.63$, MW3 $z\!=\!2.33$, MW8 $z\!=\!2.33$, SFG1 $z\!=\!3.17$, 
SFG8 $z\!=\!2.17$, VL01 $z\!=\!2.57$).
The net contribution of the pressure torques is negligible compared to the
gravitational torques at all radii.
}
\label{fig:torques_profile}
\end{figure}

%q_p and q_phi
\smallskip
We define dimensionless parameters that measure the potential
ability of the pressure and gravitational torques to make substantial
changes in the AM of a given fluid element
during the characteristic timescale $\tc$ over which the torques are acting,
\be
q_{\phi}=\tc \frac{|\vec\tau_\phi|}{|\vec l\,|}, \quad
q_{p}=\tc\frac{|\vec\tau_{\rm p}|}{|\vec l\,|} \, .
\label{eq:q}
\ee
At a distance $\sim\!f\Rv$ from the galaxy center,
the relevant timescale could be the corresponding fraction of the virial
timescale ($\tv\!=\!\Rv/\Vv$), namely $\tc\!\sim\! f\tv$.
For example, for an $\Mv\!\sim\!10^{12}\msun$ halo at $z\!\sim\!2$,
$\tv\! \sim\! 0.5\Gyr$. This means that outside the halo, at $f\!\sim\!2$
say, we would consider $\tc\!\sim\! 1\Gyr$. In the inner halo, say at
$f\!\sim\! 0.2$, we would typically use $\tc\! \sim\! 100\Myr$ for radial
inflow, but several hundred Myr for gas spiraling in along a quasi-circular 
orbit.

\begin{figure*} %6
\centering
\includegraphics[width=\linewidth]
{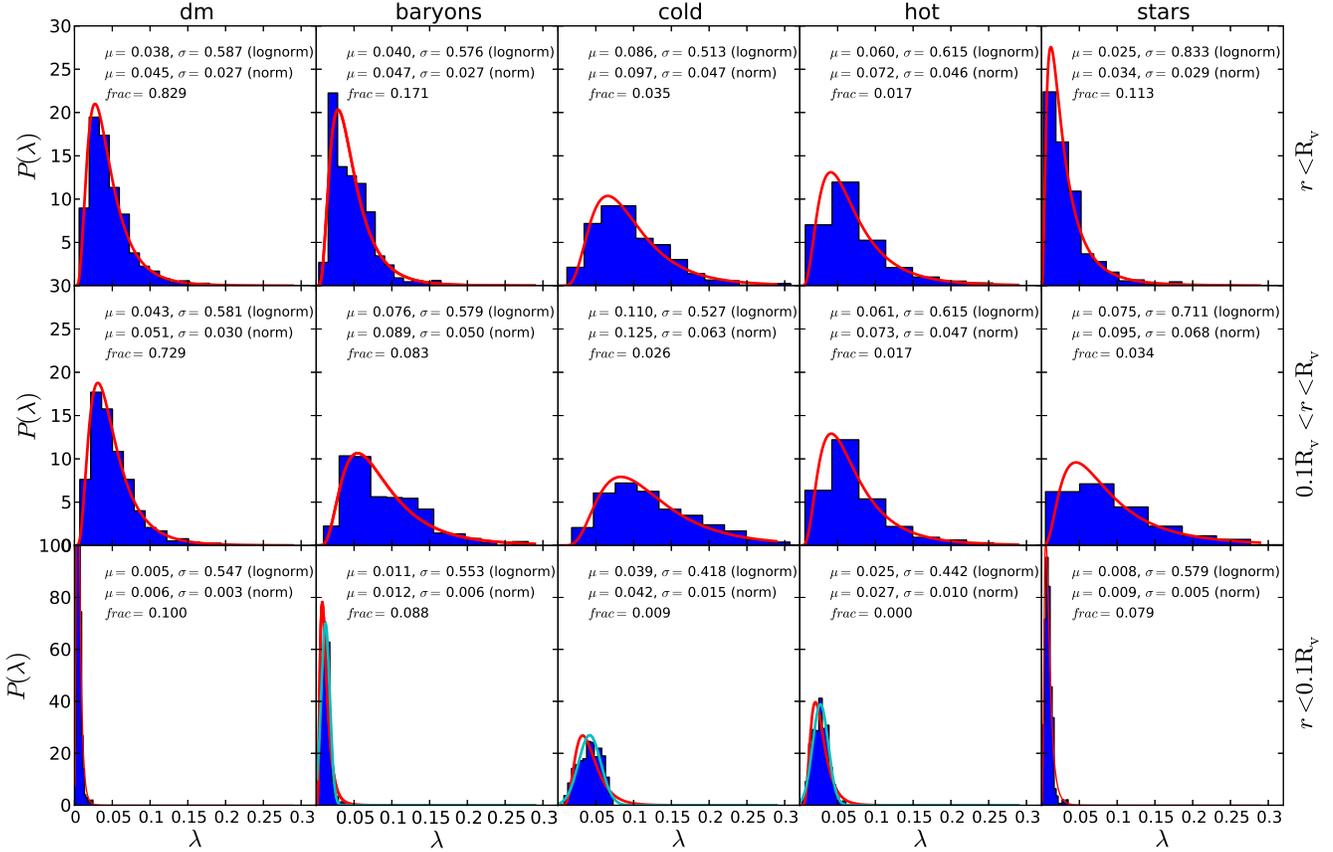}
\caption{
Spin parameter probability distribution for different components
within the halo at three radial ranges, over all galaxies and snapshots.
The ``baryon" component is the sum of the ``cold", ``hot" and ``stars"
components.
The spin $\lambda$ refers to the sAM $j$ while the direction $\hat J$
may be different for each component and for each zone.
Shown are lognormal fits (red) and Gaussian fits (light blue, shown only for
$r\!<\!0.1\Rv$), and the corresponding mean $\mu$ and standard deviation
$\sigma$i are quoted.
Also quoted is the mass fraction $frac$ of each component relative to the total
mass in the halo.
The spin of cold gas in the halo is higher than that of the dark matter
by a factor of $\sim\!3$.
It is larger by an order of magnitude than the spin of the baryons
at $r\!<\!0.1\Rv$.
}
\label{fig:spin_dist}
\end{figure*}

% computing the grav
\smallskip
The gravitational torque is calculated from the gravitational potential
at each point, computed with a softening length of 4 grid cells
via a tree-code algorithm \citep{barnes86}.
The mass distribution in each of the cubic tree nodes is represented by a 
multipole expansion to quadrupole order.
The accuracy of the force calculation is then controlled by 
the ratio $\theta$ of the size of the node exerting the
force and the distance from the center of mass of this node to the particle or
cell on which the force is exerted, which has been set to $\theta\!=\!0.8$ for
a better than 1\% accuracy. 
The spatial derivatives of the potential are calculated on a uniform grid
extracted from the AMR grid using a tri-cubic spline interpolation scheme.

% pressure torque is not important
\smallskip
\Fig{torque_ratio} shows maps of the ratio of gravitational and pressure 
torques
acting on the cold gas, $\log |\tau_{\phi}|/|\tau_{p}|$, in two galaxies. 
The maps are projections face-on in the stream plane as defined at $\Rv$, 
meant to optimize the view of the cold streams.
The visual impression is that the gravity torques dominate along the streams 
as well as in the central disc vicinity.
In the regions where the contribution of the pressure torques are comparable
to the gravitational torques (white), 
the torque itself tends to be small compared to where the gravity torques
dominate.
The regions where the pressure torques dominate (red), typically in the
boundaries of the cold streams, tend to contain a small fraction of the cold
gas, and the torques acting from all sides of the filaments tend to cancel each
other such that they do not produce substantial net angular momentum.

\smallskip
Indeed, \fig{torques_profile} compares radial profiles of $q_\phi$ and 
$q_{\rm p}$ for the cold gas. 
The torques $\vec\tau$ and the AM $\vec l$ in \equ{q} were  
integrated in shells of thickness $0.1\Rv$, and $\tc$ was set to $0.5\tv$. 
These torques were computed for the cold gas, with the dense
clumps eliminated by excluding gas densities higher than 1000 times
the mean hydrogen density. For the snapshots used, at $z\!\sim\!2-3$,
this upper limit for the density is $\sim\!10^{-2}\cmc$.
This removes fluctuations and generally lowers the $q$ values, 
but it does not affect the ratio of $q_\phi$ and $q_{\rm p}$.
The profiles shown were generated by stacking 6 snapshots from \tab{snaps}.
One can see that the net contribution of the pressure torques is negligible ---
it could make only small changes to the AM and it is much smaller
than the gravitational torques at all radii.

\smallskip
The fact that the averaged $q_\phi$ is larger than unity implies that the 
gravitational torques are capable of having a significant effect on the AM
near and outside $\Rv$.  The fact that $q_\phi$ is roughly constant in radius
implies that the torques could be more effective at larger radii, namely
outside the halo, because the time spent there is longer.
This is both because of the longer distance and the slower velocity at 
larger radii in the free-fall regime outside $\Rv$.

\smallskip
In summary for phase I, the AM is acquired by gravitational torques outside 
$\Rv$. The sAM gained by the cold gas is larger than the sAM gained by
the dark matter by a factor of 1.5-2 due to the higher quadrupole moment 
resulting from the early dissipative contraction of gas into the central 
cords of the rather thick dark-matter filaments.
It is encouraging that a similar result is obtained from different
simulations using different codes \citep[e.g.,][]{kimm11,stewart13}.

% 4 %%%%%%%%%%%%%%%%%%%%%%
\section{Phase II: AM transport through the halo}
\label{sec:II}

%------------------ 4.1
\subsection{Dark matter in the halo}
\begin{figure} %7
\centering
\includegraphics[width=\linewidth]{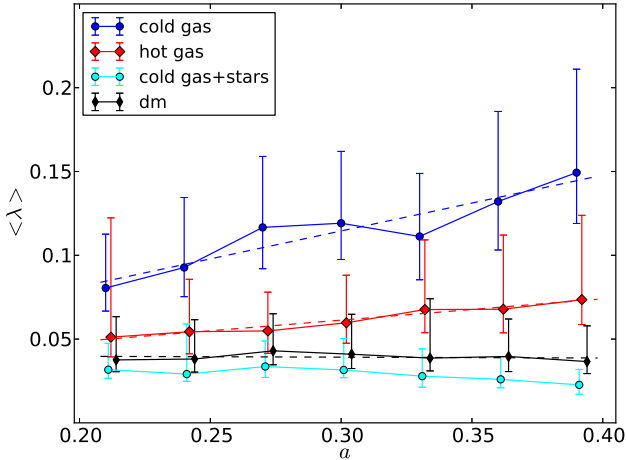}
\caption{
Evolution of the average spin parameter of different components
within the virial radius.
Cold gas (excluding the inner $0.1\Rv$) (blue),
hot gas (red), dark matter (black),
and cold gas in the halo plus stars in $r<0.1\Rv$ (cyan).
Shown are the mean and standard deviation
over all halos in bins that consist of three snapshots each.
While the spin of the cold gas in the halo is higher than that of the dark
matter by a factor of 2-3 and rising with time, the spin of the baryons in
the galaxy
is somewhat smaller than that of the halo dark matter and rather constant in
time, due to mixing and AM exchanges within the inner halo and the disc, on
zones III and IV.
The best linear fit (dashed line) for the halo cold gas is
$\lambda\!=\!0.015\!+\!0.33a$, 
and for the hot gas $\lambda\!=\!0.023\!+\!0.13a$.
}
\label{fig:spin_evol}
\end{figure}

% DM
The second phase takes place as the inflowing streams approach the
virial radius and penetrate through the outer halo, 
typically between $2\Rv$ and $0.3\Rv$.
In this regime, the dark-matter particles virialize.
The ones that flow in later with higher sAM tend to settle in 
more extended orbits, but they partly mix with the dark matter that is already 
there with lower sAM. The total spin parameter within $\Rv$ is therefore 
significantly lower than that of the recently arrived particles.
\Fig{spin_dist}, top-left panel, shows the probability distribution of the
spin parameter for the dark matter within $\Rv$ over all haloes and snapshots. 
The distribution is well fitted by a log-normal distribution with a mean  
$\la\lambda\ra\!=\!0.038$ and standard deviation $\sigma\!=\!0.59$, in fair
agreement with the findings from pure $N$-body simulations of 
more massive haloes \citep{bullock01_j}.
\Fig{spin_evol} demonstrates that the average $\lambda$ for the 
dark matter within the halo is the same at all times, 
as predicted by TTT and as seen in the earlier simulations.
Recall that \fig{spin_prof} showed that the radial profile is roughly 
$j(r)\!\propto\!r$,
consistent with the simulation results and with a toy model for halo buildup by
tidal stripping of satellites as they spiral in by dynamical friction
\citep{bullock01_j}.  

%--------------------- 4.2
\subsection{Cold gas in penetrating streams}

% cold gas
\smallskip
Contrary to the dark matter,
the streams of cold gas penetrate through the outer halo
at a velocity that is comparable to and slightly higher than $\Vv$ 
without significant mixing with the lower sAM gas that flew in earlier
and is now either in the central disc and its vicinity 
or it has been heated and removed by feedback.  
\Fig{spin_dist} shows that the distribution of $\lambda$ for the cold gas 
in the halo, excluding the inner $0.1\Rv$,
is well-fit by a log-normal distribution with a mean $\la\lambda\ra \!=\! 0.11$.
This is a factor of $\sim\!2.5$ larger than that of the dark matter in the same
volume,
partly due to the higher sAM gained by the cold gas in zone I outside the halo, 
and partly due to the mixing of the incoming dark matter with lower-sAM 
dark matter within the halo.
A similar difference between the spin of the cold gas in the halo and that of
the dark matter has been detected in other simulations 
\citep{sharma05,stewart11,kimm11}.
The previously accreted, lower-sAM cold gas, if not removed by outflows,
is largely confined to the galaxy
at $r<0.1\Rv$, as cold gas and stars, with a lower spin parameter.
This spin parameter of the galactic baryons is strongly affected by the 
very-low spin of the massive stellar bulge, 
reflecting the AM loss by the cold gas 
in the inner halo and the disc (\se{III},\se{IV}).
Evidence for this AM loss is in the fact that inside $\Rv$ the average
spin parameter of the baryons, $\la \lambda_{\rm bar}\ra \!=\! 0.040$,
is comparable to that of the dark matter, despite the fact that
most of the baryons entered the halo cold 
with a spin parameter that was higher by a factor 1.5-2.

% evolution
\smallskip
\Fig{spin_evol} reveals that the average $\lambda$ for the halo cold gas
gradually grows with time roughly linearly with expansion factor $a$, 
from about 0.08 at $z\!=\!4$ to 0.15 at $z\!=\!1.5$, while the halo spin of the
dark matter is rather constant in time.
Recall that a constant $\lambda$ is actually a growing $j$, as the
normalization factor $\Rv\Vv$ is monotonically increasing in time.
The growth in $j$ is as qualitatively expected from TTT, 
and the faster growth for the cold
gas largely emerges from the growth of the quadrupole moment of cold gas
with respect to that of the dark matter in zone I, 
as seen in \fig{quad_ratio_evol}. 
While the spin of the cold gas in the halo is 2-3 times larger than that of the
dark matter, the baryons within $0.1\Rv$, dominated by stars that 
% cold gas within the entire halo with the stars in the inner region 
formed from previously accreted cold gas, show a spin parameter that is
somewhat smaller than that of the halo dark matter, and rather constant in
time. The reduction in the spin of the baryons is partly a result
of the mixing with the baryons that entered the disc earlier with lower
specific angular momentum, and partly due to AM loss in the inner halo.

\begin{figure*} %8
\centering
\subfigure{\includegraphics[width=0.45\linewidth]{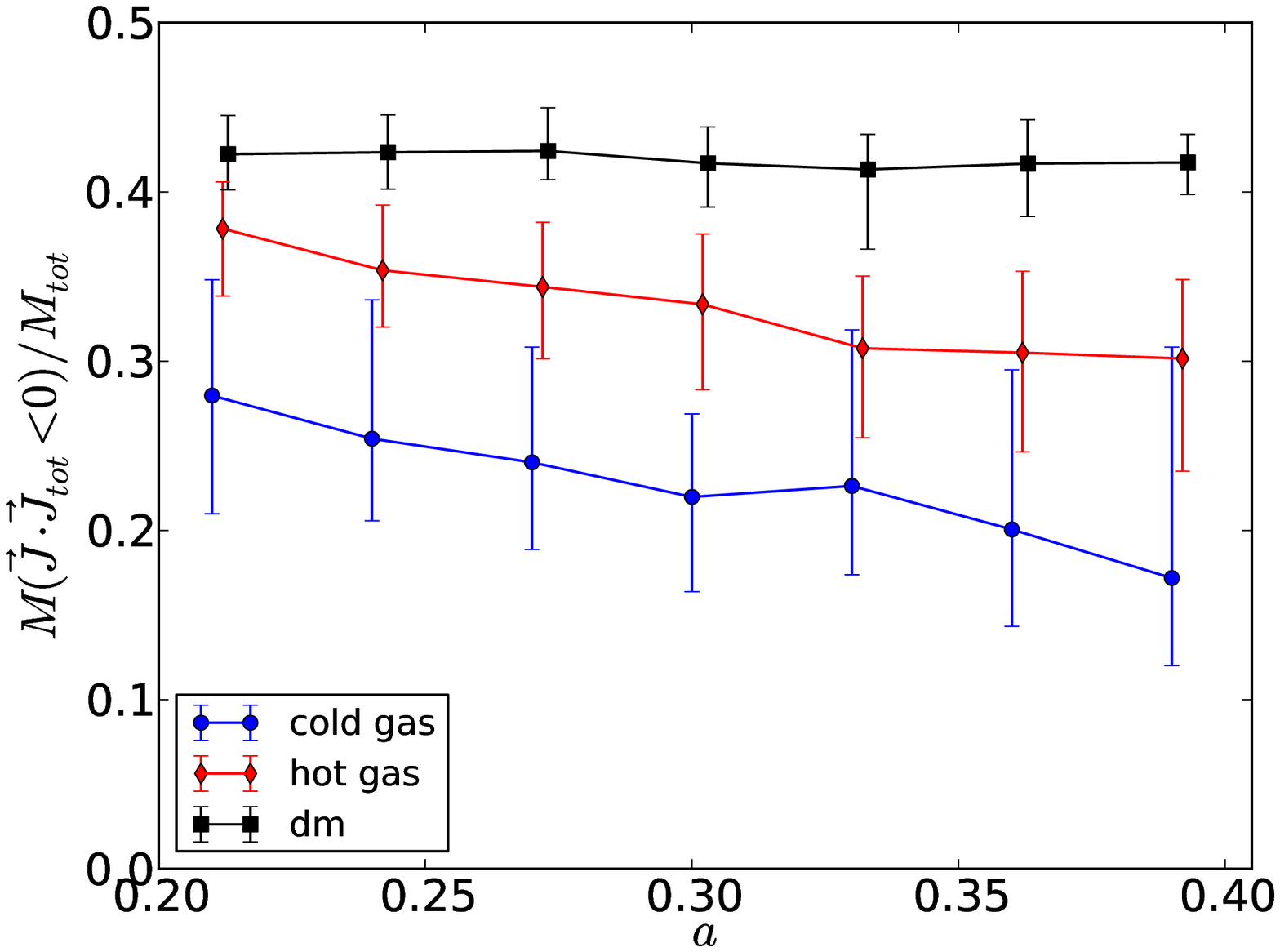}}
\subfigure{\includegraphics[width=0.45\linewidth]{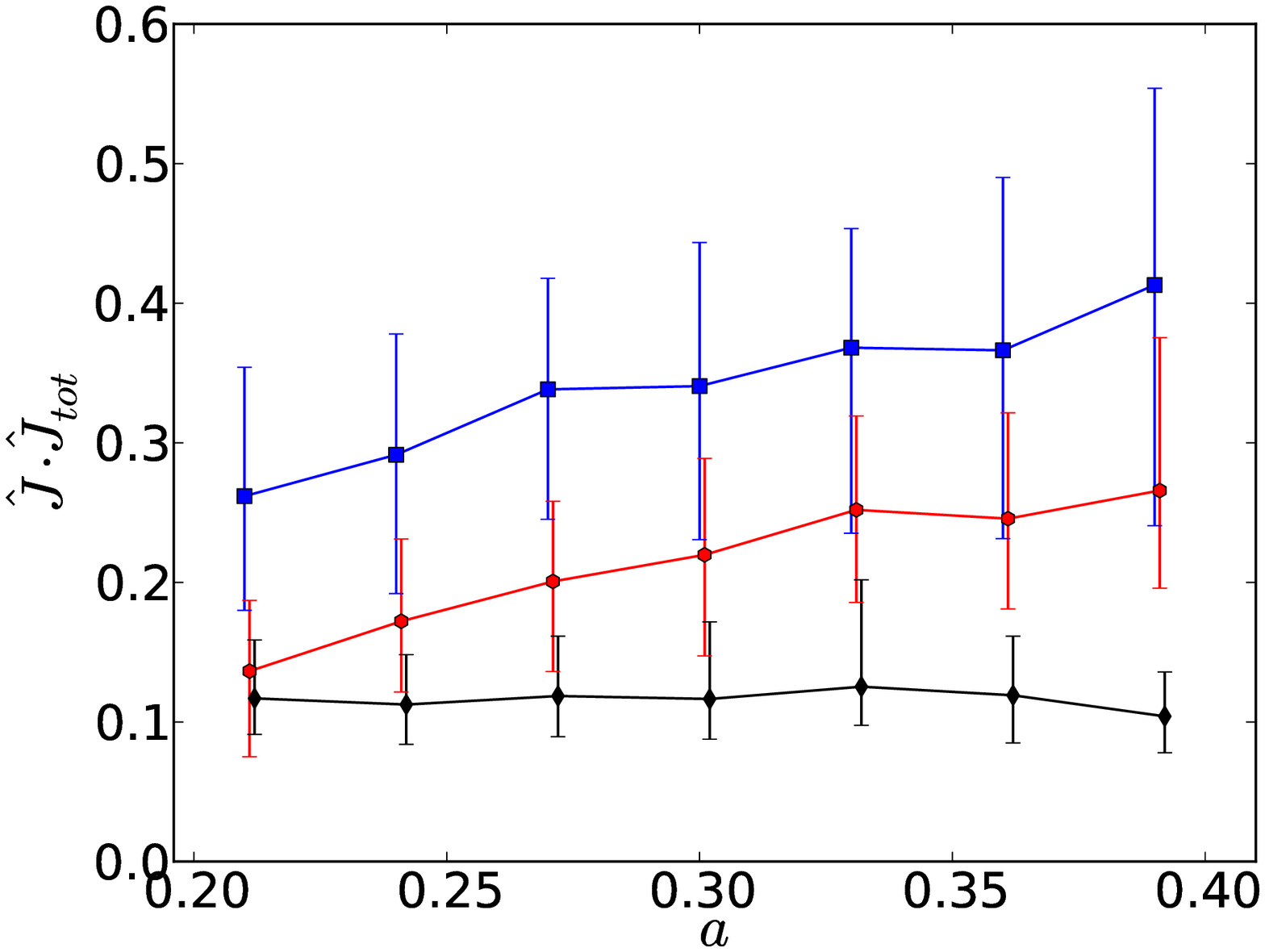}}
\caption{
Coherence of AM in the different components within the virial radius, zone II.
The quantities are mass-weighted averaged over grid cells or particles
within $(0.1-1)\Rv$, and over all snapshots and haloes.
The error bars refer to standard deviation.
{\bf Left:}
The fraction of mass with negative AM in the direction of the total AM
of that mass component, measuring incoherence.
{\bf Right:}
The cosine of the angle between the AM of the given component and the total
AM direction of that component, measuring coherence.
The AM in the cold streams is significantly more coherent than that of the
dark matter.
}
\label{fig:am_cor}
\end{figure*}

\smallskip
We notice in \fig{spin_prof} that the sAM profile for the cold gas is
roughly $j(r) \propto r^{0.5}$ throughout zone II, crudely 
consistent with the profile
expected from TTT with the AM roughly conserved during the streaming in.
% constant impact
The tendency of the cold gas to preserve its sAM in zone II is indicated by
noticing in the cold density maps shown in \fig{proj_sfg1} that the streams in
zone II tend to be straight lines, namely with constant impact parameters.
Combined with the roughly constant streaming velocity, which results from
dissipation and Lyman-alpha radiation from the streams 
\citep[e.g.,][]{goerdt10}, the sAM remains roughly constant. 
For the cold gas, the AM that has been generated in zone I is basically
being transported by the streams through zone II into the inner halo, zone III.
This will be addressed again in \se{ring_kinematics} and \se{ring_torques}, 
\fig{messy_hists} and \fig{torque_prof}.

%---------------------------- 4.3
\subsection{Coherence of AM transport}

%j in cells comparable in all components
It turns out that
the difference between the AM properties of the different components 
within the halo cannot be attributed to systematic
differences in the sAM amplitudes of individual gas cells or dark-matter 
particles
($j\!=\!bv$ with $b$ the impact parameter and $v$ the speed) --
these local sAM values for the different components are rather comparable.
The main difference in zone II 
is in the coherence of the motions of the cold gas 
and in the AM transport efficiency.
%higher velocity for the cold gas and smaller impact parameter  

% coherence
\smallskip
\Fig{am_cor} measures in two ways the level of coherence of the AM transported 
by the different components within $(0.1-1)\Rv$ as a function of time. 
The averages shown are over particles or over gas cells within each halo, 
and then averaged over all haloes.
First is the mean fraction of mass with a negative contribution to the total AM 
of that component within the halo, which serves as a measure of 
deviations from coherence. 
We see that the dark matter has a large negative fraction, with 
an average slightly above 0.4 and constant in time.
The cold gas has a much smaller negative fraction, decreasing from 0.28 at 
$z\!=\!4$ to $0.18$ at $z\!=\!1.5$, thus indicating larger coherence.
A second measure of coherency is the mean cosine of the angle between the AM 
in each cell and the total AM of that component.
The dark matter has a low mean cosine at the level of 0.1, indicating strong
incoherence.
The mean for the cold gas is much larger, increasing from 0.27 at $z\!=\!4$ to 
0.41 at $z\!=\!1.5$. 
Thus, these two measures confirm that the cold gas transports the AM in a
coherent way, while there is significant mixing for the dark matter,
and some mixing for the hot gas.

\begin{figure*} %9
\centering
\includegraphics[width=\linewidth]{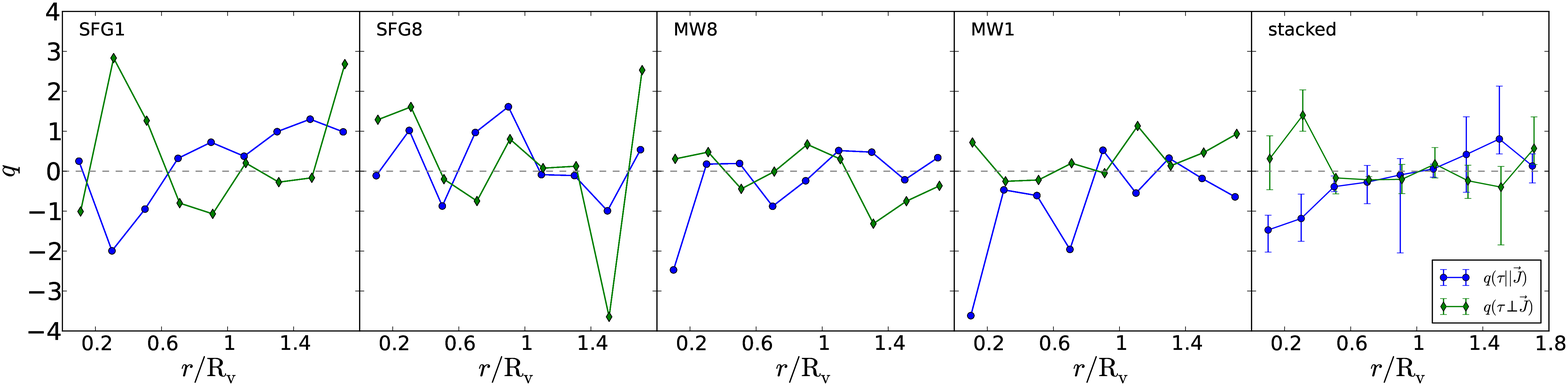}
\caption{
Profile of the torque acting on the cold gas in concentric shells,
with respect to the AM in the same shell, $\vec J$.
Shown are the component parallel to $\hat J$ (blue) and the component
perpendicular to $\hat J$ in the plane defined by $\hat J$ and
$\hat J_{\rm disc}$ (green). The former is associated with a change in AM
amplitude and the latter refers to a tendency for alignment with the disc.
Here $\tc\!=\!0.5\tv$.
The profiles are shown for 4 individual galaxies,
and the stacked profile are over 6 galaxies
(MW1 $z\!=\!1.63$, MW3 $z\!=\!2.33$, MW8 $z\!=\!2.33$,
SFG1 $z\!=\!3.17$, SFG8 $z\!=\!2.17$, VL01 $z\!=\!2.57$).
Related quantities cell-by-cell are shown in \fig{torque_prof}.
Dense clumps have been removed, to obtain smoother curves,
but the trends at small radii remain.
In the outer halo, zone II, the $q$ values for individual galaxies
can be of order unity but they fluctuate in sign, and the
stacked $q$ values there are lower than unity.
In the inner halo, zone III, the parallel and alignment components tend to be
negative and positive respectively, at the level of $\vert q \vert\!\sim\! 1$,
indicating a tendency for lowering the AM amplitude while aligning it
with the disc AM.  %big fluct at $1.7\Rv$.
}
\label{fig:q_profiles_shells}
 \end{figure*}

% visual coherence
\smallskip
The coherence of the AM in the different components is demonstrated visually 
for two of our galaxies in \fig{proj_sfg1}. Recall that it shows maps of 
density and of positive sAM in the direction of the total AM of that 
component, thus distinguishing between the parts with 
positive and negative contributions. 
For the cold gas, the AM seems to be rather coherent within each stream
throughout zone II, starting beyond $2\Rv$ and ending in the inner halo.
On the other hand, the thick streams of dark matter are well defined only
outside the virial radius and they become a rather messy mixture within $\Rv$,
and each of them consists of positive and negative contributions to the net AM
of the dark matter.

% opposite AM streams 
\smallskip
As found earlier \citep{danovich12},
\fig{proj_sfg1} illustrates that in the two examples shown 
the mass is transported mainly by three streams. 
As for the AM, 
in galaxy MW1 it is transported pretty evenly and coherently by the two 
massive cold streams, with a small negative contribution from the less massive 
stream(s).  In SFG1, most of the positive AM of the cold gas
is transported by the single most 
massive stream, a second stream has a much smaller contribution in the same
AM direction, and the third stream mostly carries AM of an opposite sign.
Thus, the different cold streams can contribute positively or negatively
to the net total AM that they bring in.
Since the measures of coherence in \fig{am_cor} refer to all the cold streams 
together, they are contaminated by the large-scale incoherence between the 
streams and thus underestimate the coherence within each stream by itself.
The counter-rotating streams can have very dramatic effects on the
central galaxy (see \se{III} and \se{IV} below).

\begin{figure*} %10
\centering
\subfigure{\includegraphics[width=\linewidth]{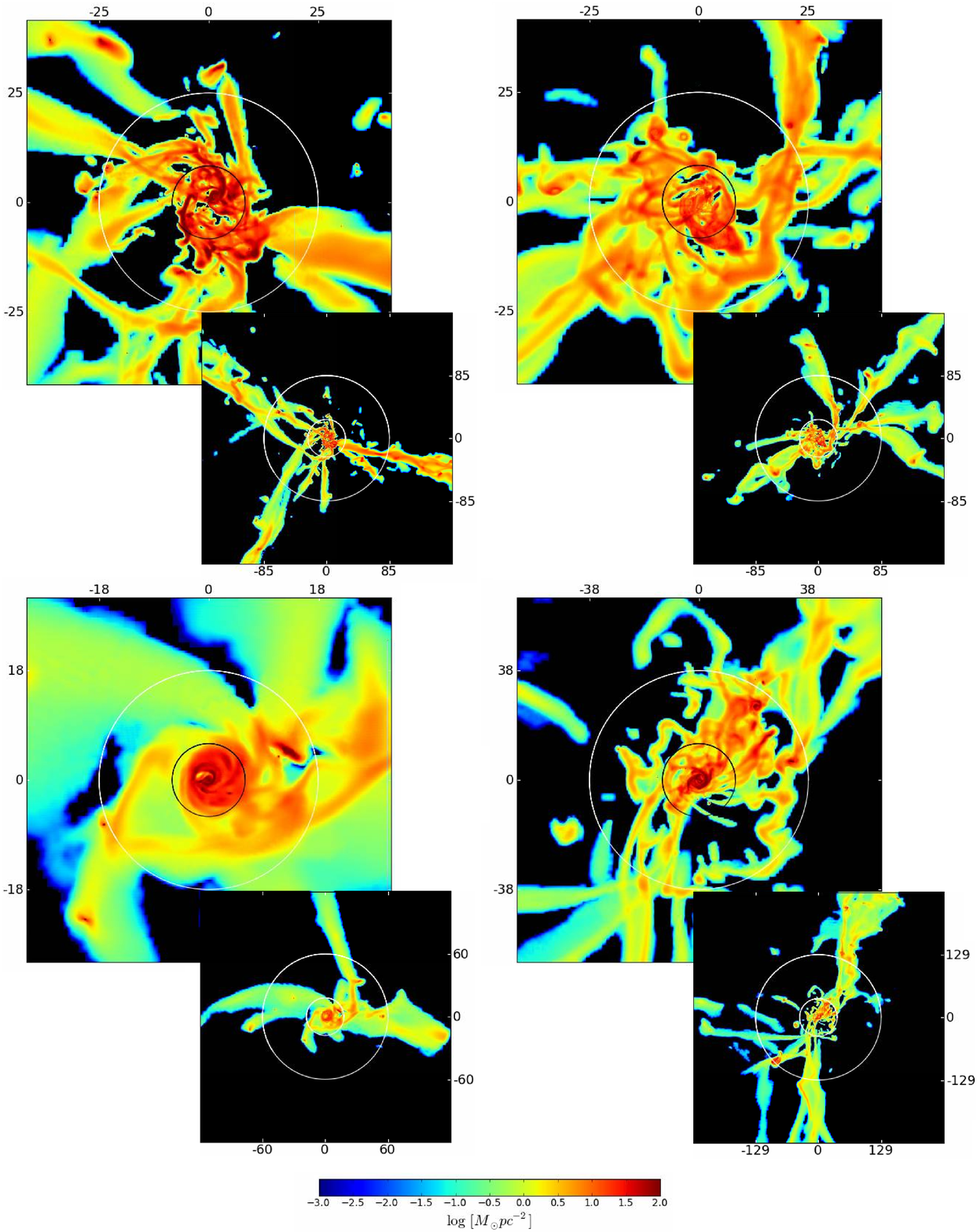}}
\caption{
Density projections of the cold dense gas at the inner halo in 4 galaxies
from \tab{snaps} 
(SFG1 $z\!=\!3.17$, SFG8 $z\!=\!2.92$, MW8 $z\!=\!2.33$, MW1 $z\!=\!1.63$ 
from top-left to bottom-right).
The projections of thickness $\Rv$ are face-on in the disc frame.
In the small zoom-out views the box is of side $4\Rv$ and the circles are 
of radii $\Rv$ and $0.3\Rv$.
In the zoom-in views the box is of side $1\Rv$ and the circles are of radii
$0.3\Rv$ and $0.1\Rv$.
The lengths on the box sides are in kpc.
The interface region between streams and disc, zone III,
shows a non-uniform, filamentary and clumpy ``ring"
where the streams settle into a rotating pattern.
}
\label{fig:messy}
\end{figure*}

\begin{figure*} %11 
\centering
\includegraphics[width=\linewidth]{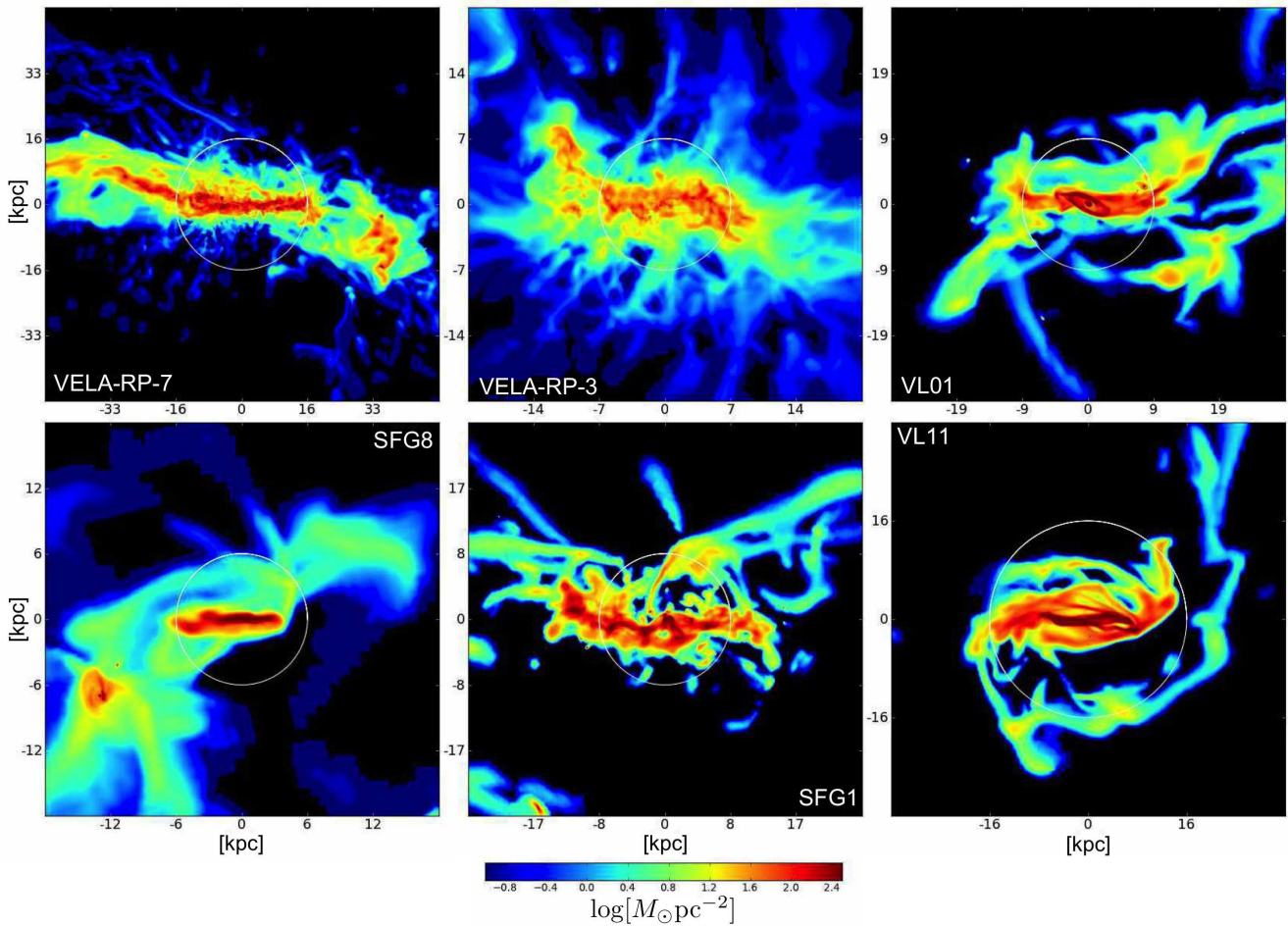}
\caption{
Projections of cold-gas density in edge-on directions of 6 discs from
\tab{snaps}
(from top-left to bottom-right:
VEL7 $z\!=\!1.17$, VEL3 $z\!=\!1.50$, VL01 $z\!=\!2.45$,
MW8 $z\!=\!2.33$, SFG1 $z\!=\!3.17$, VL11 $z\!=\!1.50$).
The box side and thickness are $0.3\Rv$. 
The circle marks $0.1\Rv$. 
These projections highlight the warps associated with the extended tilted ring.
The pictures from the simulations with higher resolution and
stronger feedback (two top-left panels) highlight the tilt of the outer ring,
which in one extreme case extends out to beyond $30\kpc$ (VEL7).
}
\label{fig:proj_edge}
\end{figure*}

\begin{figure*} %12 
\includegraphics[width=1\linewidth]{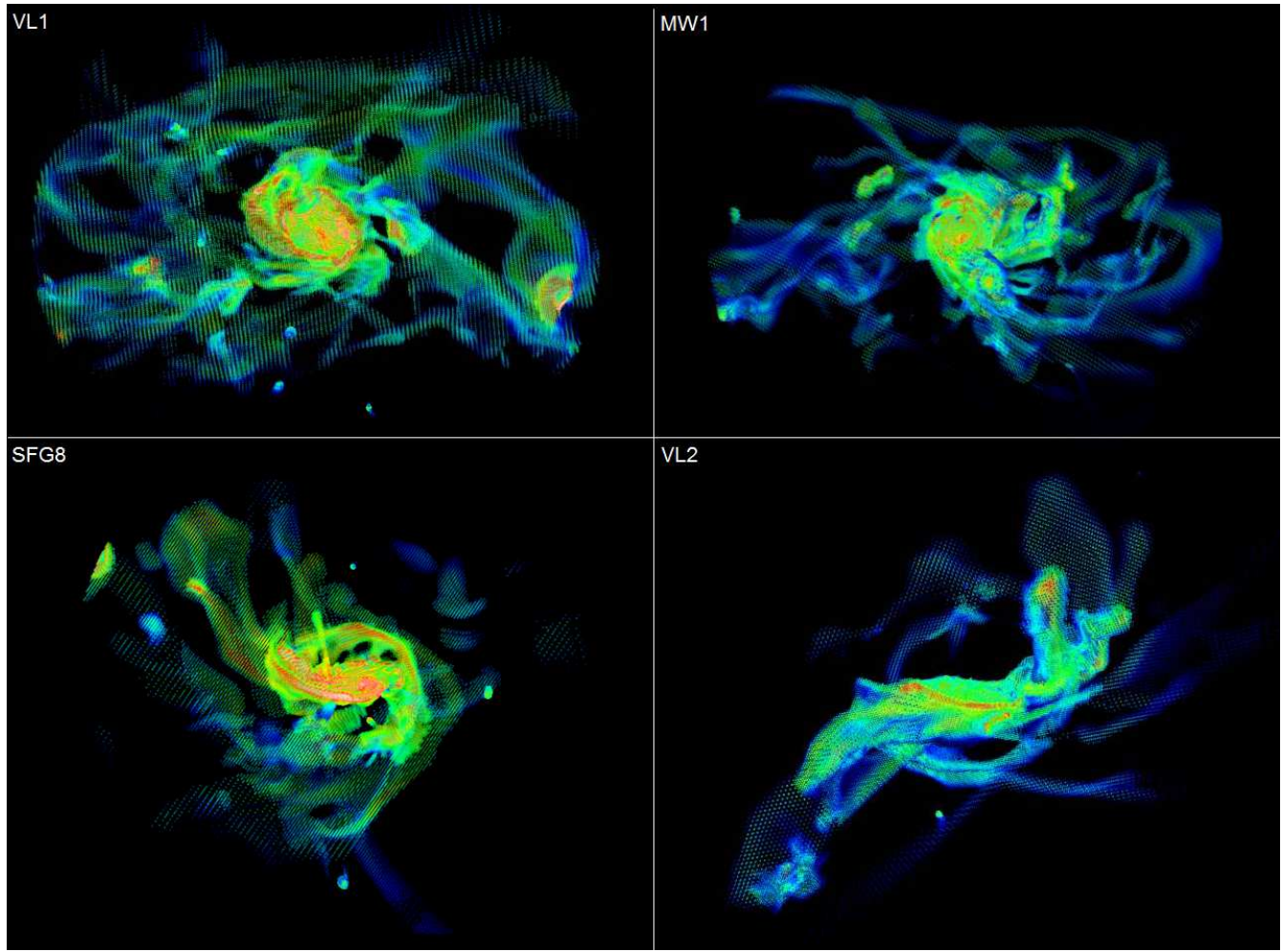}
\caption{
The extended cold-gas ring in 4 galaxies from \tab{snaps} 
(VL01 $z\!=\!2.57$, MW1 $z\!=\!1.5$, SFG8 $z\!=\!2.17$ and VL02 $z\!=\!1.33$).
The box side is $\pm 0.35\Rv$.
The cold gas density within the greater disc vicinity shows
the inner disc and the surrounding ring, where the gas streams bend, 
spiral in over less than one circular orbit, and smoothly merge with the disc.
}
\label{fig:ring_density}
\end{figure*}

\begin{figure*} %13
\vskip 4.8cm
\includegraphics{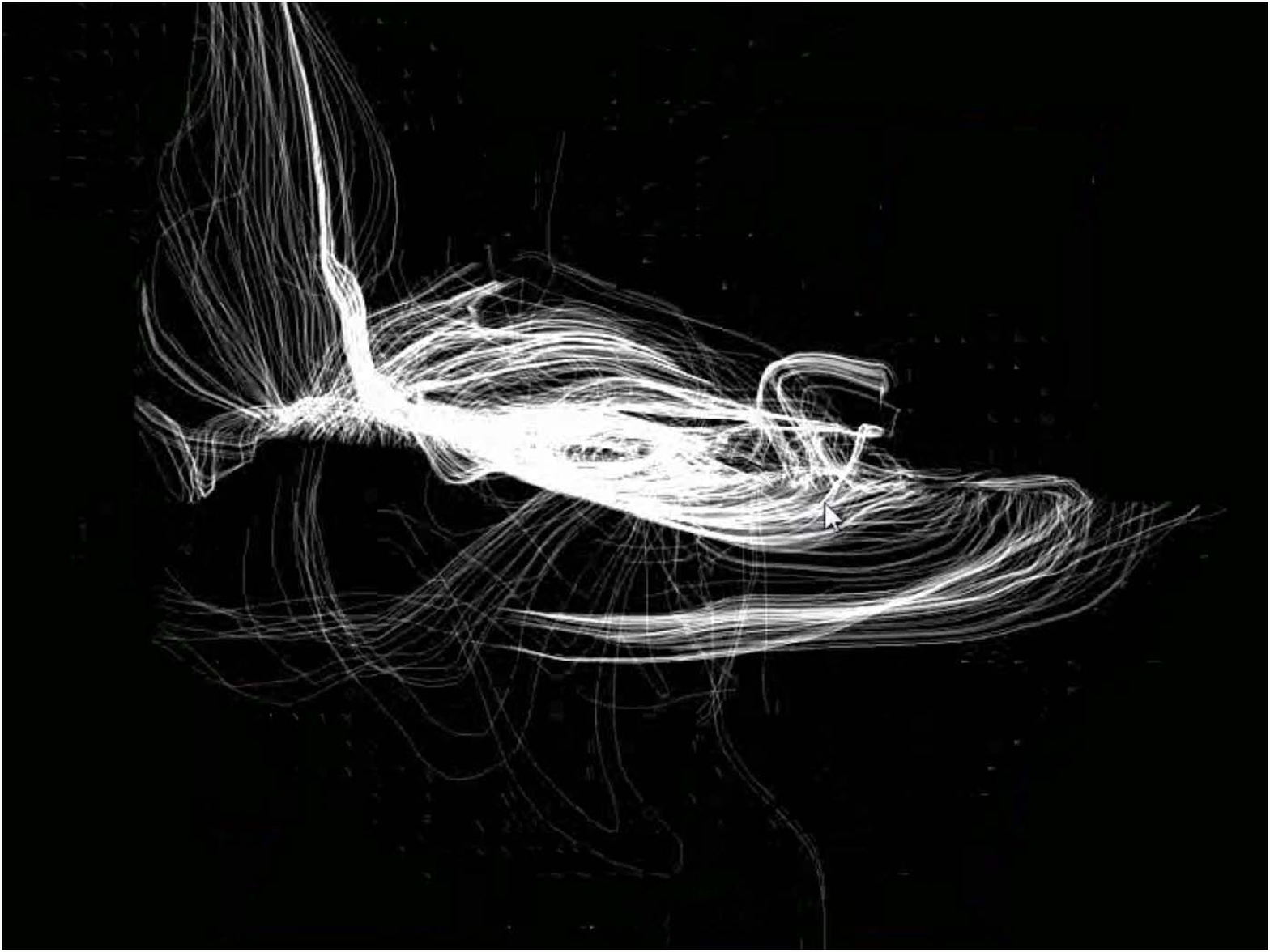}
\includegraphics{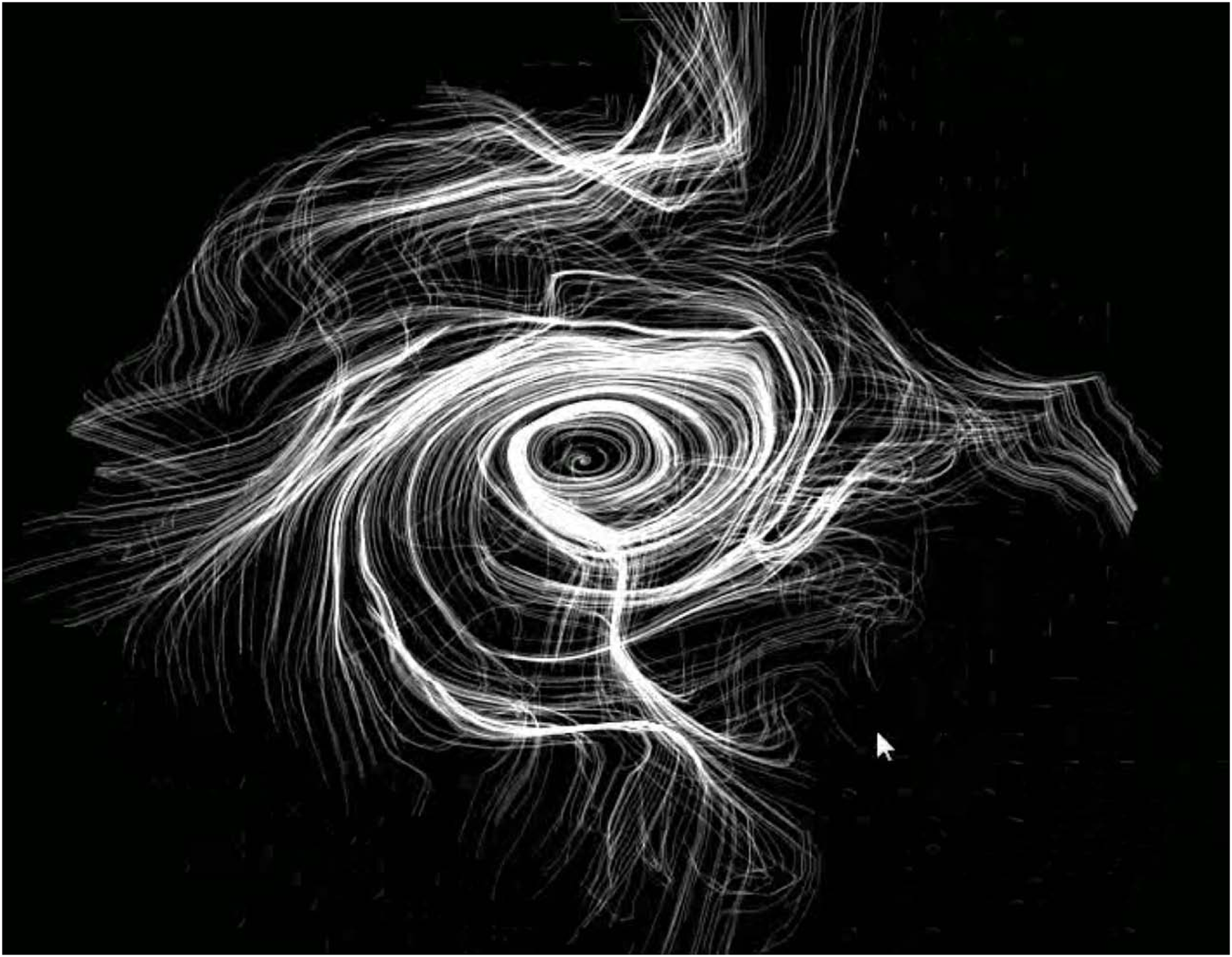}
\caption{
Streamlines of gas in the inner halo of MW3 $z=2.33$.
The box side is about $30\kpc$.
The streamlines illustrate how the inflowing streams bend and merge smoothly 
into the inner rotating disc through an extended tilted ring. 
}
\label{fig:ring_streamlines}
\end{figure*}

%----------------- 4.4
\subsection{Torques on cold streams}

We saw in \fig{torques_profile} that the gravitational torques,
with $q\!\sim\! 1\!-\!2$ in zone II, are apparently capable of making 
significant changes to the AM also in this region.
However, this refers to the total amplitude of the torque in a thin shell,
while the torque and its interplay with the AM of the shell can vary in
direction from shell to shell, so the total net effect on the stream AM as it
flows from outside $\Rv$ to the inner halo could be significantly smaller.

\smallskip
\Fig{q_profiles_shells}
shows the profiles of two components of the normalized torque $q$ acting on
shells of different radii $r$. Shown is the component parallel to $\hat J$,
which changes the amplitude of the AM.
Also shown is the perpendicular component that lies in the plane defined by
$\hat J$ and $\hat J_{\rm disc}$, which measures the tendency for
alignment with the disc AM (hereafter the ``alignment" component).
The way the disc is defined is described at the beginning of \se{IV}.
The profiles are shown for 4 individual galaxies and stacked over 6 
galaxies (6 snapshots).
We see that
the values of the components of $q$ in the shells of zone II are of order
unity, but in each individual galaxy the variation between radii is large
such that the average $q$ throughout zone II is smaller than unity.
This is consistent with the evidence discussed above for only small changes
in the AM in this zone, where the AM of the cold gas is being transported by
the streams.

%------------- 4.5
\subsection{Hot gas}

As seen in \fig{spin_dist},
the hot gas is a non-negligible component in the halo outside the galaxy,
at $(0.1-1)\Rv$, where it constitutes 22\% of the baryons (28\%
if the gas of intermediate temperatures slightly above $10^5$K is included)
and 40\% (47\%) of the gas
\citep[see also][]{keres05,db06,bdn07,ocvirk08,keres09,agertz09,cdb10}.
This gas partly consists of infalling gas that has been heated by the
virial shock and is now in quasi-static equilibrium within the halo,
and partly of outflowing gas due to feedback. 
Its average spin parameter is $\la \lambda_{\rm hot} \ra \!=\! 0.061$,
about half that of the cold gas in that region, and 42\% higher than that
of the dark matter.
The same is re-confirmed in \fig{spin_prof}, 
which emphasizes that the spin parameter of the hot gas shows
a flatter increase with radius than the dark matter and the cold gas,
converging to an sAM similar to that of the dark matter in zone I outside the
halo. 
% new
This indicates that a substantial fraction of the hot gas in the inner halo
originates from the cold-stream gas that has been heated on the way in
\citep{nelson13},
e.g., by shocks within the supersonic streams and by their interaction with the
hot CGM (Mandelker et al. in preparation).
The mixing with hot outflows is expected to reduce the hot-gas spin
as the outflows preferentially remove low-spin gas from the central dense
star-forming regions.

%\smallskip
%\adr{Need to elaborate on the hot gas. Gas that fell in hot, was heated by the
%virial shock, or ejected from the galaxy as outflow.}
%\mdb{I will try to write more.}

%%%%%%%%%%%%%%%%%%%%%%%%%%%%%%%%%%%%%%  5
\section{Phase III: AM loss due to strong torques in an extended ring}
\label{sec:III}

%   Stream dissipate in inner halo near pericenter, settle into circular orbits,
%   forming tilted outer disc at (0.1-0.3)Rv.
%   co-rotating and counter-rotating orbits. collisions.
%   Torques cause AM loss to lambda~0.04, cold gas spirals in, 
%   and outer disc settles into  the inner disc plane.

%---------------------- 5.1
\subsection{Structure of the extended ring}
\label{sec:ring_structure}

The third phase involves the cold gas in the inner halo and in the greater
vicinity of the central disc, typically at $(0.1\!-\!0.3)\Rv$.
The outer radius of this zone is determined by the impact parameter of the 
cold stream that dominates the AM. 
This region was termed ``the messy region" based on the
visual impression of the complex structure and kinematics  
seen in simulations \citep{cdb10}, or ``the AM sphere" based on the indications
for AM exchange in this region deduced from the misalignment of the disc and 
the AM of gas in the outer halo \citep{danovich12}.
Indeed, we will discuss below clear evidence for AM exchange in this zone.
For example, we see in \fig{spin_dist} that the average spin 
parameter of the cold gas at $r<0.1\Rv$ is $\lambda\! =\! 0.039$. 
While this is comparable to the spin of the dark matter in the halo,
it is smaller by a factor of 3 than the spin of the cold gas in the halo 
outside the disc.
This could be partly because of mixing of the newly arriving high-sAM gas 
with the disc gas that arrived earlier with lower sAM, 
and partly due to AM loss by torques in the greater vicinity of the disc,
which we will address below. 

% The outer disc - description
\smallskip
The cold gas streams that flow in through zone II on roughly straight lines 
with impact parameters $b$ dissipate kinetic energy near pericenter, 
at $r\! \sim\! b$, through interaction with the dense gas already there.
This makes the streams bend toward more circular orbits about the disc.
With a typical stream velocity of $1.3\Vv$, the stream with the
largest impact parameter defines the outer boundary of this region at 
$b\! \sim\! \lambda_{\rm max} \Rv$, which is typically $0.3\Rv$.
The incoming gas thus forms an extended outer ring at $(0.1\!-03)\Rv$.
The ring is in general tilted with respect to the central disc
because its incoming AM has been determined outside $\Rv$,
and is in general not aligned with the disc \citep{danovich12}. 
As we demonstrate below,
gravitational torques, exerted largely by the inner disc, make the gas in the
extended ring lose AM and thus spiral in toward the inner disc, as it gradually
aligns its AM with the spin axis of the inner disc.
This typically takes less than an orbital time.

% messy in pictures face-on
\smallskip
The large-scale structure of zone III is illustrated in \fig{messy}, 
which shows images of the projected cold gas distribution  
for 4 simulated galaxies with different masses at different redshifts.
Shown are zoom-out views in boxes containing circles of radius $2\Rv$ 
and zoom-in views in boxes of radius $0.5\Rv$, 
with the circles of radius $0.3\Rv$ and $0.1\Rv$ marked.
These projections are face-on in the disc frame, with the disc rotation
clockwise, namely the AM pointing into the page.
One can see that 2-5 streams are well defined in zone II, 
reaching zone III with 
detectable impact parameters, most of which indicate co-rotation with the disc
but some are counter-rotating.
Zone III is the interface between the incoming cold streams and the inner
more organized (though still perturbed) rotating disc.
We see that the ``ring" in zone III is far from being a uniform ring,
as the gas distribution shows large-scale deviations from circular symmetry 
involving a complex pattern of filaments and clumps that are reminiscent of 
the stream pattern in zone II.
These 4 examples also illustrate that the extent of zone III indeed scales 
with the virial radius, typically at $0.3\Rv$, which implies in turn
that it is increasing with mass and growing in time.

\smallskip
%Images of outer ring
\Fig{proj_edge} shows projections of cold-gas density in the edge-on direction 
of the disc in 6 example galaxies from our sample, highlighting the tilt of the
extended ring, which appears as pronounced extended warps.
\Fig{ring_density} shows another type of images 
of the density of cold gas within zone III in 4 simulated galaxies. 
These images were produced using the IFrIT visualization tool.
Each gas cell is treated as a particle and coloured according to the gas 
density in the cell. By an optimal selection of opacities for the different 
colours and viewing directions, the images attempt to give a certain 
three-dimensional impression of the nature of the extended ring and its 
relation to the inner disc. 
These images indicate that the gas streams in the ring spiral in over 
less than one circular orbit.
This is confirmed in \fig{ring_streamlines} that presents streamlines in the 
same region for one of the galaxies. The streamlines were generated from one
snapshot by interpolation between the velocities in grid cells.
The streamlines show how the streams bend and gradually join 
the inner rotating disc, like a stream of cars entering an expressway 
\citep[coined ``on-ramp" by][]{keres05}. 

 \begin{figure*} %14 
\centering
\includegraphics[width=\linewidth]{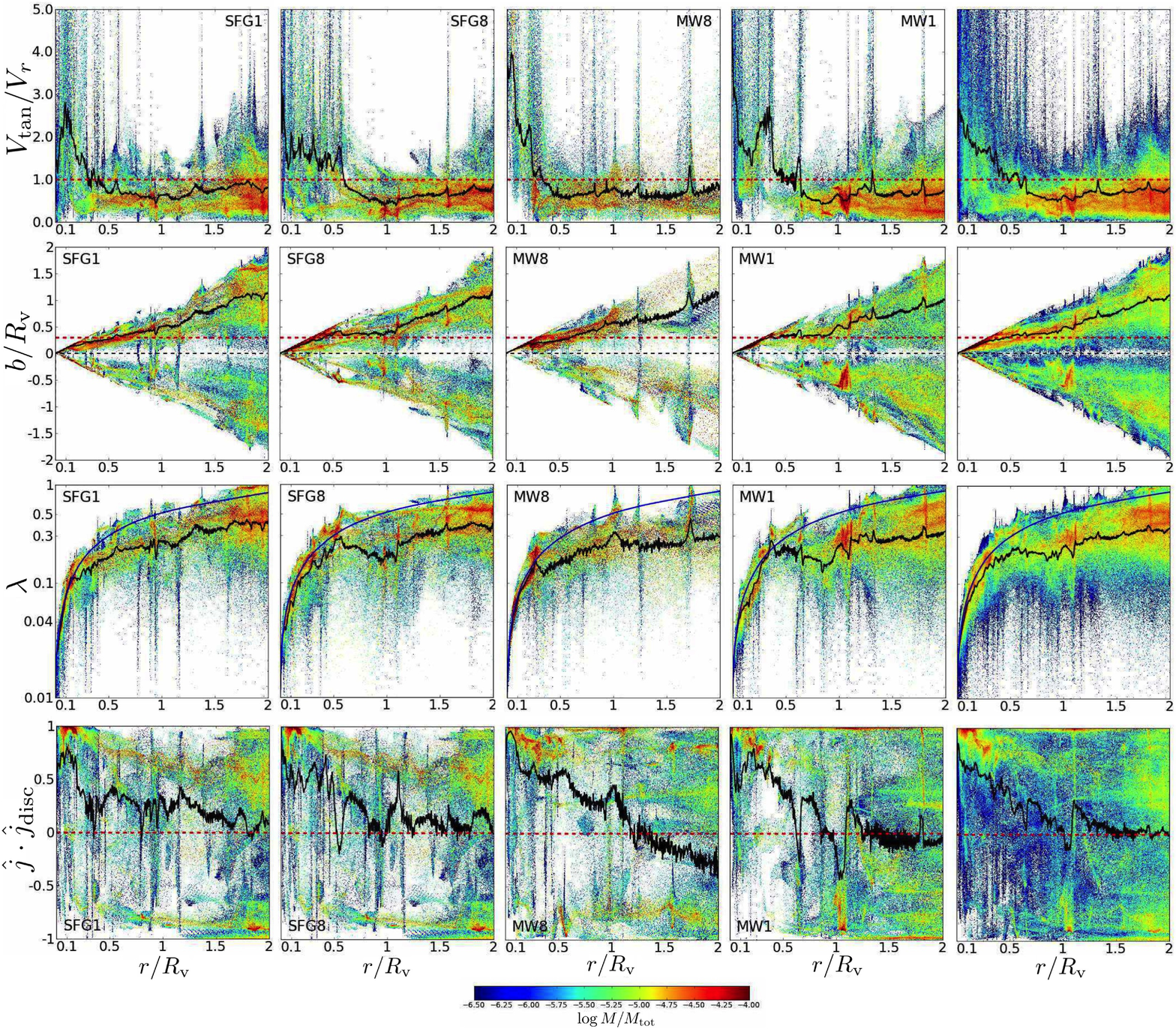}
\caption{
Kinematics in zones II and III.
Shown are 2D histograms for 4 galaxies and stacked (right column),
referring to the following parameters for the cold gas as a function of radius
{\bf Top:} Tangential to radial velocity ratio $V_{\tan}/V_r$.
{\bf Second row:} Impact parameter $b/\Rv$ 
(the red-dashed line marks $0.3\Rv$). 
{\bf Third row:} Spin parameter $\lambda$ 
(the blue line marks $(GM(<r)r)^{1/2}$).
{\bf Bottom:} Alignment of the AM with the disc AM 
$\hat{J} \cdot \hat{J}_{disc}$.
The colour corresponds to the log of the mass in each pixel.
The black line is the mass weighted mean at r. 
The snapshots from \tab{snaps} are SFG1 $z\!=\!3.17$, SFG8 $z\!=\!2.92$,
MW8 $z\!=\!2.33$, and MW1 $z\!=\!1.63$.
The outer halo in zone II is characterized by motions with a dominant radial
component, a rather constant impact parameter, a high spin parameter and an
AM direction that is poorly correlated with the disc AM.
At gradually smaller radii in the inner halo, zone III, one can see
an increasing circular velocity component, a declining impact parameter
associated with spiraling in, a declining spin parameter associated with AM
loss, and an increasing alignment with the inner disc.
}
\label{fig:messy_hists}
\end{figure*}

\begin{figure*} %15 
\centering
\includegraphics[width=\linewidth]{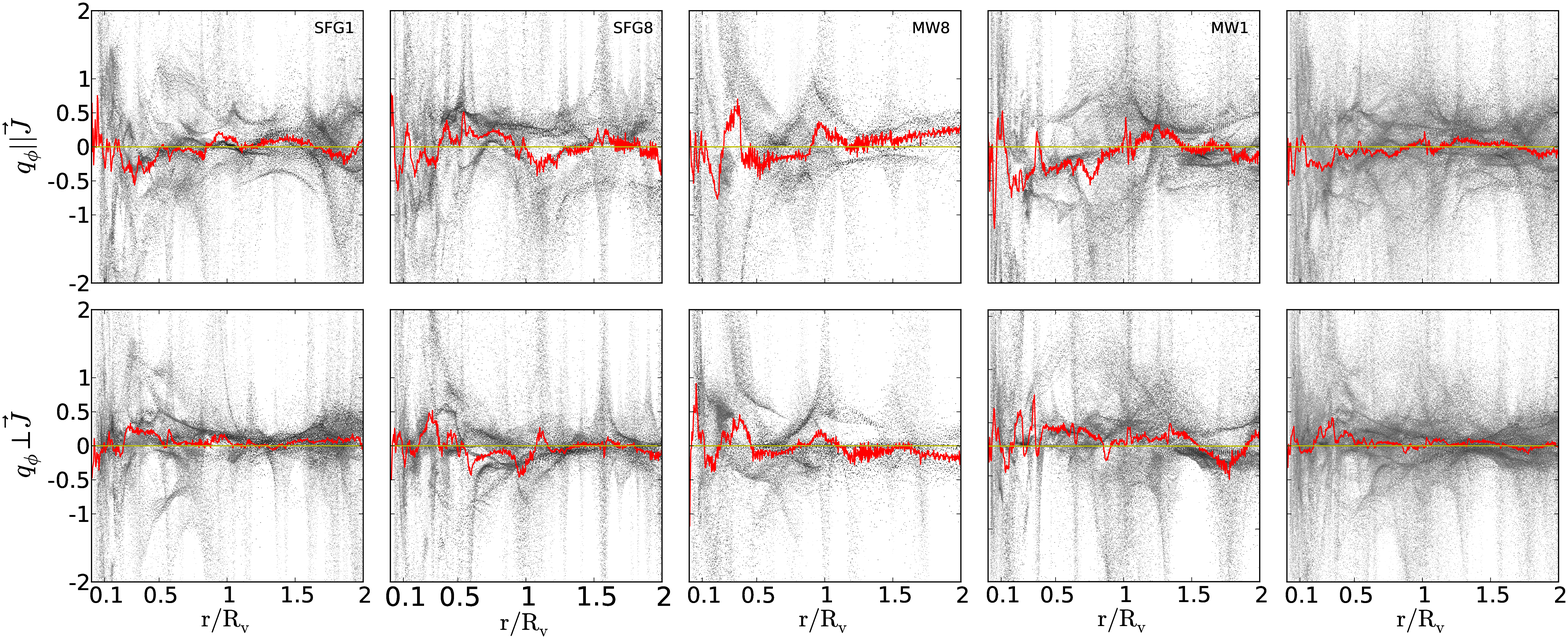}
\caption{
Radial profiles of normalized torques for the 4 galaxies shown in 
\fig{messy_hists} and stacked (right column).
{\bf Top:}
The torque parameter component in the direction of the AM of each cold
gas cell,
emphasizing the effect of the torque on the amplitude of the AM.
{\bf Bottom:}
The torque parameter component in the direction perpendicular to the AM vector
of each cold gas cell $\vec J$ and in the plane defined by $\vec J$
and the disc AM, measuring the tendency toward alignment with the disc.
The torques are ineffective in zone II, while in zone III they are capable 
of lowering the AM amplitude and increasing its alignment with the disc AM.  
These features are common to most galaxies in our sample. 
}
\label{fig:torque_prof}
\end{figure*}

%------------------ 5.2
\subsection{Kinematics of the extended ring}
\label{sec:ring_kinematics}

\Fig{messy_hists} shows the distributions of several quantities of interest
as a function of radius $r/\Rv$ for each of the 4 galaxies shown in \fig{messy}
and for the 4 galaxies stacked together (right column). 
Each grid cell is represented by a point coloured by log mass of cold gas
and the dark line marks the mass-weighted average at the given radius.
We note that in these pictures elongated horizontal features may refer to 
coherent streams, while vertical spikes may be associated with clumps.

% circular orbits
\smallskip
The upper row refers to the ratio of tangential to radial components of the
velocity $V_{\rm tan}/V_r$. 
We see that at large radii, in zones II and I,
this ratio is typically below unity, indicating rather radial inflow,
except for spikes that represent rotating satellite clumps with high 
$V_{\rm tan}$.
In the inner halo, 
the ratio is increasing toward the center, with the average
above unity at $r<(0.3-0.6)\Rv$, indicating an increasing tendency for circular
motion in zone III.

% impact parameter
\smallskip
The second row in \fig{messy_hists} is the impact parameter of the cold gas. 
It is computed using $b\!=\!|\vec{r}\!\times\! \hat{v}|$, 
where $\vec{r}$ is the position vector of the gas cell 
and $\hat{v}$ is the unit vector in the direction of 
the total velocity of the gas cell. 
The impact parameter is defined as positive or negative depending 
on the orientation of each cell's 
AM vector with respect to the disc AM vector. 
In this case the dark line is the mass-weighted average of the absolute value
of the impact parameter. 
The average tends to be rather flat or slowly declining with decreasing radius
in the outer halo, zone II, following a dominant stream that is represented 
by a horizontal concentration of red points. 
In zone III, the average curve following the dominant stream is declining 
linearly with decreasing radius, close to the boundary that refers to 
circular motion, $b\!=\!r$.
The outer radius of zone III is well defined in these plots.

% counter-rotation  zone III
\smallskip
We can also see in the second row of \fig{messy_hists} that
while most of the gas mass tends to be co-rotating with the disc, 
there is a significant fraction that is counter-rotating, with a negative
impact parameter, originating from counter-rotating streams.
We find that on average about $30\%$ of the instreaming gas mass
in the inner halo, $(0.1\!-\!0.3)\Rv$, is counter-rotating with respect to 
the net instreaming angular momentum (AM) in the same volume.
This component carries a negative contribution with an amplitude that is on
average $43\%$ of the positive component, namely $75\%$ of the net total 
inflowing AM. The counter rotation can thus have very important effects on 
the outer ring and on the inner disc (see \se{IV}).

%Streams of lower impact parameters settle into a smaller outer disc,

% spin
\smallskip
The third row of \fig{messy_hists}
shows the sAM of the cold gas, normalized as a spin parameter $\lambda$. 
The smooth blue line is an estimate of the maximum $\lambda$ possible 
for particles in circular orbits at the given radius under the assumption
that the mass distribution is spheri-symmetric.
We see that the 
mean is bellow the maximum for circular orbits, but there is some gas
slightly above the limit, indicating deviations from spherical symmetry.
The steep decline of the average curve in zone III, following most of the mass, 
is gradually approaching the boundary for circular orbits.

% alignment
\smallskip
The bottom row of \fig{messy_hists} shows the cosine of the angle between
the cold gas AM and the disc spin axis, with unity corresponding to perfect
alignment. 
Horizontal filaments of concentrated red points refer to streams.
The average shows poor alignment in zone II, with the cosine smaller than 0.4, 
implying uncorrelated orientations for the stream AM and the disc AM
\citep[as found in][]{danovich12}.
There is a marked increase in the alignment with the inner disc in zone III, 
from below 0.4 to 0.7-0.9.
The streams with misaligned AM seen in zone II either tend toward alignment 
or disappear in zone III.

%------------------- 5.3
\subsection{Torques in the ring}
\label{sec:ring_torques}

The gas reaching zone III with a non-negligible impact parameter and the
correspondingly high AM needs to lose AM in order to spiral in and change its
direction for alignment with the inner disc.  
We saw in \fig{torque_ratio} and \fig{torques_profile} 
that the gravitational torques dominate the AM exchange.
In zone III they should be strong torques, exerted locally, presumably
mostly by the inner disc, and possibly by dynamical friction.

\begin{figure} %16 
\centering
\includegraphics[width=1\linewidth]{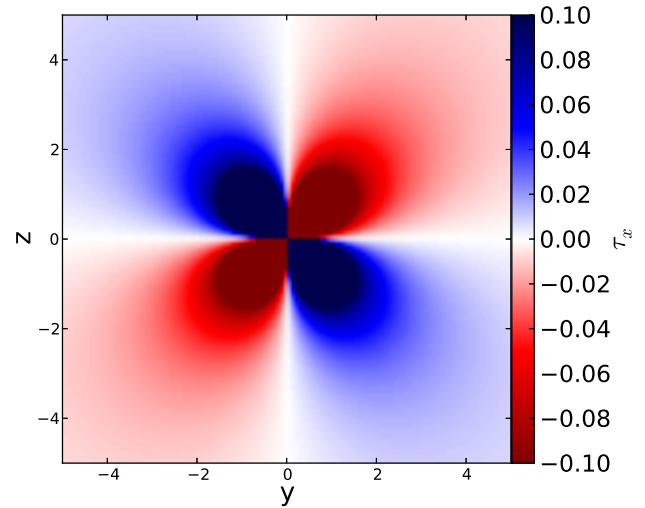}
\caption{
Gravitational torques exerted by an idealized uniform axi-symmetric
disc.
The disc is in the $x\!-\!y$ plane with its axis of symmetry along the $z$
axis.
The map shows the $x$ component of the torque (pointing out of the paper)
in the $x\!=\!0$ plane, perpendicular to the disc plane through its center.
The values of $G$ and the disc radius and mass are set to unity.
The torque shows a characteristic quadrupole pattern. The $z$ component of
the torque vanishes.
\label{fig:quad_disc}
}
\end{figure}

\begin{figure*} %17
\centering
\includegraphics[width=\linewidth]{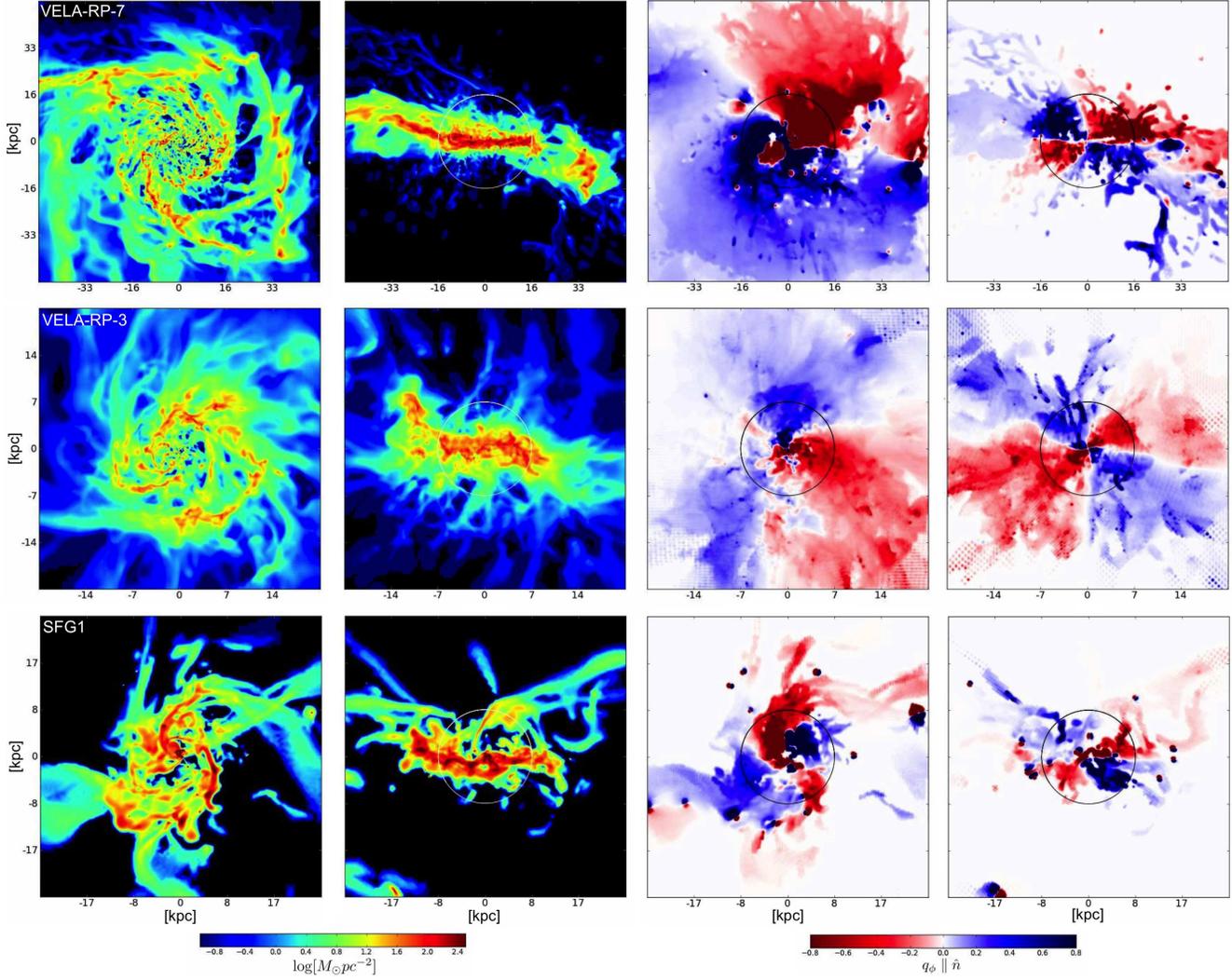}
\caption{
Torques in the the greater disc vicinity, zone III.
Shown for the cold gas in 3 galaxies
(VEL7 $z\!=\!1.17$, VEL3 $z\!=\!1.50$, SFG1 $z\!=\!3.17$)
are face-on and edge-on projected density maps (left columns),
and the corresponding normalized torque component
perpendicular to the plane shown, with $\tc\!=\!0.2\tv$ (right columns).
The box side is $0.6\Rv$ and the thickness of the slice is $0.3\Rv$.
The circle marks $0.1\Rv$.
In the edge-on views the torque shows a pattern that bears resemblance
to the quadrupole pattern produced by an idealized disc, \fig{quad_disc},
but strong torques are also seen in the $z$ direction, indicating deviations
of the torque source from an idealized disc.
}
\label{fig:torque_dir}
\end{figure*}

\smallskip
\Fig{torque_prof} shows the distributions of the torque components parallel 
and perpendicular to the local AM vector of the cold gas in grid cells, as a
function of radius.
It is to be compared to \fig{q_profiles_shells}, 
which refers to the torque components acting on the cold gas averaged over 
shells with respect to the AM in the shell. 
Four panels refer to the 4 galaxies shown in \fig{messy_hists} and the
right row shows the stacked distributions.
The sign of the parallel torque indicates whether it acts to increase or 
decrease the AM in that cell. 
The time assumed here for $q$ in \equ{q} is $\tc\!=\!0.2\tv$, 
a characteristic crossing time for zone III.
This serves as a lower limit for the actual time available for the torques to
act during the spiraling in of the streams in the ring.
The vertical features of high density of points 
correspond to neighborhoods of satellite clumps, 
where the force is directed toward the clump center but the radius vector
is measured with respect to the galaxy at the halo center.  
Horizontal features correspond to streams.
In the outer halo, zone II, the average parallel component is typically 
much smaller than unity.
In the inner region, there is a tendency for the parallel component to
become negative, namely acting to reduce the AM amplitude in each cell.
The horizontal filaments seen in zone II tend to bend toward
negative values upon entry to zone III, indicating streams that are
being increasingly torqued in the inner region. 

% shrinkage time
\smallskip
The timescale for spiraling in of the extended ring is 
the timescale for a change of order unity in the AM magnitude,
namely $t_\tau\! \sim\! |q_{\parallel}|^{-1} \tc$.
With $q_{\parallel}\! \sim\! -0.3$ and $\tc\!=\!0.2\tv$ in \fig{torque_prof}, 
this is $t_{\tau} \!\sim\! 0.6\tv$, namely on the order
of the characteristic orbital time in zone III.
A similar conclusion can be obtained from the torque profiles shown
in \fig{q_profiles_shells}, where in zone III 
$q_{\parallel} \!\sim\! -1$ and where $\tc\!=\!0.5\tv$, 
so $t_\tau \!\sim\! 0.5\tv$.
This is consistent with the visual impression from \fig{ring_density}
and \fig{ring_streamlines}.

% alignment time
\smallskip
The bottom panels of \fig{torque_prof} show 
the perpendicular component of $q$ with respect to the local AM $\vec J$, 
chosen in the plane defined by $\hat J$ and $\hat J_{\rm disc}$, 
thus measuring the torque that tends to change the alignment of 
the local AM with the disc AM (see \se{IV}).
In the outer halo, the alignment torque is typically rather small,
consistent with what we saw in \fig{q_profiles_shells}.
In the inner halo, it shows larger fluctuations that on average are 
positive, namely, they act to align the outer ring with the inner disc,
as seen in \fig{messy_hists}.
A similar tendency for alignment is seen in the torque profiles shown
in \fig{q_profiles_shells}, where in zone III the perpendicular torque
is $q_{\perp}\!\sim\! +1$. 

%-------------------------------- 5.4
\subsection{On the origin of the torques}

% toy model disc exerting torques
The AM exchange in the ring in zone III may largely arise from the
gravitational torques exerted by the inner disc.
To explore this possibility, we appeal to a simple toy model consisting of a 
uniform axisymmetric disc with radius $R$ and mass $M$
about an origin in the $x\!-\!y$ plane.
The gravitational potential generated by the disc at a point $(r,\theta)$,
expanded to its monopole and quadrupole terms, is
\be
\phi(r,\theta)=
-\frac{GM}{r}+\frac{\alpha G M R^2}{r^3} P_2(cos\theta)
+O\left(\frac{R^4}{r^5}\right) \, ,
\ee
where $\theta$ is the spherical polar angle measured from the $z$ axis,
$\alpha$ is a numerical factor, and $P_2(x)\!=\!(3x^2-1)/2$ is the
second Legendre polynomial. 
The monopole and quadrupole terms are sufficient for a good approximation far 
enough from the disc.
The only contribution of the force $\vec{F}\!=\!-\nabla \phi$ to the torque
is through its polar-angle component 
$F_{\theta}\!=\!-r^{-1} \partial \phi/ \partial \theta$. 
The resulting torque, acting in the azimuthal direction ($\hat\varphi$), is
\be
\tau_{\varphi}=rF_{\theta}
\propto \frac{\alpha GMR^2}{r^3}\cos\theta\sin\theta \, .
\ee
The torque component in an edge-on direction, e.g., the Cartesian direction 
$x$ in the disc plane, is
\be
\tau_{x}
%=\vec{\tau}_{\varphi}\cdot\hat{x}
=\tau_{\varphi}\cdot (-\sin\varphi) 
\propto - \frac{\alpha GMR^2}{r^3}\cos\theta\sin\theta\sin\varphi \, .
\ee
\Fig{quad_disc} shows the resulting pattern of the amplitude of this
torque $\tau_{x}$ as viewed edge-on (namely projected in the $y-z$ plane).
The quadrupolar pattern, with alternating signs in the borders between 
quadrants where the torque reverses its direction, is very characteristic.

% what we learn from the toy
\smallskip
From this simplified analysis, we learn that the torques due to
an axisymmetric disc do not affect the AM component parallel to the disc axis
of symmetry.
It tends to add an AM component in a direction perpendicular to the disc axis, 
inducing rotation in a plane perpendicular to the disc 
(namely about an axis that lies in the disc plane), and thus moving mass 
toward the disc plane.
In the case of an extended outer ring that is tilted with respect to the inner
disc, the inner disc applies torques perpendicular to the AM
of the ring particles, which results in a precession of their AM vectors.
Being composed of collisional gas, the differential precession in the extended
ring is associated with dissipation, which causes the outer ring to gradually 
align itself with the inner disc. 
We learn that the torques exerted by the inner disc work in general to align 
the ring and the disc.
As long as they are not yet aligned, namely in the outer parts of zone III,
these torques can also reduce the AM amplitude and induce the spiraling in of
the extended ring.

\begin{figure} %18
\centering
\includegraphics[width=1\linewidth]{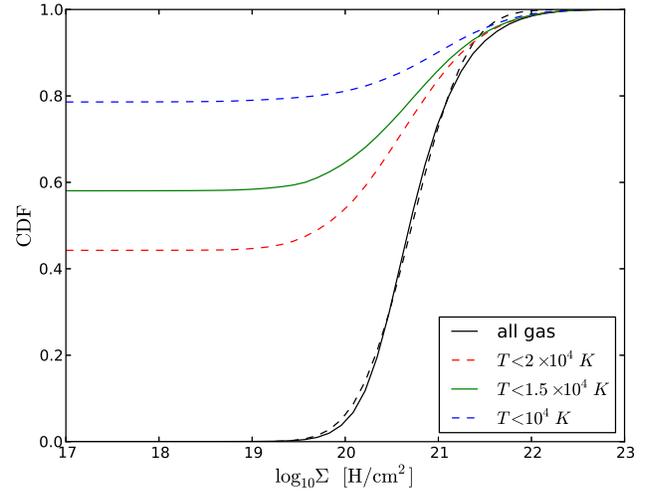}
\caption{
Cumulative distributions (CDF) of the HI column density along random lines
of sight through the extended ring, averaged over all snapshots of all
galaxies.
Each of 400 lines of sight has a circular aperture of radius $300\pc$.
The lines of sight are chosen at random orientations
and with random points of intersection with the ring at projected radii
in the range $(0.1\!-\!0.3)\Rv$, averaged over all snapshots in the sample of
simulations.
The HI gas is crudely approximated by all gas with $n\!>\!0.01 {\rm H}\cmc$ and
$T\!<\!1.5\times 10^4$K (green),
with upper and lower limits provided by $T\!<\!10^4$K (dashed blue)
and $T\!<\!1.5\times 10^4$K (dashed red), respectively.
The CDF of all the gas
is shown for reference (black) and fitted by a Gaussian PDF
(dashed black).
While about 58\% of the lines of sight are of low HI column densities,
about 30\% are above a column density of $2\times 10^{20} {\rm H}\cms$
and are thus observable as damped-Lyman-alpha systems.
}
\label{fig:los_surface_density}
\end{figure}

% torque directions in sims
\smallskip
\Fig{torque_dir} shows maps of gas density and torque magnitude in 
directions face-on and edge-on with respect to the discs in 3 simulated
galaxies.
The edge-on projections show a quadrupolar pattern that is rather similar 
to the pattern shown in \fig{quad_disc} for the 
idealized case where the source is a pure disc.
This is consistent with the conjecture that the inner disc is an important 
source of torque in zone III, with a significant contribution to the
reorientation of the instreaming outer ring. 
However, the face-on projections also show a non-vanishing torque component
along the disc axis of symmetry, which indicates that the source of the
torque in zone III deviates from a pure disc.
This is not surprising given that the high-redshift discs are very turbulent 
and perturbed, undergoing violent disc instability and frequent mergers,
and given the complex structure in zone III itself.

%--------------------- 5.5
\subsection{Observing the extended ring}

We find that the rotating ring-like structure typically extends to
$0.3\Rv$, e.g., about $30\kpc$ for massive galaxies at $z\!\sim\!2$,
and its radius scales with the virial radius.
This is a refinement of earlier results (based on a few simulations with lower
resolution) that indicated a somewhat more extended rotating pattern filling
a large fraction of the halo \citep{stewart11,stewart13}.
We find that the cold gas in the outer halo is predominantly flowing in 
along rather straight coherent streams, with a certain tangential velocity 
component associated with the impact parameter of the stream.

\smallskip
The extended ring tends to be tilted relative to the disc plane, with the tilt
increasing with radius.
The mean tilt angle between the disc and the spin axis of the cold dense gas
at $r\!=\!(0.1\!-\!0.3)\Rv$ is found to be $\sim\! 47^\circ$, namely largely 
uncorrelated with the orientation of the inner disc.

\smallskip
The internal structure of the extended ring is expected to be
filamentary and clumpy, with the cold gas actually occupying only a fraction 
of the disc area.
\Fig{los_surface_density} shows an estimate of the cumulative distribution of 
HI column density along random lines of sight through the extended ring, 
at projected radii in the range $(0.1\!-\!0.3)\Rv$, stacked for all snapshots 
in the sample.
Following \citet{goerdt12}, 
the HI gas is crudely approximated by the cold gas below $1.5\times 10^4K$,
with upper and lower limits provided by $T\!<\!10^4$K and 
$T\!<2\!\times\! 10^4$K, respectively. 
A threshold density of $n\! >\! 0.01 {\rm H}\cmc$ is applied in order to crudely
select cells that are expected to be shielded against ionizing radiation 
\citep[justified based on Figure 2 of][]{goerdt12}.
Following the division into different column density classes in 
\citet{fumagalli11}, we see that $\sim\!30\%$ of the lines of sight are
observable as damped Lyman-$\alpha$ absorbers  
($N_{HI}\! >\! 10^{20.3} {\rm H}\cms$). 
Then 
$\sim\!11\%$ are Super Lyman-limit systems 
($N_{HI}\! =\! 10^{19}\!-\!10^{20.3} {\rm H}\cms$),
and less than a percent are Lyman-limit systems
($N_{HI}\! =\! 10^{17.2}\!-\!10^{19} {\rm H}\cms$).
About $58\%$ of the lines of sight have HI column densities below 
$10^{17} {\rm H} \cms$, most of which are totally devoid of HI in our
simulations. 

\smallskip
The metallicity of the cold dense gas within $r\!=\!(0.1\!-\!0.3)\Rv$
averaged over all galaxies and snapshots is about $Z\!\simeq\!0.1$ solar.
%$[O/H]+12\sim 7.7$ (compared to the solar value of 8.66).
This is in the ball park of the observed metallicities in
damped Lyman-alpha systems \citep{fumagalli14b}.

\smallskip
Our results concerning the detectability of the outer ring
should be regarded as a preliminary proof of concept.
The detectability may depend on mass and on redshift, and may be affected by
different levels of feedback. Our analysis here is rather simplified, not
involving proper radiative transport \citep[unlike, e.g.,][]{fumagalli11}.
By limiting the analysis to dense, self-shielded gas, we may miss the 
contribution of ionized gas to Lyman-limit systems.
We defer detailed predictions for the observability of the outer ring
to future work.

%%%%%%%%%%%%%%%%%%%%%%%%%%%%%%%%%%% 6
 \section{Phase IV: AM evolution in the inner disc}
\label{sec:IV}

%-------
\subsection{Spin in the galaxy: VDI and feedback}

Phase IV involves final adjustments of the AM within the inner disc. 
Processes responsible for substantial changes in the AM are feedback and VDI.
Here we only discuss these in a qualitative way, deferring a more quantitative
analysis of the AM evolution in the disc to another paper.  

\begin{figure*} %19 
\centering
\includegraphics[width=1\linewidth]{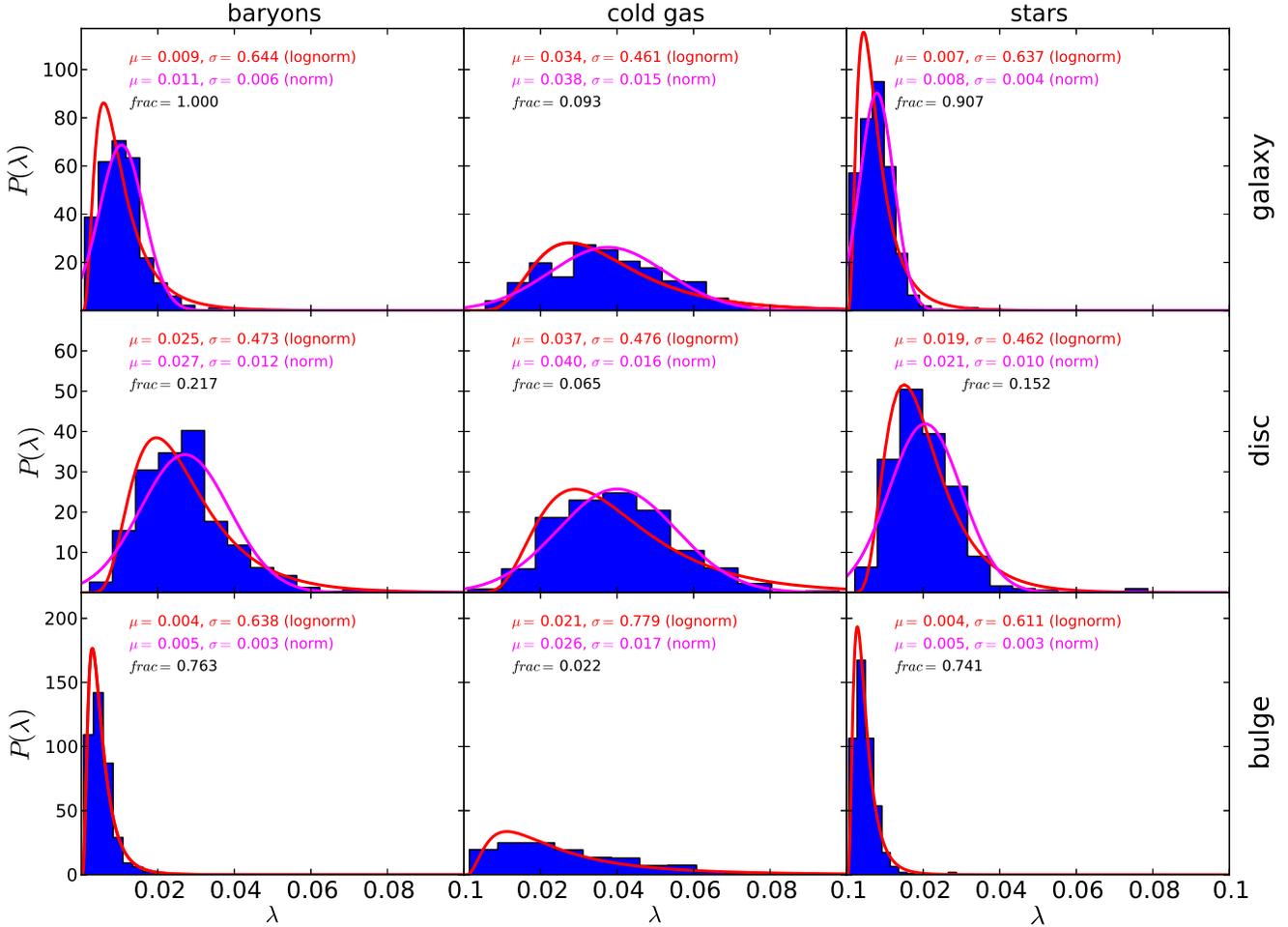}
\caption{
Galaxy spin parameter distributions over all galaxies and snapshots.
The columns refer to all the baryons, and divided into
cold gas and stars (the hot gas is negligible).
The rows refer to the whole galaxy, and divided into disc and bulge.
The distributions were fitted by a log-normal (red) and a normal (magenta)
distribution, with the quoted mean and standard deviation.
The mass fraction in each component is quoted.
The disc is confined to a cylinder of radius $\Rd$ (\tab{gals})
and height as defined for each galaxy at a given redshift in
\citet{mandelker14}
Only galaxies with a minor-to-major axial ratio smaller than 0.5 are
considered.
Disc gas and stars are chosen to satisfy the kinematic criterion
$j_z \leq 0.8 j_{\rm max}$ while the rest of the stars are assigned to
the bulge.
The average spin of the baryons in the disc is $\sim\!0.027$,
only moderately smaller than that of the dark matter in the halo,
while the spin of the baryons in the galaxy including the bulge
is only $\sim\!0.01$.
In simulations with stronger feedback, the galaxy spin is typically
higher by $\sim\!50\%$.
}
\label{fig:disc_spin}
\end{figure*}

\smallskip % spin distribution
\Fig{disc_spin} shows the probability distribution of the spin parameter
for the baryons, cold gas and stars within the whole galaxy, the disc and 
the bulge components.
Following \citet{mandelker14},
the disc is confined to a cylinder of height $\pm 1\kpc$ 
and radius $\Rd$ (listed in \tab{gals})
that encompasses 85\% of the gas mass out to $0.15\Rv$.
The disc is also selected kinematically to consist of gas cells or stellar 
particles for which $j_z\!\geq\!0.8\,r\,v$, where $r$ is the radial distance 
from the center and $v$ is the cell or particle speed. 
This limits the disc cells or particles
to those where the sAM component perpendicular to the disc is larger than 
80\% of the maximum possible for the given energy at the particle position.
 
\smallskip %disc spin
We see that the mean spin parameter of the gas disc is 
$\lambda_{\rm cold}\!=\!0.040$, higher than that of the stellar disc, 
$\lambda_{\rm stars}\!=\!0.019$. 
The spin of the baryons (gas and stars) in the disc is naturally in between, 
$\lambda_{\rm bar}\!=\!0.025$,
closer to the stellar disc that constitutes $\sim\!70\%$ of the mass.
These disc spins underestimate the spins obtained in simulations with
stronger feedback by $\sim\!33\%$ (see \se{feedback}).

\smallskip % disc vs halo spin
The spin parameter of the disc is thus lower than that of the cold gas in
the halo by a factor of 3-4, 
consistent with substantial AM loss by torques in zone III, 
as discussed in \se{III}.
It is interesting to note that the disc spin is comparable to and somewhat 
smaller than that of the dark matter in the halo \citep[see e.g.,][]{dutton12}. 
This result is not very different from the result obtained by 
the simplistic cylindrical contraction model despite the very different 
AM evolution tracks for the cold gas and 
for the dark matter, as discussed in the current paper. 
This is good news for semi-analytic model of galaxy formation, which commonly
adopt the disc sizes based on the simplistic model.

\smallskip  % Gaussian fit
It can be seen in \fig{disc_spin} that,
unlike the other components in the galaxy and in the halo,
the distribution of $\lambda$ for the gas disc is better fit by a Gaussian
than by a log-normal distribution.

% feedback
\smallskip
The sAM of the gas disc is expected to be larger than that of the stellar 
disc for two reasons.
First, the gas is a more direct tracer of the recently arrived mass, which is
expected to come in with a higher sAM, while the stars are made of baryons that 
arrived earlier with lower sAM.
Second, the gas that arrived earlier was subject to outflows generated
by stellar, supernova and AGN feedback. These mechanisms preferentially 
remove gas from the dense central regions where the star-formation rate (SFR)
is high and the 
AGN activity takes place, and therefore the outflows tend to increase the 
effective sAM of the remaining gas in the disc 
\citep[e.g.,][]{maller02,governato10,brook11,brooks11,guedes11}.

% VDI
\smallskip
The high-redshift massive discs with high gas surface density 
undergo violent disc instability, in which the discs develop large-scale
elongated transient perturbations and small-scale giant clumps.
These perturbations exert mutual torques, which preferentially drive AM out
carried by a small fraction of the disc mass.
In response, most of the disc mass flows in, 
as clump migration and as off-clump gas inflow. 
This leads to the formation of a central bulge of relatively low AM.  
The timescale for inflow is on the order of one or two disc orbital times
\citep{dsc09,dekel13}.
It has been argued by \citet{db14} that
the original spin parameter of the gas disc, 
which determines the initial gas surface
density in the disc, determines whether the gas inflow rate is faster
or slower than the SFR. If the spin is relatively low, then
the gas fraction is high,
the inflow is faster than the SFR, 
it is gaseous and dissipative, and it leads to compaction into a 
gas-rich, high-SFR ``blue nugget".
If the spin is relatively high, then the gas fraction is lower, 
the SFR is faster, and the result is a more extended stellar disc.
This is expected to lead to a bi-modal distribution of star-forming galaxies,
the extended clumpy discs and the compact blue nuggets, with the latter more
frequent at higher redshifts.
The extended discs had more AM to begin with, and have not lost much during
their VDI phase, so they are expected to be of high AM.
The blue nuggets had less AM initially, and have lost much of it during
compaction, being carried away by an outer, co-planar,
gaseous and star-forming ring \citep{genzel13,zolotov15},
not to be confused with the tilted extended ring at larger radii discussed in
\se{III}.
However, feedback is likely to remove more low-AM central gas from the
nuggets, thus partly compensating for the AM loss and bringing the mean sAM up. 

\smallskip % feedback in zone IV
The final disc spin parameter in phase IV is sensitive to the strength of the
feedback processes. We briefly mention in the next section
preliminary estimates that indicate an increase of $\sim\!50\%$
in the spin of the inner disc when stronger feedback is incorporated,
while the effects on the outer ring and gas in the halo and outside it
are much smaller.

%--------------
\subsection{Counter rotation} % counter rotation - disc flips - zone IV

We saw in \se{III} that the amplitude of the counter-rotating component of 
the instreaming gas in $(0.1-0.3)\Rv$ is $\sim\!75\%$ of the amplitude of the 
net AM of that component. 
In order to evaluate the effect of the counter-rotating streams on the disc, we
measure $\Jneg$, the AM component in the direction anti-parallel to the disc
AM, for the instreaming gas at radial distances and velocities that will
bring it to the disc in less than an outer-disc orbital time.
This time is typically $\torb\! \sim\! 250 \Myr$ in our galaxies at 
$z\! \sim\! 2$, involving most of the instreaming gas at $0.1\!-\!0.5\Rv$.
For the galaxies at $z\!=\!2\!-\!3$,
the median of $\Jneg/\Jd$ (where $\Jd$ is the AM of the gas disc)
is about 10\%, but about 20\% of the galaxies have $\Jneg/\Jd\! \sim\! 1$.
Indeed, in a period of $250\Myr$, about $15\!-\!20\%$ of the discs flip their
orientations, namely change their spin direction by more than $90^\circ$.
This substantial fraction of counter rotation can affect the galaxy in 
several different ways.
The gaseous streams can be important in stimulating the violent
disc instability in high-redshift galaxies, which seems to be nonlinear
with a Toomre parameter significantly above unity 
(Inoue et al. in preparation).
In turn, it can trigger a dramatic wet compaction event into a compact 
star-forming ``blue nugget", followed by quenching to a compact  
``red nugget" that eventually makes the core of an elliptical galaxy
\citep{db14,zolotov15}.
Indeed, our simulations indicate that the frequency of spin flips 
is suspiciously similar to the fraction of galaxies that
start their compaction event in a similar period.

\smallskip
Alternatively,
the stellar component of the streams could lead to the formation of
a stellar disc with two counter-rotating components relative to each other,
as observed and as analyzed in simulations \citep{navarro13}.

%%%%%%%%%%%%%%%%%%%%%%%%%%% 7
\section{Possible Effects of Stronger Feedback}
\label{sec:feedback}

%---------------
\subsection{Simulations with different feedback strength}

% limitations of gen 1
The simulations used here are state-of-the-art in terms of high-resolution
AMR hydrodynamics and the treatment of key physical processes at the
subgrid level. In particular, they properly trace the cosmological streams
that feed galaxies at high redshift, including mergers and smooth flows,
and they resolve the violent disc instability that governs the high-$z$ disc
evolution and plays an important role in bulge formation
\citep{cdb10,ceverino12,ceverino14_e,mandelker14}.
The AMR code is superior to the common SPH codes
for properly tracing some of
the high-resolution hydrodynamical processes involved
\citep[e.g.,][]{agertz07,scannapieco12,bauer12},
and it seems to be comparable in its performance to new codes using a moving
unstructured grid \citep{bauer12}.
However, like other simulations,
the current simulations are not yet doing the most accurate possible job
in treating the star formation and feedback processes.
For example, the code used here
assumes a somewhat high SFR efficiency per free-fall
time, it does not follow in detail the formation of molecules
and the effect of metallicity on SFR \citep{kd12},
and it does not explicitly include radiative stellar feedback
\citep{murray10,kd10,hopkins12c,dk13}
or AGN feedback \citep{silk98,hopkins06,ciotti07,booth09,cattaneo09,
fabian12,degraf15}.
Therefore, the early SFR is overestimated,
while the suppression of SFR in small galaxies is underestimated,
resulting in excessive early star formation prior to $z\!\sim\!3$,
by a factor of order 2.
As a result, the typical gas fraction and SFR at $z\! \sim\! 2$ are
lower by a factor of $\sim\! 2$ than the average observed values in
star-forming galaxies \citep{cdb10,daddi10,tacconi10}.
Furthermore, the simulated mass-loading factor of the galactic outflow rate 
with respect to the SFR is typically $\eta\!\lsim\!1$, not 
reproducing some of the observed strong outflows with mass loading factors of 
$\eta\!\gsim\!1$ \citep{steidel10,genzel11,dk13}.
This leads to a stellar-to-halo mass  fraction of $\sim\! 0.1$ within the
virial radius, a factor of $\sim\!2\!-\!3$ higher than the observationally
indicated value
\citep[e.g.,][]{perezgonzalez08,moster13,behroozi13}.

\smallskip % gen 3
We are now in a process of analyzing a new suite of galaxies simulated 
with higher resolution
($17.5\!-\!30\pc$), lower SFR efficiency and stronger stellar feedback
including radiative feedback. They are described in
accompanying papers \citep{ceverino14,moody14,zolotov15}.
The agreement with the gas fraction, SFR and stellar-to-halo mass fraction
deduced from observations is significantly better.
Preliminary tests of the AM evolution in these simulations indicate that
our findings from the current simulations are robust outside the halo
and in most of the halo volume. The stronger feedback makes only
a small difference in the inner halo, and a somewhat stronger but still
small difference in the disc, with no qualitative change.
We provide here a brief, preliminary report on these differences,
deferring a detailed study of the sensitivity of galaxy AM to feedback 
strength, including AGN feedback, to future work.

%-------------
\subsection{Robustness to feedback in the halo}

\smallskip % fig 1
Starting with \fig{spin_prof},
the spin profiles in the simulations with stronger feedback are
similar within $\sim\!10\%$ for all the component at all redshifts,
outside the halo, in the outer halo and in the inner halo.
%except that the spin of the cold gas near $r=0.1\Rv$ is lower by $\sim\!20\%$.

\smallskip % fig 6
In \fig{spin_dist}, which displays the distributions of spin parameters for
different components,
the simulations with stronger feedback reveal     
a total baryonic mass in the halo that is lower by $\sim\!35\%$, %36\%
while its spin is higher by $\sim\!40\%$, %42\%, 
indicating either more low-spin outflow from the halo 
or less high-spin inflow into the halo (DeGraf et al., in preparation).
The baryonic spins in the volumes
$r\!=\!(0.1\!-\!1)\Rv$ and in $r\!<\!\Rv$ become higher
by only 10\% and 17\% respectively, indicating that the larger 
difference in $r\!<\!\Rv$ is largely
due to a higher coherence between the directions of the AM vectors of the
different components in the simulations with stronger feedback.
In addition, due to the stronger feedback and lower SFR efficiency,
the mass fraction of cold gas in the halo is doubled at
the expense of stellar fraction in the halo and galaxy. 
However, the spin of the cold gas in the halo vary by only $\sim\!10\%$,
while the stellar spin becomes lower.

\smallskip % fig 7
In \fig{spin_evol}, showing the evolution of the average spin parameter for
different components, the only noticeable effect of stronger feedback seems to 
be that the spin of cold gas in the halo plus stars in the inner halo becomes 
larger, by a factor of $\sim\!2$. A comparison to the smaller differences
seen in \fig{spin_dist} indicates again that this strong effect is largely 
due to a higher coherence between the directions of the AM vectors of the 
different components in the simulations with stronger feedback.

\smallskip % fig 11
Moving to the inner halo, we included 
in \fig{proj_edge} pictures of two snapshots from the simulations
with higher resolution and stronger feedback (two top-left panels).
These images demonstrate that the outer ring is a common feature there too.
The pictures highlight the tilt of the outer ring,
which in one extreme case extends out to beyond $30\kpc$ (VEL7).

\smallskip % fig 15
In \fig{torque_prof}, which addresses the normalized torque components parallel 
and perpendicular to the AM, we find that the same general features 
are revealed by the simulations with stronger feedback.
The tendency for lowering the AM amplitude of the outer ring in
the inner halo becomes somewhat stronger, possibly due to a somewhat
larger extent of the disc.

%-----------------
\subsection{Feedback effects in the inner galaxy}

\smallskip % disc  fig 19
Somewhat more noticeable effects of the boosted feedback are limited to the 
inner galaxy, zone IV.
The hot component is negligible there, so the baryons consist of stars and
cold gas.
In \fig{spin_dist}, the stronger feedback is making the baryon spin within  
$0.1\Rv$ become higher by $\sim\!10\%$. % 8\%
However, in \fig{disc_spin}, the stronger feedback causes a larger increase of
$\sim\!50\%$ inside the disc radius $\Rd$.
One can see in \tab{gals} that $\Rd$ is typically smaller than $0.1\Rv$.
This implies that most of the spin increase by feedback is within the disc,
inside $\Rd$, typically at radii of one to a few kpc.
This is consistent with low-spin outflows from the central regions.
The bulge-to-total stellar mass ratio in the galaxy decreases due to the
stronger feedback by about 20\%.
The changes in the spins of the other components in \fig{disc_spin}
are typically less than 20\%.

\smallskip
Based on these preliminary comparisons to the simulations with stronger
feedback, we conclude that our results concerning the AM-buildup scenario
based on the current simulations are qualitatively robust.
The spin estimates seem to be accurate to within $\sim\!10\%$ outside and 
inside the halo, including in the outer ring. 
On the other hand, the spin of the inner galaxy may be underestimated,
by a factor of $\sim\!33\%$, with the bulge-to-total ratio overestimated 
by $\sim\!20\%$.

\begin{figure} %20   
\vskip 5.6cm
\includegraphics{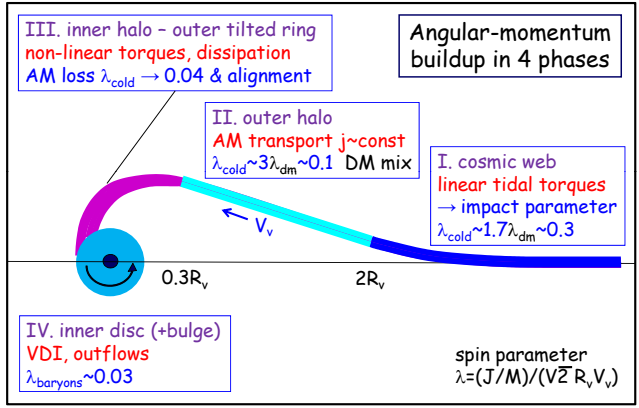}
\caption{
Origin of angular momentum in galaxies fed by cold streams in four steps.
Phase I is the linear TTT phase, occurring in the cosmic web outside the halo,
where the thin, elongated cold streams acquire more sAM than the dark matter,
expressed as impact parameters.
Phase II is the transport phase, where the streams flow roughly along straight
lines down to $\sim\!0.3\Rv$.
Phase III is the strong-torque phase, in a tilted, inflowing, extended ring
at $(0.1\!-\!0.3)\Rv$.
Phase IV involves the inner inner disc and the growing bulge, where the final
rearrangement of AM occurs due to VDI torques and outflows.
}
\label{fig:fig_am}
 \end{figure}

%%%%%%%%%%%%%%%%%%%%%%%%%%%%%%%%%%%%% 8
\section{Discussion and Conclusion}
\label{sec:conc}

% short-comings in the standard picture 1
The common understanding of AM acquisition by galaxies has been based on the
following simplified conjectures
\citep[e.g.,][]{fe80,mmw98,bullock01_j}:
(a) The origin of AM is outside the halo, prior to maximum expansion,
and can be understood in terms of TTT.
%where a second-order expansion of the potential about the proto-halo center 
%is sufficient.
(b) The baryons and the dark matter initially occupy the same Lagrangian
volume and reach a well-defined maximum expansion near which they acquire most
of the AM. Therefore they obtain the same sAM by TTT.
(c) In TTT, a second-order expansion of the gravitational potential about the
proto-halo centre is sufficient.
(d) The gas is heated by a virial shock and it initially shares 
with the dark matter the same mass and AM distribution within the halo.
(e) The gas feeds the galaxy by wide-angle cylindrical infall 
and it therefore conserves AM during the infall.
(f) The AM of the infalling gas determines the density profile of the disc.
Most analytic and semi-analytic modeling of galaxy formation has been based 
on these conjectures.
Here, we addressed the origin of AM in high-redshift galaxies
using high-resolution zoom-in cosmological simulations.
We analyzed the AM evolution in the context of our developing understanding 
of massive galaxy formation at the nodes of the cosmic web.
We found that most of the common simplifying assumptions are actually invalid
at high redshift, 
and that the picture of AM evolution is quite different than originally 
thought.  Nevertheless, at the end of the day,
the AM distribution in the final disc is not extremely different from
the results of the toy modeling based on the simplified assumptions.

% failures of common picture 2
\smallskip
% new
The basic ideas of TTT,
and the effective gain of AM prior to maximum expansion in phase I outside
the halo, seem to be qualitatively valid. 
However, the cold gas, confined to elongated pre-contracted thin streams, 
gains sAM higher than that of the dark matter by a factor 1.5-2 
because of the high quadrupole moment associated with the thin gas streams
while the streams are still near maximum expansion along their major axes.
TTT is thus valid but it has to be applied using the inertia tensor of the cold
gas after its early contraction to the central cords of the filaments.
The dark matter, on the other hand, streams along much thicker filaments and
between them, so, in practice, it accretes from many directions, and thus 
obeys better the standard TTT applied in Lagrangian space.
In phase II in the outer halo, while the dark matter virializes in the halo
and partly mixes with the lower sAM mass that arrived earlier, 
the cold gas flows in as coherent streams, roughly conserving AM while
transporting it to the inner halo, and only there it mixes with the baryons 
that arrived earlier.  As a result, the spin parameter of the cold gas in the 
halo is $\sim\!3$ times larger than that of the dark matter.
In phase III, at the greater disc vicinity in the inner halo,
the cold gas settles into an extended non-uniform rotating ring, where 
torques, partly exerted by the inner disc, make the cold-gas 
AM decrease in amplitude 
and align with the inner disc, namely the cold-gas AM is not conserved, and is
shared with the rest of the gas, the stars, and the dark matter. 
Finally, within the inner disc, outflows increase the effective sAM,
and VDI drives AM out and thus makes the density profile more centrally
concentrated. This scenario of AM buildup is very different from the common
simplified cylindrical infall model. 

% success of toy model, analytic and SAM 3
\smallskip
The encouraging surprise is that the final distribution of spin parameter of 
the baryons in high-z massive discs is not extremely different from that of 
the dark matter in the haloes, both roughly resembling a lognormal distribution 
with means of $\la \lambda \ra \!\simeq\! 0.03$ and $0.04$ respectively. 
The stellar disc component has a 
somewhat lower average spin, $\la \lambda \ra \!\simeq\! 0.02$,
while for the gas component it is somewhat larger, 
$\la \lambda \ra \!\simeq\! 0.04$. 
This implies that the disc sizes predicted by analytic and semi-analytic models
are adequate crude approximations with a systematic error smaller than a factor
of 2.

% the new picture 4
\smallskip
Our analysis reveals four phases in the acquisition of AM by the baryons
that make high-redshift massive galaxies, as illustrated in \fig{fig_am}. 
%I
In phase I, well outside the halo, the cold streams gain AM
by linear tidal torques from the cosmic web, exerted on the cold gas after its
early contraction to the central cords of the filaments feeding the halo. 
This can also be viewed as
asymmetric transverse flows from voids and pancakes into the streams
\citep{pichon11}. 
The sAM of the cold gas in narrow streams is higher by a factor 1.5-2 than that
gained by the dark matter.
The AM of the streams is expressed as impact parameters with respect to the
galaxy center.
The dominant stream typically develops an impact parameter of $\sim\!0.3\Rv$, 
while the second and third streams have smaller impact parameters, possibly
counter-rotating.
%II
Phase II involves the coherent transport of cold gas mass and AM 
into the halo and through its outer regions. While the dark matter mixes,
the cold streams remain coherent and keep a roughly constant velocity
and impact parameter, thus largely preserving AM. The spin parameter of the 
cold gas in the halo is $\sim\!3$ times larger than that of the dark matter.
% III
The pericenter of the dominant stream dictates the radius of zone III,
where the streams dissipate, bend, and settle into an extended ring,
a perturbed non-uniform rotating pattern in the greater vicinity of the inner 
disc.\footnote{We use the terminology of ``outer tilted ring" in order to
distinguish this new feature from the ``inner disc", 
but the outer ring could sometimes be referred to as an ``outer tilted disc", 
as the ``ring" spirals in into the inner disc and the borders between the 
zones are not always well defined.}
The plane of the outer ring is dictated by the original AM of the dominant
stream, which could be tilted with respect to the inner disc 
\citep{danovich12}.
Strong torques, largely exerted by the inner disc, cause AM exchange that 
allows inflow toward the inner disc within less than an orbital time
and a gradual alignment with the inner disc.  
The AM lost from the inspiraling cold gas can go to outflowing gas and to 
the dark matter.
% IV
Finally, there are AM adjustments within the inner disc due to VDI and
outflows, which have not been addressed in detail here. 

% observability
\smallskip
The potential observable properties of the inflowing cold gas streams 
through the outer halo have been discussed elsewhere,
in absorption \citep{fumagalli11,faucher11,goerdt12,fumagalli14a},
and in Lyman-alpha emission \citep{loeb09,goerdt10}.
The most practical observable prediction is the existence of an extended
rotating ring, extending from $\sim\!10$ to $\sim\!30\kpc$.
The extended ring tends to be tilted relative to the inner disc plane, 
with the tilt increasing with radius, reaching an orientation 
that is only poorly correlated with the disc orientation.

% observability
\smallskip
The internal structure of the extended ring is expected to be
filamentary, clumpy, and the gas actually occupies only a fraction of the 
disc area. 
About 58\% of the lines of sight are of very low HI column densities.
However, about 30\% of the lines of sight are above a column density of 
$2\times 10^{20} {\rm H}\cms$ and are thus observable as damped Lyman-alpha 
systems.
The metallicity of the cold gas in the extended ring is predicted to be
about $Z\!\simeq\!0.1$ solar, in the ball park of the observed metallicities in
damped Lyman-alpha systems \citep{fumagalli14b}.
These estimates are valid on average for the mass and 
redshift ranges studied here and for the feedback used, and should therefore
be considered as a preliminary proof of concept.

% Preliminary observations  
\smallskip
There are preliminary observational indications consistent with an extended
ring.
\citet{bouche13} detected in a $z\!=\!2.3$ galaxy of an estimated virial mass
$10^{11}\msun$
a low-metallicity cold-gas absorption system at a projected distance of
$26\kpc$ from the center, namely
about a third of the virial radius, with a radial velocity of $180\kms$ 
relative to the center-of-mass velocity, namely consistent with
a close-to-circular motion.  
A detection of low-metallicity CGM gas at $54\kpc$ from the center of a
$z\!=\!2.4$ galaxy is also reported by
\citet{crighton_hennawi13}.
Observed MgII absorption lines at large radii in intermediate-redshift
galaxies ($z<1$) indicate velocities consistent with rotation in terms of 
magnitude and in terms of residing in one side of the galaxy systemic 
velocity, typically
aligned with one arm of the rotation curve \citep{steidel02,kacprzak10}.

% counter-rotation gas Conclusion
\smallskip
We found significant counter rotation in the incoming streams of cold gas,
at an average level of 75\% of its net AM amplitude, which in 15-20\% of the
cases flips the gas-disc spin in one disc orbital time. 
This can have important effects on stimulating VDI and triggering a compaction
into a ``blue nugget" followed by quenching to a ``red nugget" 
\citep{zolotov15}. Disc shrinkage due to counter rotation has been demonstrated
in idealized simulations of proto-stellar discs \citep{dyda15}.

% counter-rotation stellar  Conclusion
Counter-rotating stellar components are expected in the disc and its greater
environment, reminiscent of the distinct cold streams with opposite 
impact parameters.
Mutually counter-rotating streams tend to dissipate at inner radii
and coalesce into one coherently rotating disc, 
but stars formed prior to this coalescence may
end up in mutually counter-rotating discs in the same galaxy.
These could dominate at different radii and be tilted relative to each other,
or they may reside in the same disc structure.
Observational evidence for such counter-rotating discs are reported and
discussed 
\citep[][and references therein]{rubin92, prada99,navarro13,corsini14}.

% feedback Conclusion
\smallskip
Preliminary comparisons of the results from the current simulations and 
from a new suite of simulations with higher resolution, lower SFR efficiency 
and an additional radiative stellar feedback reveal that our main 
results are robust \se{feedback}. 
The stronger feedback tends to reduce the bulge-to-disc
ratio by $\sim\!20\%$ and sometimes make the disc more extended, 
with a spin parameter higher by $\sim\!50\%$ for the galaxy within the disc
radius $\Rd$, but only by $\sim\!8\%$ inside $0.1\Rv$.
However, the feedback-driven winds tend to flow perpendicular to the disc 
and have only weak interaction with the instreaming cold gas
(DeGraf et al., in preparation). They therefore have a rather weak
effect on the AM of the outer ring in the inner halo,
and an even weaker 
effect on the instreaming cold gas in the halo and outside it.
The sensitivity of the galaxy AM to feedback strength, including AGN feedback, 
is to be addressed in detail in a future study.
We note that feedback is likely to have a stronger affect on the AM 
in less massive galaxies, dwarfs or high-redshift building blocks of 
massive galaxies \citep[e.g.,][]{ds86,maller02,governato10}.

% future work
\smallskip
In this work we have made only limited steps toward a comprehensive study
of the presented picture for AM buildup in high-redshift massive galaxies.
Future work, analytic and using simulations, should complete the missing 
details.
For example, the acquisition of AM by linear tidal torques in phase I can be
performed in detail, analyzing the role played by the cosmic web structure
and the pre-collapse of gas into the filament cords.
The penetration of the supersonic streams through the outer halo, 
including the effects of shocks and of hydrodynamical, thermal and 
gravitational instabilities, would add to the
understanding of the transport of mass and AM in this zone.
The structure and kinematics of the extended ring in the messy interface
region should 
be investigated in more detail. Finally, the AM-changing processes within 
the inner disc, including SFR, feedback and VDI, should be addressed 
more thoroughly with high-resolution simulations that incorporate more
realistic feedback.

%%%%%%%%%%%%%%%%%%%%%%%%%%%%%%%%%
\section*{Acknowledgments}

We acknowledge stimulating discussions with F. Bournaud, J. Bullock, 
J. Devriendt, M. Fall, C. Pichon, D. Pogosian, and F. van den Bosch.
This research has been supported by ISF grant 24/12,
by GIF grant G-1052-104.7/2009,
by a DIP grant,
by NSF grant AST-1010033,
by the I-CORE Program of the PBC and the ISF grant 1829/12.
DC is a Juan de la Cierva fellow and he has been partly supported 
by MINECO grant AYA2012-31101.
The simulations were performed in the astro cluster at HU, in the National
Energy Research Scientific Computing Center (NERSC) at Lawrence Berkeley 
National Laboratory, and in NASA Advanced Supercomputing (NAS) at NASA Ames 
Research Center.

%%%%%%%%%%%%%%%%%%%%%%%%%%%%%%%%%%%%
\bibliographystyle{mn2e}
\bibliography{flows}

\begin{thebibliography}{}

\bibitem[\protect\citeauthoryear{{Agertz}, {Moore}, {Stadel}, {Potter},
  {Miniati}, {Read}, {Mayer} \& {et al.}}{{Agertz} et~al.}{2007}]{agertz07}
{Agertz} O.,  {Moore} B.,  {Stadel} J.,  {Potter} D.,  {Miniati} F.,  {Read}
  J.,  {Mayer} L.,    {et al.} 2007, \mnras, 380, 963

\bibitem[\protect\citeauthoryear{{Agertz}, {Teyssier} \& {Moore}}{{Agertz}
  et~al.}{2009}]{agertz09}
{Agertz} O.,  {Teyssier} R.,    {Moore} B.,  2009, \mnras, 397, L64

\bibitem[\protect\citeauthoryear{{Algorry}, {Navarro}, {Abadi}, {Sales},
  {Steinmetz} \& {Piontek}}{{Algorry} et~al.}{2013}]{navarro13}
{Algorry} D.~G.,  {Navarro} J.~F.,  {Abadi} M.~G.,  {Sales} L.~V.,  {Steinmetz}
  M.,    {Piontek} F.,  2013, ArXiv:1311.1215

\bibitem[\protect\citeauthoryear{{Barnes} \& {Hut}}{{Barnes} \&
  {Hut}}{1986}]{barnes86}
{Barnes} J.,  {Hut} P.,  1986, \nat, 324, 446

\bibitem[\protect\citeauthoryear{{Barro}, {Faber}, {Perez-Gonzalez}, {Koo},
  {Williams}, {Kocevski}, {...}, {Dekel} \& {et al.,}}{{Barro}
  et~al.}{2013}]{barro13}
{Barro} G.,  {Faber} S.~M.,  {Perez-Gonzalez} P.~G.,  {Koo} D.~C.,  {Williams}
  C.~C.,  {Kocevski} D.~D.,  {...} {Dekel} A.,    {et al.,} 2013, \apj, 765,
  104

\bibitem[\protect\citeauthoryear{{Barro}, {Faber}, {Perez-Gonzalez}, {Trump},
  {Koo}, {Wuyts}, {Guo}, {Dekel} \& {et al.,}}{{Barro} et~al.}{2014}]{barro14a}
{Barro} G.,  {Faber} S.~M.,  {Perez-Gonzalez} P.~G.,  {Trump} J.~R.,  {Koo}
  D.~C.,  {Wuyts} S.,  {Guo} Y.,  {Dekel} A.,    {et al.,} 2014,
  arXiv:1311:0000

\bibitem[\protect\citeauthoryear{{Bauer} \& {Springel}}{{Bauer} \&
  {Springel}}{2012}]{bauer12}
{Bauer} A.,  {Springel} V.,  2012, \mnras, 423, 2558

\bibitem[\protect\citeauthoryear{{Behroozi}, {Wechsler} \& {Conroy}}{{Behroozi}
  et~al.}{2013}]{behroozi13}
{Behroozi} P.~S.,  {Wechsler} R.~H.,    {Conroy} C.,  2013, \apjl, 762, L31

\bibitem[\protect\citeauthoryear{{Bett}, {Eke}, {Frenk}, {Jenkins}, {Helly} \&
  {Navarro}}{{Bett} et~al.}{2007}]{bett07}
{Bett} P.,  {Eke} V.,  {Frenk} C.~S.,  {Jenkins} A.,  {Helly} J.,    {Navarro}
  J.,  2007, \mnras, 376, 215

\bibitem[\protect\citeauthoryear{{Bett}, {Eke}, {Frenk}, {Jenkins} \&
  {Okamoto}}{{Bett} et~al.}{2010}]{bett10}
{Bett} P.,  {Eke} V.,  {Frenk} C.~S.,  {Jenkins} A.,    {Okamoto} T.,  2010,
  \mnras, 404, 1137

\bibitem[\protect\citeauthoryear{{Binney}}{{Binney}}{1977}]{binney77}
{Binney} J.,  1977, \apj, 215, 483

\bibitem[\protect\citeauthoryear{{Birnboim} \& {Dekel}}{{Birnboim} \&
  {Dekel}}{2003}]{bd03}
{Birnboim} Y.,  {Dekel} A.,  2003, \mnras, 345, 349

\bibitem[\protect\citeauthoryear{{Birnboim}, {Dekel} \& {Neistein}}{{Birnboim}
  et~al.}{2007}]{bdn07}
{Birnboim} Y.,  {Dekel} A.,    {Neistein} E.,  2007, \mnras, 380, 339

\bibitem[\protect\citeauthoryear{{Booth} \& {Schaye}}{{Booth} \&
  {Schaye}}{2009}]{booth09}
{Booth} C.~M.,  {Schaye} J.,  2009, \mnras, 398, 53

\bibitem[\protect\citeauthoryear{{Bouch{\'e}}, {Murphy}, {Kacprzak},
  {P{\'e}roux}, {Contini}, {Martin} \& {Dessauges-Zavadsky}}{{Bouch{\'e}}
  et~al.}{2013}]{bouche13}
{Bouch{\'e}} N.,  {Murphy} M.~T.,  {Kacprzak} G.~G.,  {P{\'e}roux} C.,
  {Contini} T.,  {Martin} C.~L.,    {Dessauges-Zavadsky} M.,  2013, Science,
  341, 50

\bibitem[\protect\citeauthoryear{{Bournaud}, {Dekel}, {Teyssier}, {Cacciato},
  {Daddi}, {Juneau} \& {Shankar}}{{Bournaud} et~al.}{2011}]{bournaud11}
{Bournaud} F.,  {Dekel} A.,  {Teyssier} R.,  {Cacciato} M.,  {Daddi} E.,
  {Juneau} S.,    {Shankar} F.,  2011, \apjl, 741, L33

\bibitem[\protect\citeauthoryear{{Bournaud}, {Elmegreen} \&
  {Elmegreen}}{{Bournaud} et~al.}{2007}]{bournaud07c}
{Bournaud} F.,  {Elmegreen} B.~G.,    {Elmegreen} D.~M.,  2007, \apj, 670, 237

\bibitem[\protect\citeauthoryear{{Brook}, {Governato}, {Ro{\v s}kar},
  {Stinson}, {Brooks}, {Wadsley}, {Quinn} \& {et al.}}{{Brook}
  et~al.}{2011}]{brook11}
{Brook} C.~B.,  {Governato} F.,  {Ro{\v s}kar} R.,  {Stinson} G.,  {Brooks}
  A.~M.,  {Wadsley} J.,  {Quinn} T.,    {et al.} 2011, \mnras, 415, 1051

\bibitem[\protect\citeauthoryear{{Brooks}, {Solomon}, {Governato}, {McCleary},
  {MacArthur}, {Brook}, {Jonsson}, {Quinn} \& {Wadsley}}{{Brooks}
  et~al.}{2011}]{brooks11}
{Brooks} A.~M.,  {Solomon} A.~R.,  {Governato} F.,  {McCleary} J.,  {MacArthur}
  L.~A.,  {Brook} C.~B.~A.,  {Jonsson} P.,  {Quinn} T.~R.,    {Wadsley} J.,
  2011, \apj, 728, 51

\bibitem[\protect\citeauthoryear{{Bullock}, {Dekel}, {Kolatt}, {Kravtsov},
  {Klypin}, {Porciani} \& {Primack}}{{Bullock} et~al.}{2001}]{bullock01_j}
{Bullock} J.~S.,  {Dekel} A.,  {Kolatt} T.~S.,  {Kravtsov} A.~V.,  {Klypin}
  A.~A.,  {Porciani} C.,    {Primack} J.~R.,  2001, \apj, 555, 240

\bibitem[\protect\citeauthoryear{{Cattaneo}, {Dekel}, {Devriendt}, {Guiderdoni}
  \& {Blaizot}}{{Cattaneo} et~al.}{2006}]{cattaneo06}
{Cattaneo} A.,  {Dekel} A.,  {Devriendt} J.,  {Guiderdoni} B.,    {Blaizot} J.,
   2006, \mnras, 370, 1651

\bibitem[\protect\citeauthoryear{{Cattaneo}, {Faber}, {Binney}, {Dekel},
  {Kormendy} \& {Mushotzky}}{{Cattaneo} et~al.}{2009}]{cattaneo09}
{Cattaneo} A.,  {Faber} S.~M.,  {Binney} J.,  {Dekel} A.,  {Kormendy} J.,
  {Mushotzky} R.,  2009, \nat, 460, 213

\bibitem[\protect\citeauthoryear{{Ceverino}, {Dekel} \& {Bournaud}}{{Ceverino}
  et~al.}{2010}]{cdb10}
{Ceverino} D.,  {Dekel} A.,    {Bournaud} F.,  2010, \mnras, 404, 2151

\bibitem[\protect\citeauthoryear{{Ceverino}, {Dekel}, {Mandelker}, {Bournaud},
  {Burkert}, {Genzel} \& {Primack}}{{Ceverino} et~al.}{2012}]{ceverino12}
{Ceverino} D.,  {Dekel} A.,  {Mandelker} N.,  {Bournaud} F.,  {Burkert} A.,
  {Genzel} R.,    {Primack} J.,  2012, \mnras, pp --2211

\bibitem[\protect\citeauthoryear{{Ceverino}, {Dekel}, {Tweed} \&
  {Primack}}{{Ceverino} et~al.}{2014}]{ceverino14_e}
{Ceverino} D.,  {Dekel} A.,  {Tweed} D.,    {Primack} J.,  2014,
  arXiv:1409.2622

\bibitem[\protect\citeauthoryear{{Ceverino} \& {Klypin}}{{Ceverino} \&
  {Klypin}}{2009}]{ceverino+k09}
{Ceverino} D.,  {Klypin} A.,  2009, \apj, 695, 292

\bibitem[\protect\citeauthoryear{{Ceverino}, {Klypin}, {Klimek},
  {Trujillo-Gomez}, {Churchill}, {Primack} \& {Dekel}}{{Ceverino}
  et~al.}{2014}]{ceverino14}
{Ceverino} D.,  {Klypin} A.,  {Klimek} E.~S.,  {Trujillo-Gomez} S.,
  {Churchill} C.~W.,  {Primack} J.,    {Dekel} A.,  2014, \mnras, 442, 1545

\bibitem[\protect\citeauthoryear{{Ciotti} \& {Ostriker}}{{Ciotti} \&
  {Ostriker}}{2007}]{ciotti07}
{Ciotti} L.,  {Ostriker} J.~P.,  2007, astro-ph/0703057

\bibitem[\protect\citeauthoryear{{Codis}, {Pichon}, {Devriendt}, {Slyz},
  {Pogosyan}, {Dubois} \& {Sousbie}}{{Codis} et~al.}{2012}]{codis12}
{Codis} S.,  {Pichon} C.,  {Devriendt} J.,  {Slyz} A.,  {Pogosyan} D.,
  {Dubois} Y.,    {Sousbie} T.,  2012, \mnras, 427, 3320

\bibitem[\protect\citeauthoryear{{Corsini}}{{Corsini}}{2014}]{corsini14}
{Corsini} E.~M.,  2014, ArXiv:1403.1263

\bibitem[\protect\citeauthoryear{{Crighton}, {Hennawi} \&
  {Prochaska}}{{Crighton} et~al.}{2013}]{crighton_hennawi13}
{Crighton} N.~H.~M.,  {Hennawi} J.~F.,    {Prochaska} J.~X.,  2013, \apjl, 776,
  L18

\bibitem[\protect\citeauthoryear{{Daddi}, {Bournaud}, {Walter}, {Dannerbauer},
  {Carilli}, {Dickinson}, {Elbaz}, {Morrison} \& {et al.}}{{Daddi}
  et~al.}{2010}]{daddi10}
{Daddi} E.,  {Bournaud} F.,  {Walter} F.,  {Dannerbauer} H.,  {Carilli} C.~L.,
  {Dickinson} M.,  {Elbaz} D.,  {Morrison} G.~E.,    {et al.} 2010, \apj, 713,
  686

\bibitem[\protect\citeauthoryear{{Danovich}, {Dekel}, {Hahn} \&
  {Teyssier}}{{Danovich} et~al.}{2012}]{danovich12}
{Danovich} M.,  {Dekel} A.,  {Hahn} O.,    {Teyssier} R.,  2012, \mnras, 422,
  1732

\bibitem[\protect\citeauthoryear{{DeGraf}, {Dekel}, {Gabor} \&
  {Bournaud}}{{DeGraf} et~al.}{2014}]{degraf15}
{DeGraf} C.,  {Dekel} A.,  {Gabor} J.,    {Bournaud} F.,  2014, arXiv:1412.3819

\bibitem[\protect\citeauthoryear{{Dekel} \& {Birnboim}}{{Dekel} \&
  {Birnboim}}{2006}]{db06}
{Dekel} A.,  {Birnboim} Y.,  2006, \mnras, 368, 2

\bibitem[\protect\citeauthoryear{{Dekel}, {Birnboim}, {Engel}, {Freundlich},
  {Goerdt}, {Mumcuoglu}, {Neistein}, {Pichon}, {Teyssier} \& {Zinger}}{{Dekel}
  et~al.}{2009}]{dekel09}
{Dekel} A.,  {Birnboim} Y.,  {Engel} G.,  {Freundlich} J.,  {Goerdt} T.,
  {Mumcuoglu} M.,  {Neistein} E.,  {Pichon} C.,  {Teyssier} R.,    {Zinger} E.,
   2009, \nat, 457, 451

\bibitem[\protect\citeauthoryear{{Dekel} \& {Burkert}}{{Dekel} \&
  {Burkert}}{2014}]{db14}
{Dekel} A.,  {Burkert} A.,  2014, arXiv:1310.1074

\bibitem[\protect\citeauthoryear{{Dekel} \& {Krumholz}}{{Dekel} \&
  {Krumholz}}{2013}]{dk13}
{Dekel} A.,  {Krumholz} M.~R.,  2013, \mnras, 432, 455

\bibitem[\protect\citeauthoryear{{Dekel}, {Sari} \& {Ceverino}}{{Dekel}
  et~al.}{2009}]{dsc09}
{Dekel} A.,  {Sari} R.,    {Ceverino} D.,  2009, \apj, 703, 785

\bibitem[\protect\citeauthoryear{{Dekel} \& {Silk}}{{Dekel} \&
  {Silk}}{1986}]{ds86}
{Dekel} A.,  {Silk} J.,  1986, \apj, 303, 39

\bibitem[\protect\citeauthoryear{{Dekel}, {Zolotov}, {Tweed}, {Cacciato},
  {Ceverino} \& {Primack}}{{Dekel} et~al.}{2013}]{dekel13}
{Dekel} A.,  {Zolotov} A.,  {Tweed} D.,  {Cacciato} M.,  {Ceverino} D.,
  {Primack} J.~R.,  2013, \mnras, 435, 999

\bibitem[\protect\citeauthoryear{{Dijkstra} \& {Loeb}}{{Dijkstra} \&
  {Loeb}}{2009}]{loeb09}
{Dijkstra} M.,  {Loeb} A.,  2009, ArXiv:

\bibitem[\protect\citeauthoryear{{Doroshkevich}}{{Doroshkevich}}{1970}]{doroshkevich70}
{Doroshkevich} A.~G.,  1970, Astrophysics, 6, 320

\bibitem[\protect\citeauthoryear{{Dutton} \& {van den Bosch}}{{Dutton} \& {van
  den Bosch}}{2012}]{dutton12}
{Dutton} A.~A.,  {van den Bosch} 2012, \mnras, 421, 608

\bibitem[\protect\citeauthoryear{{Dyda}, {Lovelace}, {Ustyugova}, {Romanova} \&
  {Koldoba}}{{Dyda} et~al.}{2015}]{dyda15}
{Dyda} S.,  {Lovelace} R.~V.~E.,  {Ustyugova} G.~V.,  {Romanova} M.~M.,
  {Koldoba} A.~V.,  2015, \mnras, 446, 613

\bibitem[\protect\citeauthoryear{{Fabian}}{{Fabian}}{2012}]{fabian12}
{Fabian} A.~C.,  2012, \araa, 50, 455

\bibitem[\protect\citeauthoryear{{Fall}}{{Fall}}{1983}]{fall83}
{Fall} S.~M.,  1983, in {Athanassoula} E.,  ed., Internal Kinematics and
  Dynamics of Galaxies Vol.~100 of IAU Symposium, Galaxy formation - some
  comparisons between theory and observation.
pp 391--398

\bibitem[\protect\citeauthoryear{{Fall} \& {Efstathiou}}{{Fall} \&
  {Efstathiou}}{1980}]{fe80}
{Fall} S.~M.,  {Efstathiou} G.,  1980, \mnras, 193, 189

\bibitem[\protect\citeauthoryear{{Fall} \& {Romanowsky}}{{Fall} \&
  {Romanowsky}}{2013}]{fall13}
{Fall} S.~M.,  {Romanowsky} A.~J.,  2013, \apjl, 769, L26

\bibitem[\protect\citeauthoryear{{Faucher-Gigu{\`e}re} \& {Kere{\v
  s}}}{{Faucher-Gigu{\`e}re} \& {Kere{\v s}}}{2011}]{faucher11}
{Faucher-Gigu{\`e}re} C.-A.,  {Kere{\v s}} D.,  2011, \mnras, 412, L118

\bibitem[\protect\citeauthoryear{{Fumagalli}, {Hennawi}, {Prochaska}, {Kasen},
  {Dekel}, {Ceverino} \& {Primack}}{{Fumagalli} et~al.}{2014}]{fumagalli14a}
{Fumagalli} M.,  {Hennawi} J.~F.,  {Prochaska} J.~X.,  {Kasen} D.,  {Dekel} A.,
   {Ceverino} D.,    {Primack} J.,  2014, \apj, 780, 74

\bibitem[\protect\citeauthoryear{{Fumagalli}, {O'Meara}, {Prochaska}, {Kanekar}
  \& {Wolfe}}{{Fumagalli} et~al.}{2014}]{fumagalli14b}
{Fumagalli} M.,  {O'Meara} J.~M.,  {Prochaska} J.~X.,  {Kanekar} N.,    {Wolfe}
  A.~M.,  2014, arXiv:1404.2599

\bibitem[\protect\citeauthoryear{{Fumagalli}, {Prochaska}, {Kasen}, {Dekel},
  {Ceverino} \& {Primack}}{{Fumagalli} et~al.}{2011}]{fumagalli11}
{Fumagalli} M.,  {Prochaska} J.~X.,  {Kasen} D.,  {Dekel} A.,  {Ceverino} D.,
   {Primack} J.~R.,  2011, \mnras, 418, 1796

\bibitem[\protect\citeauthoryear{{Genzel}, {Burkert}, {Bouch{\'e}}, {Cresci},
  {F{\"o}rster Schreiber}, {Shapley}, {Shapiro}, {Tacconi} \& {et
  al.,}}{{Genzel} et~al.}{2008}]{genzel08}
{Genzel} R.,  {Burkert} A.,  {Bouch{\'e}} N.,  {Cresci} G.,  {F{\"o}rster
  Schreiber} N.~M.,  {Shapley} A.,  {Shapiro} K.,  {Tacconi} L.~J.,    {et
  al.,} 2008, \apj, 687, 59

\bibitem[\protect\citeauthoryear{{Genzel}, {F{\"o}rster Schreiber}, {Lang},
  {Tacchella}, {Tacconi}, {Wuyts} \& {et al.}}{{Genzel}
  et~al.}{2013}]{genzel13}
{Genzel} R.,  {F{\"o}rster Schreiber} N.~M.,  {Lang} P.,  {Tacchella} S.,
  {Tacconi} L.~J.,  {Wuyts} S.,    {et al.} 2013, arXiv:1310.3838

\bibitem[\protect\citeauthoryear{{Genzel}, {Newman}, {Jones}, {F{\"o}rster
  Schreiber}, {Shapiro}, {Genel}, {Lilly} \& {et al.}}{{Genzel}
  et~al.}{2011}]{genzel11}
{Genzel} R.,  {Newman} S.,  {Jones} T.,  {F{\"o}rster Schreiber} N.~M.,
  {Shapiro} K.,  {Genel} S.,  {Lilly} S.~J.,    {et al.} 2011, \apj, 733, 101

\bibitem[\protect\citeauthoryear{{Genzel}, {Tacconi}, {Eisenhauer},
  {F{\"o}rster Schreiber}, {Cimatti}, {Daddi}, {Bouch{\'e}} \& {et
  al.}}{{Genzel} et~al.}{2006}]{genzel06}
{Genzel} R.,  {Tacconi} L.~J.,  {Eisenhauer} F.,  {F{\"o}rster Schreiber}
  N.~M.,  {Cimatti} A.,  {Daddi} E.,  {Bouch{\'e}} N.,    {et al.} 2006, \nat,
  442, 786

\bibitem[\protect\citeauthoryear{{Goerdt}, {Dekel}, {Sternberg}, {Ceverino},
  {Teyssier} \& {Primack}}{{Goerdt} et~al.}{2010}]{goerdt10}
{Goerdt} T.,  {Dekel} A.,  {Sternberg} A.,  {Ceverino} D.,  {Teyssier} R.,
  {Primack} J.~R.,  2010, \mnras, 407, 613

\bibitem[\protect\citeauthoryear{{Goerdt}, {Dekel}, {Sternberg}, {Gnat} \&
  {Ceverino}}{{Goerdt} et~al.}{2012}]{goerdt12}
{Goerdt} T.,  {Dekel} A.,  {Sternberg} A.,  {Gnat} O.,    {Ceverino} D.,  2012,
  \mnras, 424, 2292

\bibitem[\protect\citeauthoryear{{Governato}, {Brook}, {Mayer}, {Brooks},
  {Rhee}, {Wadsley}, {Jonsson}, {Willman}, {Stinson}, {Quinn} \&
  {Madau}}{{Governato} et~al.}{2010}]{governato10}
{Governato} F.,  {Brook} C.,  {Mayer} L.,  {Brooks} A.,  {Rhee} G.,  {Wadsley}
  J.,  {Jonsson} P.,  {Willman} B.,  {Stinson} G.,  {Quinn} T.,    {Madau} P.,
  2010, \nat, 463, 203

\bibitem[\protect\citeauthoryear{{Guedes}, {Callegari}, {Madau} \&
  {Mayer}}{{Guedes} et~al.}{2011}]{guedes11}
{Guedes} J.,  {Callegari} S.,  {Madau} P.,    {Mayer} L.,  2011, \apj, 742, 76

\bibitem[\protect\citeauthoryear{{Hahn}, {Teyssier} \& {Carollo}}{{Hahn}
  et~al.}{2010}]{hahn10}
{Hahn} O.,  {Teyssier} R.,    {Carollo} C.~M.,  2010, \mnras, 405, 274

\bibitem[\protect\citeauthoryear{{Hopkins}, {Hernquist}, {Cox}, {Robertson} \&
  {Springel}}{{Hopkins} et~al.}{2006}]{hopkins06}
{Hopkins} P.~F.,  {Hernquist} L.,  {Cox} T.~J.,  {Robertson} B.,    {Springel}
  V.,  2006, \apjs, 163, 50

\bibitem[\protect\citeauthoryear{{Hopkins}, {Kere{\v s}}, {Murray}, {Quataert}
  \& {Hernquist}}{{Hopkins} et~al.}{2012}]{hopkins12c}
{Hopkins} P.~F.,  {Kere{\v s}} D.,  {Murray} N.,  {Quataert} E.,    {Hernquist}
  L.,  2012, \mnras, 427, 968

\bibitem[\protect\citeauthoryear{{Hoyle}}{{Hoyle}}{1951}]{hoyle51}
{Hoyle} F.,  1951, in Problems of Cosmical Aerodynamics {The Origin of the
  Rotations of the Galaxies}.
p.~195

\bibitem[\protect\citeauthoryear{{Immeli}, {Samland}, {Westera} \&
  {Gerhard}}{{Immeli} et~al.}{2004}]{immeli04a}
{Immeli} A.,  {Samland} M.,  {Westera} P.,    {Gerhard} O.,  2004, \apj, 611,
  20

\bibitem[\protect\citeauthoryear{{Kacprzak}, {Churchill}, {Ceverino},
  {Steidel}, {Klypin} \& {Murphy}}{{Kacprzak} et~al.}{2010}]{kacprzak10}
{Kacprzak} G.~G.,  {Churchill} C.~W.,  {Ceverino} D.,  {Steidel} C.~C.,
  {Klypin} A.,    {Murphy} M.~T.,  2010, \apj, 711, 533

\bibitem[\protect\citeauthoryear{{Kennicutt}
  Jr.}{{Kennicutt}}{1998}]{kennicutt98}
{Kennicutt} Jr. R.~C.,  1998, \apj, 498, 541

\bibitem[\protect\citeauthoryear{{Kere{\v s}}, {Katz}, {Fardal}, {Dav{\'e}} \&
  {Weinberg}}{{Kere{\v s}} et~al.}{2009}]{keres09}
{Kere{\v s}} D.,  {Katz} N.,  {Fardal} M.,  {Dav{\'e}} R.,    {Weinberg} D.~H.,
   2009, \mnras, 395, 160

\bibitem[\protect\citeauthoryear{{Kere{\v s}}, {Katz}, {Weinberg} \&
  {Dav{\'e}}}{{Kere{\v s}} et~al.}{2005}]{keres05}
{Kere{\v s}} D.,  {Katz} N.,  {Weinberg} D.~H.,    {Dav{\'e}} R.,  2005,
  \mnras, 363, 2

\bibitem[\protect\citeauthoryear{{Kimm}, {Devriendt}, {Slyz}, {Pichon},
  {Kassin} \& {Dubois}}{{Kimm} et~al.}{2011}]{kimm11}
{Kimm} T.,  {Devriendt} J.,  {Slyz} A.,  {Pichon} C.,  {Kassin} S.~A.,
  {Dubois} Y.,  2011, arXiv:1106.0538

\bibitem[\protect\citeauthoryear{{Komatsu}, {Dunkley}, {Nolta}, {Bennett},
  {Gold}, {Hinshaw}, {Jarosik}, {Larson} \& {et al.}}{{Komatsu}
  et~al.}{2008}]{komatsu08_wmap5}
{Komatsu} E.,  {Dunkley} J.,  {Nolta} M.~R.,  {Bennett} C.~L.,  {Gold} B.,
  {Hinshaw} G.,  {Jarosik} N.,  {Larson} D.,    {et al.} 2008, ArXiv:

\bibitem[\protect\citeauthoryear{{Kravtsov}}{{Kravtsov}}{2003}]{krav03}
{Kravtsov} A.~V.,  2003, \apjl, 590, L1

\bibitem[\protect\citeauthoryear{{Kravtsov}, {Klypin} \& {Khokhlov}}{{Kravtsov}
  et~al.}{1997}]{krav97}
{Kravtsov} A.~V.,  {Klypin} A.~A.,    {Khokhlov} A.~M.,  1997, \apjs, 111, 73

\bibitem[\protect\citeauthoryear{{Krumholz} \& {Dekel}}{{Krumholz} \&
  {Dekel}}{2010}]{kd10}
{Krumholz} M.~R.,  {Dekel} A.,  2010, \mnras, 406, 112

\bibitem[\protect\citeauthoryear{{Krumholz} \& {Dekel}}{{Krumholz} \&
  {Dekel}}{2012}]{kd12}
{Krumholz} M.~R.,  {Dekel} A.,  2012, \apj, 753, 16

\bibitem[\protect\citeauthoryear{{Maller} \& {Dekel}}{{Maller} \&
  {Dekel}}{2002}]{maller02}
{Maller} A.~H.,  {Dekel} A.,  2002, \mnras, 335, 487

\bibitem[\protect\citeauthoryear{{Maller}, {Dekel} \& {Somerville}}{{Maller}
  et~al.}{2002}]{maller_somer02}
{Maller} A.~H.,  {Dekel} A.,    {Somerville} R.,  2002, \mnras, 329, 423

\bibitem[\protect\citeauthoryear{{Mandelker}, {Dekel}, {Ceverino}, {Tweed},
  {Moody} \& {Primack}}{{Mandelker} et~al.}{2014}]{mandelker14}
{Mandelker} N.,  {Dekel} A.,  {Ceverino} D.,  {Tweed} D.,  {Moody} C.~E.,
  {Primack} J.,  2014, arXiv:1311.0013

\bibitem[\protect\citeauthoryear{{Mo}, {Mao} \& {White}}{{Mo}
  et~al.}{1998}]{mmw98}
{Mo} H.~J.,  {Mao} S.,    {White} S.~D.~M.,  1998, \mnras, 295, 319

\bibitem[\protect\citeauthoryear{{Moody}, {Guo}, {Mandelker}, {Ceverino},
  {Mozena}, {Koo}, {Dekel} \& {Primack}}{{Moody} et~al.}{2014}]{moody14}
{Moody} C.~E.,  {Guo} Y.,  {Mandelker} N.,  {Ceverino} D.,  {Mozena} M.,  {Koo}
  D.~C.,  {Dekel} A.,    {Primack} J.,  2014, arXiv:1405.5266

\bibitem[\protect\citeauthoryear{{Moster}, {Naab} \& {White}}{{Moster}
  et~al.}{2013}]{moster13}
{Moster} B.~P.,  {Naab} T.,    {White} S.~D.~M.,  2013, \mnras, 428, 3121

\bibitem[\protect\citeauthoryear{{Murray}, {Quataert} \& {Thompson}}{{Murray}
  et~al.}{2010}]{murray10}
{Murray} N.,  {Quataert} E.,    {Thompson} T.~A.,  2010, \apj, 709, 191

\bibitem[\protect\citeauthoryear{{Navarro}, {Abadi} \& {Steinmetz}}{{Navarro}
  et~al.}{2004}]{navarro04}
{Navarro} J.~F.,  {Abadi} M.~G.,    {Steinmetz} M.,  2004, \apjl, 613, L41

\bibitem[\protect\citeauthoryear{{Nelson}, {Vogelsberger}, {Genel}, {Sijacki},
  {Kere{\v s}}, {Springel} \& {Hernquist}}{{Nelson} et~al.}{2013}]{nelson13}
{Nelson} D.,  {Vogelsberger} M.,  {Genel} S.,  {Sijacki} D.,  {Kere{\v s}} D.,
  {Springel} V.,    {Hernquist} L.,  2013, \mnras, 429, 3353

\bibitem[\protect\citeauthoryear{{Noguchi}}{{Noguchi}}{1999}]{noguchi99}
{Noguchi} M.,  1999, \apj, 514, 77

\bibitem[\protect\citeauthoryear{{Ocvirk}, {Pichon} \& {Teyssier}}{{Ocvirk}
  et~al.}{2008}]{ocvirk08}
{Ocvirk} P.,  {Pichon} C.,    {Teyssier} R.,  2008, \mnras, 390, 1326

\bibitem[\protect\citeauthoryear{{Peebles}}{{Peebles}}{1969}]{peebles69}
{Peebles} P.~J.~E.,  1969, \apj, 155, 393

\bibitem[\protect\citeauthoryear{{P{\'e}rez-Gonz{\'a}lez}, {Rieke}, {Villar},
  {Barro}, {Blaylock}, {Egami}, {Gallego} \& {et al.}}{{P{\'e}rez-Gonz{\'a}lez}
  et~al.}{2008}]{perezgonzalez08}
{P{\'e}rez-Gonz{\'a}lez} P.~G.,  {Rieke} G.~H.,  {Villar} V.,  {Barro} G.,
  {Blaylock} M.,  {Egami} E.,  {Gallego} J.,    {et al.} 2008, \apj, 675, 234

\bibitem[\protect\citeauthoryear{{Pichon}, {Pogosyan}, {Kimm}, {Slyz},
  {Devriendt} \& {Dubois}}{{Pichon} et~al.}{2011}]{pichon11}
{Pichon} C.,  {Pogosyan} D.,  {Kimm} T.,  {Slyz} A.,  {Devriendt} J.,
  {Dubois} Y.,  2011, \mnras, 418, 2493

\bibitem[\protect\citeauthoryear{{Porciani}, {Dekel} \& {Hoffman}}{{Porciani}
  et~al.}{2002a}]{porciani02a}
{Porciani} C.,  {Dekel} A.,    {Hoffman} Y.,  2002a, \mnras, 332, 325

\bibitem[\protect\citeauthoryear{{Porciani}, {Dekel} \& {Hoffman}}{{Porciani}
  et~al.}{2002b}]{porciani02b}
{Porciani} C.,  {Dekel} A.,    {Hoffman} Y.,  2002b, \mnras, 332, 339

\bibitem[\protect\citeauthoryear{{Powell}, {Slyz} \& {Devriendt}}{{Powell}
  et~al.}{2011}]{powell11}
{Powell} L.~C.,  {Slyz} A.,    {Devriendt} J.,  2011, \mnras, 414, 3671

\bibitem[\protect\citeauthoryear{{Prada} \& {Guti{\'e}rrez}}{{Prada} \&
  {Guti{\'e}rrez}}{1999}]{prada99}
{Prada} F.,  {Guti{\'e}rrez} C.~M.,  1999, \apj, 517, 123

\bibitem[\protect\citeauthoryear{{Prada}, {Klypin}, {Simonneau},
  {Betancort-Rijo}, {Patiri}, {Gottl{\"o}ber} \& {Sanchez-Conde}}{{Prada}
  et~al.}{2006}]{prada06}
{Prada} F.,  {Klypin} A.~A.,  {Simonneau} E.,  {Betancort-Rijo} J.,  {Patiri}
  S.,  {Gottl{\"o}ber} S.,    {Sanchez-Conde} M.~A.,  2006, \apj, 645, 1001

\bibitem[\protect\citeauthoryear{{Rees} \& {Ostriker}}{{Rees} \&
  {Ostriker}}{1977}]{ro77}
{Rees} M.~J.,  {Ostriker} J.~P.,  1977, \mnras, 179, 541

\bibitem[\protect\citeauthoryear{{Romanowsky} \& {Fall}}{{Romanowsky} \&
  {Fall}}{2012}]{romanowsky12}
{Romanowsky} A.~J.,  {Fall} S.~M.,  2012, \apjs, 203, 17

\bibitem[\protect\citeauthoryear{{Rubin}, {Graham} \& {Kenney}}{{Rubin}
  et~al.}{1992}]{rubin92}
{Rubin} V.~C.,  {Graham} J.~A.,    {Kenney} J.~D.~P.,  1992, \apjl, 394, L9

\bibitem[\protect\citeauthoryear{{Scannapieco}, {Wadepuhl}, {Parry}, {Navarro},
  {Jenkins}, {Springel}, {Teyssier} \& {et al.}}{{Scannapieco}
  et~al.}{2012}]{scannapieco12}
{Scannapieco} C.,  {Wadepuhl} M.,  {Parry} O.~H.,  {Navarro} J.~F.,  {Jenkins}
  A.,  {Springel} V.,  {Teyssier} R.,    {et al.} 2012, \mnras, 423, 1726

\bibitem[\protect\citeauthoryear{{Sharma} \& {Steinmetz}}{{Sharma} \&
  {Steinmetz}}{2005}]{sharma05}
{Sharma} S.,  {Steinmetz} M.,  2005, \apj, 628, 21

\bibitem[\protect\citeauthoryear{{Silk}}{{Silk}}{1977}]{silk77}
{Silk} J.,  1977, \apj, 211, 638

\bibitem[\protect\citeauthoryear{{Silk} \& {Rees}}{{Silk} \&
  {Rees}}{1998}]{silk98}
{Silk} J.,  {Rees} M.~J.,  1998, \aap, 331, L1

\bibitem[\protect\citeauthoryear{{Steidel}, {Erb}, {Shapley}, {Pettini},
  {Reddy}, {Bogosavljevi{\'c}}, {Rudie} \& {Rakic}}{{Steidel}
  et~al.}{2010}]{steidel10}
{Steidel} C.~C.,  {Erb} D.~K.,  {Shapley} A.~E.,  {Pettini} M.,  {Reddy} N.,
  {Bogosavljevi{\'c}} M.,  {Rudie} G.~C.,    {Rakic} O.,  2010, \apj, 717, 289

\bibitem[\protect\citeauthoryear{{Steidel}, {Kollmeier}, {Shapley},
  {Churchill}, {Dickinson} \& {Pettini}}{{Steidel} et~al.}{2002}]{steidel02}
{Steidel} C.~C.,  {Kollmeier} J.~A.,  {Shapley} A.~E.,  {Churchill} C.~W.,
  {Dickinson} M.,    {Pettini} M.,  2002, \apj, 570, 526

\bibitem[\protect\citeauthoryear{{Stewart}, {Brooks}, {Bullock}, {Maller},
  {Diemand}, {Wadsley} \& {Moustakas}}{{Stewart} et~al.}{2013}]{stewart13}
{Stewart} K.~R.,  {Brooks} A.~M.,  {Bullock} J.~S.,  {Maller} A.~H.,  {Diemand}
  J.,  {Wadsley} J.,    {Moustakas} L.~A.,  2013, \apj, 769, 74

\bibitem[\protect\citeauthoryear{{Stewart}, {Kaufmann}, {Bullock}, {Barton},
  {Maller}, {Diemand} \& {Wadsley}}{{Stewart} et~al.}{2011}]{stewart11}
{Stewart} K.~R.,  {Kaufmann} T.,  {Bullock} J.~S.,  {Barton} E.~J.,  {Maller}
  A.~H.,  {Diemand} J.,    {Wadsley} J.,  2011, \apj, 738, 39

\bibitem[\protect\citeauthoryear{{Tacconi}, {Genzel}, {Neri}, {Cox}, {Cooper},
  {Shapiro}, {Bolatto}, {Bouch{\'e}} \& {et al.}}{{Tacconi}
  et~al.}{2010}]{tacconi10}
{Tacconi} L.~J.,  {Genzel} R.,  {Neri} R.,  {Cox} P.,  {Cooper} M.~C.,
  {Shapiro} K.,  {Bolatto} A.,  {Bouch{\'e}} N.,    {et al.} 2010, \nat, 463,
  781

\bibitem[\protect\citeauthoryear{{Tillson}, {Devriendt}, {Slyz}, {Miller} \&
  {Pichon}}{{Tillson} et~al.}{2012}]{tillson12}
{Tillson} H.,  {Devriendt} J.,  {Slyz} A.,  {Miller} L.,    {Pichon} C.,  2012,
  arXiv:1211.3124

\bibitem[\protect\citeauthoryear{{White}}{{White}}{1984}]{white84}
{White} S.~D.~M.,  1984, \apj, 286, 38

\bibitem[\protect\citeauthoryear{{Zolotov}, {Dekel} \& {et al.,}}{{Zolotov}
  et~al.}{2014}]{zolotov15}
{Zolotov} A.,  {Dekel} A.,    {et al.,} 2014, arXiv:1412.4783

\end{thebibliography}

%%%%%%%%%%%%%%%%%%%%%%%%%%%%%
\bsp

\label{lastpage}

\end{document}